\newcommand \Mpc {h^{-1}{\rm Mpc}}
\newcommand \kpc {h^{-1}{\rm kpc}}

\newcommand \arcs{\hbox{$^{\prime\prime}$}}

\newcommand \kms {{\rm km~s}^{-1}}
\newcommand \msun {h^{-1} M_\odot}
\newcommand \mlsun {h M_\odot/L_\odot}
\newcommand \beqn {\begin{equation}}
\newcommand \eeqn {\end{equation}}

\newcommand \ncziinew {515 }

\documentclass{aastex}
\usepackage{emulateapj5}
\usepackage{epsfig}
\usepackage{natbib}

\begin{document}

\title{CAIRNS: The Cluster And Infall Region Nearby Survey
II. Environmental Dependence of Infrared Mass-to-Light Ratios}

\author{Kenneth Rines\altaffilmark{1,2}, Margaret J. Geller\altaffilmark{3},
Antonaldo Diaferio\altaffilmark{4}, Michael J. Kurtz\altaffilmark{3}, and Thomas H. Jarrett\altaffilmark{5}} 
\email{krines@cfa.harvard.edu}

\altaffiltext{1}{Yale Center for Astronomy and Astrophysics, Yale University, P.O. Box 208121, New Haven CT 06520-8121; krines@astro.yale.edu}
\altaffiltext{2}{Harvard-Smithsonian Center for Astrophysics, 60 Garden St,
Cambridge, MA 02138}
\altaffiltext{3}{Smithsonian Astrophysical Observatory; mgeller, mkurtz@cfa.harvard.edu}
\altaffiltext{4}{Universit\`a degli Studi di Torino,
Dipartimento di Fisica Generale ``Amedeo Avogadro'', Torino, Italy; diaferio@ph.unito.it}
\altaffiltext{5}{IPAC/Caltech 100-22 Pasadena, CA 91225; jarrett@ipac.caltech.edu}

\begin{abstract}

CAIRNS (Cluster And Infall Region Nearby Survey) is a spectroscopic
survey of the infall regions surrounding nine nearby rich clusters of
galaxies.  In Paper I, we used redshifts within $\sim 10\Mpc$ of the
centers of the clusters to determine the mass profiles of the clusters
based on the phase space distribution of the galaxies.  Here, we use
2MASS photometry and an additional \ncziinew redshifts to investigate
the environmental dependence of near-infrared mass-to-light ratios.
In the virial regions, the halo occupation function is non-linear; the
number of bright galaxies per halo increases more slowly than the mass
of the halo.  On larger scales, the light contained in galaxies is
less clustered than the mass in rich clusters.  Specifically, the
mass-to-light ratio inside the virial radius is a factor of
$1.8\pm0.3$ larger than that outside the virial radius.  This
difference could result from changing fractions of baryonic to total
matter or from variations in the efficiency of galaxy formation or
disruption with environment.  The average mass-to-light ratio $M/L_K =
53\pm 5 h$ implies $\Omega _m = 0.18\pm 0.03$ (statistical) using the
luminosity density based on 2dFGRS data.  These results are difficult
to reconcile with independent methods which suggest higher $\Omega
_m$.  Reconciling these values by invoking bias requires that the
typical value of $M/L_K$ changes significantly at densities of
$\lesssim$3$\rho_c$.

\end{abstract}


\section{Introduction}

The relative distribution of matter and light in the universe is one
of the outstanding problems in astrophysics.  Clusters of galaxies,
the largest gravitationally relaxed objects in the universe, are
important probes of the distribution of mass and
light. \citet[][]{zwicky1933} first computed the mass-to-light ratio
of the Coma cluster using the virial theorem and found that dark
matter dominates the cluster mass.  Recent determinations using the
virial theorem yield mass-to-light ratios of $M/L_{B_j}\sim 250
\mlsun$ \citep[][and references therein]{g2000}.  Equating the
mass-to-light ratio in clusters to the global value provides an
estimate of the mass density of the universe \citep{oort58}; this
estimate is subject to significant systematic error introduced by
differences in galaxy populations between cluster cores and lower
density regions \citep{cye97,g2000}.  Indeed, some numerical
simulations suggest that cluster mass-to-light ratios exceed the
universal value \citep[][but see also Ostriker et
al.~2003]{diaferio1999,kk1999,bahcall2000,benson00}.

Determining the global matter density from cluster mass-to-light
ratios therefore requires knowledge of the dependence of mass-to-light
ratios on environment.  \citet{bld95} show that mass-to-light ratios
increase with scale from galaxies to groups to clusters.  Ellipticals
have larger overall values of $M/L_B$ than spirals, presumably a
result of younger, bluer stellar populations in spirals.
At the scale of cluster virial radii, mass-to-light ratios appear to
reach a maximum value.  Some estimates of the mass-to-light ratio on
very large scales ($>$10$\Mpc$) are available \citep[see references
in][]{bld95}, but the systematic uncertainties are large.

There are few estimates of mass-to-light ratios on scales between
cluster virial radii and scales of 10$\Mpc$
\citep[][]{elt97,small98,kaiserxx,rines2000,rines01a,bg03,katgert03,kneib03}.
On these scales, many galaxies near clusters are bound to the cluster
but not yet in equilibrium \citep{gunngott}.  These cluster infall
regions have received relatively little scrutiny because they are
mildly nonlinear, making their properties very difficult to predict
analytically.  However, these scales are exactly the ones in which
galaxy properties change dramatically \citep[][and references
therein]{ellingson01,lewis02,gomez03,treu03,balogh03}.  Variations in
the mass-to-light ratio with environment could have important physical
implications; they could be produced either by a varying dark matter
fraction or by variations in the efficiency of star formation with
environment.  In blue light, however, higher star formation rates in
field galaxies compared to cluster galaxies could produce lower
mass-to-light ratios outside cluster cores resulting only from the
different contributions of young and old stars to the total
luminosity \citep{bahcall2000}.

Because clusters are not in equilibrium outside the virial radius,
neither X-ray observations nor Jeans analysis provide secure mass
determinations at these large radii. There are now two methods of
approaching this problem: weak gravitational lensing \citep{kaiserxx}
and kinematics of the infall region \citep[][hereafter DG97 and
D99]{dg97,diaferio1999}. \citet{kaiserxx} analyzed the weak lensing
signal from a supercluster at $z \approx 0.4$; the mass-to-light ratio
($M/L_B$=280 $\pm$40 $h$ for early-type galaxy light) is constant on
scales up to $6~\Mpc$. \citet{wilson01} finds similar results for weak
lensing in blank fields; \citet{gray02} obtain similar results
for a different supercluster.  Recently, \citet{kneib03} used weak
lensing to estimate the mass profile of CL0024+1654 to a radius of
3.25$\Mpc$.  \citet{kneib03} conclude that the mass-to-light ratio is
roughly constant on these scales.

          Galaxies in cluster infall regions produce sharp features in
redshift surveys \citep{kg82,shectman82,dgh86,kais87,ostriker88,rg89}.
Early investigations of this infall pattern focused on its use as a
direct indicator of the global matter density $\Omega_m$.
Unfortunately, random motions caused by galaxy-galaxy interactions and
substructure within the infall region smear out this cosmological
signal \citep[DG97, ][]{vh98}.  Instead of sharp peaks in redshift
space, infall regions around real clusters typically display a
well-defined envelope in redshift space which is significantly denser
than the surrounding environment \citep[][hereafter Paper I, and
references therein]{cairnsi}.
 
DG97 analyzed the dynamics of infall regions with numerical simulations
and found that in the outskirts of clusters, random motions due to
substructure and non-radial motions make a substantial contribution to
the amplitude of the caustics which delineate the infall regions
\citep[see also][and references therein]{vh98}. DG97
showed that the amplitude of the caustics is a measure of the escape
velocity from the cluster; identification of the caustics therefore
allows a determination of the mass profile of the cluster on scales
$\lesssim 10\Mpc$.

DG97 and D99 show that nonparametric measurements of caustics yield
cluster mass profiles accurate to $\sim$50\% on scales of up to 10
$h^{-1}$ Mpc.  This method assumes only that galaxies trace the
velocity field. Indeed, simulations suggest that little or no velocity
bias exists on linear and mildly non-linear scales
\citep{kauffmann1999a,kauffmann1999b}.  \citet[][hereafter
GDK]{gdk99}, applied the kinematic method of D99 to the
infall region of the Coma cluster.  GDK reproduced the X-ray derived
mass profile and extended direct determinations of the mass profile to
a radius of $10~\Mpc$.  The caustic method has also been applied to
the Shapley Supercluster
\citep{rqcm}, A576 \citep[][hereafter R00]{rines2000}, AWM7
\citep{kg2000}, the Fornax cluster \citep{drink}, A1644
\citep{tustin}, A2199 \citep{rines02}, and six other nearby clusters
(Paper I).  \citet{bg03} applied the caustic technique to an ensemble
cluster created by stacking redshifts around 43 clusters from the 2dF
Galaxy Redshift Survey.  R00 found an enclosed mass-to-light ratio of
$M/L_R \sim 300 h$ within 4$~\Mpc$ of A576.  \citet{rines01a} used
2MASS photometry and the mass profile from GDK to compute the
mass-to-light profile of Coma in the K-band.  They found a roughly
flat profile with a possible decrease in $M/L_K$ with radius by no
more than a factor of 3.  \citet{bg03} find a decreasing ratio of mass
density to total galaxy number density.  For early-type galaxies only,
the number density profile is consistent with a constant mass-to-light
(actually mass-to-number) ratio.

Here, we calculate the infrared mass-to-light profile within the
turnaround radius for the CAIRNS clusters (Paper I), a sample of nine
nearby rich, X-ray luminous clusters.  We use photometry from the Two
Micron All Sky Survey \citep[2MASS,][]{twomass} and add several new
redshifts to obtain complete or nearly complete surveys of galaxies up
to 1-2 magnitudes fainter than $M^*_{K_s}$ \citep[as determined
by][hereafter K01 and C01]{twomasslfn,twomdflfn}.  Infrared light is a
better tracer of stellar mass than optical light
\citep[][]{gpb96,zibetti02}; it is relatively insensitive to dust
extinction and recent star formation.  Despite these advantages, there
are very few measurements of infrared mass-to-light ratios in clusters
\citep{tustin,rines01a,lin03}.

Mass-to-light ratios within virial regions (where the masses
are more accurate than in the infall regions) provide interesting
constraints on the distribution of dark matter and stellar mass
\citep[see also][hereafter L03]{lin03}.  The virial
masses in our sample span an order of magnitude in mass.  
More massive clusters have larger mass-to-light ratios.

Cluster virial regions also provide potentially important constraints
on the halo occupation distribution \citep[e.g.,][and references
therein]{peacock00,berlind02,berlind03}, the number of galaxies in a
halo of a given mass \citep[see ][for a recent
review]{2002PhR...372....1C}.  The motions of galaxies and hot gas
yield estimates of the dynamical mass independent of the number of
galaxies (provided enough galaxies are present to yield a virial
mass).  Our mass profiles in Paper I are among the first to extend
significantly beyond $r_{200}$.  Thus, they should provide accurate
estimates of $r_{200}$.  Also, the recent release of 2MASS allows us
to count galaxies based on their near-infrared light, which is close
to selecting galaxies by stellar mass.  Thus, both the masses and
galaxy numbers are better defined than the few previous direct
estimates of the halo occupation function
\citep{peacock00,marinoni02,lin04}.

We describe the cluster sample, the near-infrared photometry, and the
spectroscopic observations in $\S$ 2.  
We discuss the galaxy properties (luminosity functions and broadband
colors) within and outside the virial radius and compare both
populations to field galaxies in $\S 4$.  We calculate the number
density and luminosity density profiles in $\S 5$ and compare them to
simple theoretical models.  We compute radial profiles of the
mass-to-light ratio in $\S 5$.  In $\S 6$ we constrain the halo
occupation distribution for the CAIRNS clusters and explore the
dependence of mass-to-light ratios on halo mass.  We discuss possible
systematic uncertainties and the implications of our results in $\S 7$
and conclude in $\S 8$.  We assume a cosmology of $H_0 = 100~h~\kms,
\Omega_m = 0.3, \Omega _\Lambda = 0.7$ except as noted in $\S 7$.

\section{Observations}

\subsection{The CAIRNS Cluster Sample}

We selected the CAIRNS parent sample from all nearby
($cz<15,000~\kms$), Abell richness class $R\geq1$ \citep[][]{aco1989},
X-ray luminous ($L_X>2.5 \times 10^{43} h^{-2}$erg s$^{-1}$) galaxy
clusters with declination $\delta>-15^\circ$. Using X-ray data from
the X-ray Brightest Abell Clusters catalog \citep{xbacs}, the parent
cluster sample contains 14 systems.  We selected a representative
sample of 8 of these 14 clusters (Table \ref{sample}).  The cluster properties listed in Table
\ref{sample} are from Paper I.  The 6 clusters meeting the
selection criteria but not targeted in CAIRNS are: A193, A426, A2063,
A2107, A2147, and A2657.  The 8 CAIRNS clusters span a variety of
morphologies, from isolated clusters (A496, A2199) to major mergers
(A168, A1367).  

\begin{table*}[th] \footnotesize
\begin{center}
\caption{\label{sample} \sc CAIRNS Parent Population}
\begin{tabular}{lcccccccc}
\tableline
\tableline
\tablewidth{0pt}
Cluster &\multicolumn{2}{c}{X-ray Coordinates} & $cz_\odot$ & $\sigma
 _p$ & $L_X$ & $T_X$ & Richness  \\ 
 & RA (J2000) & DEC (J2000) &  $\kms$ & $\kms$ & $10^{43}
 h^{-2}$~ergs~s$^{-1}$& keV  \\ 
\tableline
A119 & 00 56 12.9 & -01 14 06 & 13268 &698$^{+36}_{-31}$ & 8.1 & 5.1 & 1  \\
A168 & 01 15 08.8 & +00 21 14 & 13395 &579$^{+36}_{-30}$ & 2.7  & 2.6 & 2  \\
A496 & 04 33 35.2 & -13 14 45 & 9900 &721$^{+35}_{-30}$ & 8.9  & 4.7 & 1  \\
A539 & 05 16 32.1 & +06 26 31 & 8717 &734$^{+53}_{-44}$ & 2.7  & 3.0 & 1 \\ 
A576 & 07 21 31.6 & +55 45 50 & 11510 &1009$^{+41}_{-36}$ & 3.5 & 3.7 & 1  \\
A1367 & 11 44 36.2 & +19 46 19 & 6495 &782$^{+56}_{-46}$ & 4.1  & 3.5 & 2  \\ 
Coma & 12 59 31.9 & +27 54 10 & 6973 &1042$^{+33}_{-30}$ & 18.0  & 8.0 & 2  \\
A2199 & 16 28 39.5 & +39 33 00 & 9101 &796$^{+38}_{-33}$ & 9.1  & 4.7 & 2  \\
\tableline
A194 & 01 25 50.4 & -01 21 54 & 5341 &495$^{+41}_{-33}$ & 0.4  & 2.6 & 0  \\
\tableline
\end{tabular}
\end{center}
\end{table*}

\begin{table*}[th] \footnotesize
\begin{center}
\caption{\label{centers} \sc CAIRNS Hierarchical Centers}
\begin{tabular}{lccccc}
\tableline
\tableline
\tablewidth{0pt}
Cluster &\multicolumn{2}{c}{Hierarchical Center} & $cz_{cen}$ & $cz_{CMB}$ & $\Delta R$  \\ 
 & RA (J2000) & DEC (J2000) & $\kms$ & $\kms$ & $\kpc$  \\ 
\tableline
A119 & 00 56 10.1 & -01 15 20 & 13278 & 12948 & 56  \\
A168 & 01 15 00.7 & +00 15 31 & 13493 & 13176 & 239  \\
A496 & 04 33 38.6 & -13 15 47 & 9831 & 9786 & 24  \\
A539 & 05 16 37.0 & +06 26 57 & 8648 & 8650 & 33 \\ 
A576 & 07 21 32.0 & +55 45 21 & 11487 & 11561 & 16  \\
A1367 & 11 44 49.1 & +19 46 03 & 6509 & 6837 & 61  \\ 
Coma & 13 00 00.7 & +27 56 51 & 7096 & 7365 & 153  \\
A2199 & 16 28 47.0 & +39 30 22 & 9156 & 9181 & 86  \\
\tableline
A194 & 01 25 48.0 & -01 21 34 & 5317 & 5011 & 11  \\
\tableline
\end{tabular}
\end{center}
\end{table*}

The redshift limit is set by the small aperture of the 1.5-m
Tillinghast telescope used for the vast majority of our spectroscopic
observations. The richness minimum guarantees that the systems contain
sufficiently large numbers of galaxies to sample the velocity
distribution.  The X-ray luminosity minimum guarantees that the
systems are real clusters and not superpositions of galaxy groups
\citep[cf. the discussion of A2197 in][]{rines01a, rines02}.  
Three additional clusters with smaller X-ray luminosities (A147, A194
and A2197) serendipitously lie in the survey regions of A168 and
A2199.  A147 and A2197 lie at nearly identical redshifts to A168 and
A2199; their dynamics are probably dominated by the more massive
cluster \citep{rines02}.  A194, however, is cleanly separated from
A168 and we therefore analyze it as a ninth system.  The inclusion of
A194 extends the parameter space covered by the CAIRNS sample.  The
X-ray temperature of A194 listed in \citep{xbacs} is an extrapolation
of the $L_X - T_X$ relation; in Table \ref{sample} we therefore list the
direct temperature estimate of \citet{1998PASJ...50..187F} from {\em
ASCA} data. \citet{1998PASJ...50..187F} lists X-ray temperatures for 6
of the 8 CAIRNS clusters which agree with those listed in
\citet{xbacs}. 

In Paper I, we applied a hierarchical clustering analysis (described
in D99) to the redshift catalogs to determine the central coordinates
and redshift of the largest system of galaxies in each cluster.  Table
\ref{centers} lists these hierarchical centers and their projected
separations from the X-ray peaks.  We adopt these hierarchical centers
as the cluster centers.

\subsection{2MASS Photometry}

2MASS is an all-sky survey with uniform, complete photometry
\citep{twomasscalib} in three infrared bands (J, H, and K$_s$, a
modified version of the K filter truncated at longer wavelengths). We
use photometry from the final extended source catalog
\citep[XSC,][]{twomassxsc}.  The 2MASS XSC computes magnitudes in the
$K_s$-band using several different methods, including aperture
magnitudes (using a circular aperture with radius 7\arcs), isophotal
magnitudes which include light within the elliptical isophote
corresponding to $\mu_{K_s}$=20 mag/arcsec$^2$, Kron magnitudes, and
extrapolated ``total'' magnitudes \citep{twomassxsc}.  The sky
coverage of the catalog is complete except for small regions around
bright stars.

The 2MASS isophotal magnitudes omit $\sim$15\% of the total flux of
individual galaxies (K01).  C01 compare 2MASS photometry from the
Second Incremental Data Release (2IDR) with deeper infrared photometry
from \citet{loveday00}.  They find that Kron magnitudes are slightly
fainter than the total magnitudes in deeper surveys \citep[see
also][]{andreon02} and that 2MASS extrapolated total magnitudes are
slightly brighter than Kron (roughly total) magnitudes from the deeper
survey.
2MASS is a relatively shallow survey and thus likely misses many low
surface brightness galaxies \cite{andreon02,bell03}.  In this work we
focus on bright galaxies (which typically have high surface
brightness) so this bias is less important than, e.g., estimates of
the luminosity density or stellar mass density.

Except where stated otherwise, we use the $K_s$-band survey
extrapolated ``total'' magnitudes.  Galactic extinction is usually
negligible in the near-infrared.  We correct for Galactic extinction
by using the value in the center of the cluster.  We make K
corrections and evolutionary corrections of $<$0.15 magnitudes based
on \citet{pogg97}.  Because these corrections are small and not
strongly dependent on the galaxy model at the redshifts of the CAIRNS
clusters, we apply a uniform correction for all galaxies in a given
cluster interpolated from the model Elliptical SED with solar
metallicity and a star-formation e-folding time of 1 Gyr.

We reprocess two galaxies in A576 and two galaxies near A2199 using
the methodology of the 2MASS Large Galaxy Atlas \citep{lga}.  The
galaxies in A576 (CGCG 261-056 NED01 and CGCG 261-056 NED 02) are
bright ellipticals near the cluster center and also close to a bright
star.  One of the galaxies near A2199, UGC 10459, is an extremely flat
edge-on disk galaxy.  The other, NGC 6175, shows two nuclei aligned
NW-SE.  The SE component is brighter in $K_s$ band.

\subsection{Spectroscopy}

The 2MASS photometry allows selection of complete,
near-infrared-selected samples extending $\sim$1-2 magnitudes fainter
than the $M_{K_s}^*=-23.77 + 5 \mbox{log}~h$ determined for the field
galaxy luminosity function in 2MASS extrapolated magnitudes (C01).  We
define $K_s$-selected samples according to these magnitude limits
within the smaller of the turnaround radius $r_t$ (the radius within
which the average density is 3.5$\rho_c$) or the limiting radius
$r_{max}$ of the caustic pattern (our membership criterion) in each
cluster (see Paper I).  Table
\ref{ksamples} lists these radii and the apparent and absolute
magnitude limits of these catalogs for the 9 clusters.  Our redshift
catalogs are 99.7\% complete for cluster galaxy candidates brighter
than $M_{K_s}=-23 + 5 \mbox{log} h$ and 97.6\% complete for candidates
brighter than $M_{K_s}^*+1$.

Most of the galaxies in these samples have redshifts  in the
redshift catalogs from the CAIRNS project (Paper I).  Between 2002
June and 2003 September, we collected new redshifts for 
\ncziinew galaxies with the FAST
spectrograph \citep{fast} on the 1.5-m Tillinghast telescope of the
Fred Lawrence Whipple Observatory (FLWO).  FAST is a high throughput,
long slit spectrograph with a thinned, backside illuminated,
antireflection coated CCD detector.  The slit length is 180\arcs; our
observations used a slit width of 3\arcs ~and a 300 lines mm$^{-1}$
grating.  This setup yields spectral resolution of 6-8 \AA ~and covers
the wavelength range 3600-7200 \AA.  We obtain redshifts by
cross-correlation with spectral templates of emission-dominated and
absorption-dominated galaxy spectra created from FAST observations
\citep{km98}.  The typical uncertainty in the redshifts is 30$~\kms$.
Table \ref{cziinew} lists the new redshifts.
The additional redshifts make no
significant difference to the locations of the caustics or to the
resulting mass profiles.  We thus use the caustics and mass profiles
from Paper I.

\begin{deluxetable}{lcrr}
\tablecolumns{4}
\tablewidth{0pc}
\tablecaption{New Spectroscopic Data\label{cziinew}}
\small
\tablehead{
\colhead{}    RA & DEC & cz & $\sigma _{cz}$  \\
\colhead{}    (J2000) & (J2000) & ($\kms$)  & ($\kms$) \\
}
\startdata
 00:46:06.30 & -01:43:43.0 & 4074 & 16   \\ 
 00:46:34.60 & -01:37:07.0 & 16888 & 40   \\ 
 00:47:12.60 & -01:59:31.0 & 38906 & 40   \\ 
 00:50:46.10 & -03:17:53.0 & 17204 & 23   \\ 
 00:53:05.60 & -03:38:45.0 & 15993 & 41   \\ 
\enddata
\tablecomments{The complete version of this table is in the
 electronic edition of the Journal.  The printed edition contains only
 a sample.}
\end{deluxetable}
 
An important difference between the FAST spectra collected for CAIRNS
and those collected for other, larger redshift surveys
\citep{2df,sdss} is that CAIRNS suffers no incompleteness due to fiber
placement constraints.

We calculate the maximum fraction $f_{noz}$ of light missing from our
catalogs if we assume that all galaxies without redshifts and brighter
than the magnitude limit are cluster members (Table \ref{ksamples}).
In other words, we evaluate the potential observational bias which
results if every galaxy without a redshift were a cluster member.  For
this extreme case, the total luminosity within $r_{max}$ is
underestimated by the fraction $f_{noz}< 0.10$ for
all clusters.  The new redshifts in Table \ref{cziinew} contribute
significantly to the completeness of these catalogs.  Because the
galaxies without redshifts are almost entirely faint galaxies at large
distances from the cluster center, $f_{noz}$ is a very conservative
upper limit on the fraction of light missing within the completeness
limits (the surface number density of member galaxies decreases with
radius and the fraction of background galaxies increases with apparent
magnitude).

Assuming that the luminosity function in clusters and infall regions
is identical to that in the field (we test this assumption in $\S
3.1$), we calculate the fraction $f_{L}$ of total galaxy light
contained in galaxies brighter than our completeness limits.  This
fraction is greater than 60\% for all clusters.  From repeated
measurements, apparent magnitudes in the 2MASS XSC have an uncertainty
of $\sim 0.14$ magnitude at $K_s=13.4-13.5$; the galaxy catalogs
probably suffer incompleteness fainter than $K_s \approx 14$.  Thus,
the 2MASS XSC provides accurate magnitudes for galaxies within our
completeness limits, but it is difficult to use 2MASS galaxy counts at
much fainter magnitudes to estimate the contribution of fainter
galaxies to the total cluster/infall region light.  Note that the
field luminosity function of C01 that we adopt here has a steeper
faint-end slope than the luminosity function calculated from Kron
magnitudes.  If we adopt the Kron magnitude faint-end slope of C01,
$f_L$ increases by 7-15\% (the best sampled clusters have the
smallest changes).  We discuss this issue further in $\S 3.1$ and $\S
7.2$.

\begin{figure*}
\figurenum{1}
\label{kcomplete}
\epsscale{0.8}
\plotone{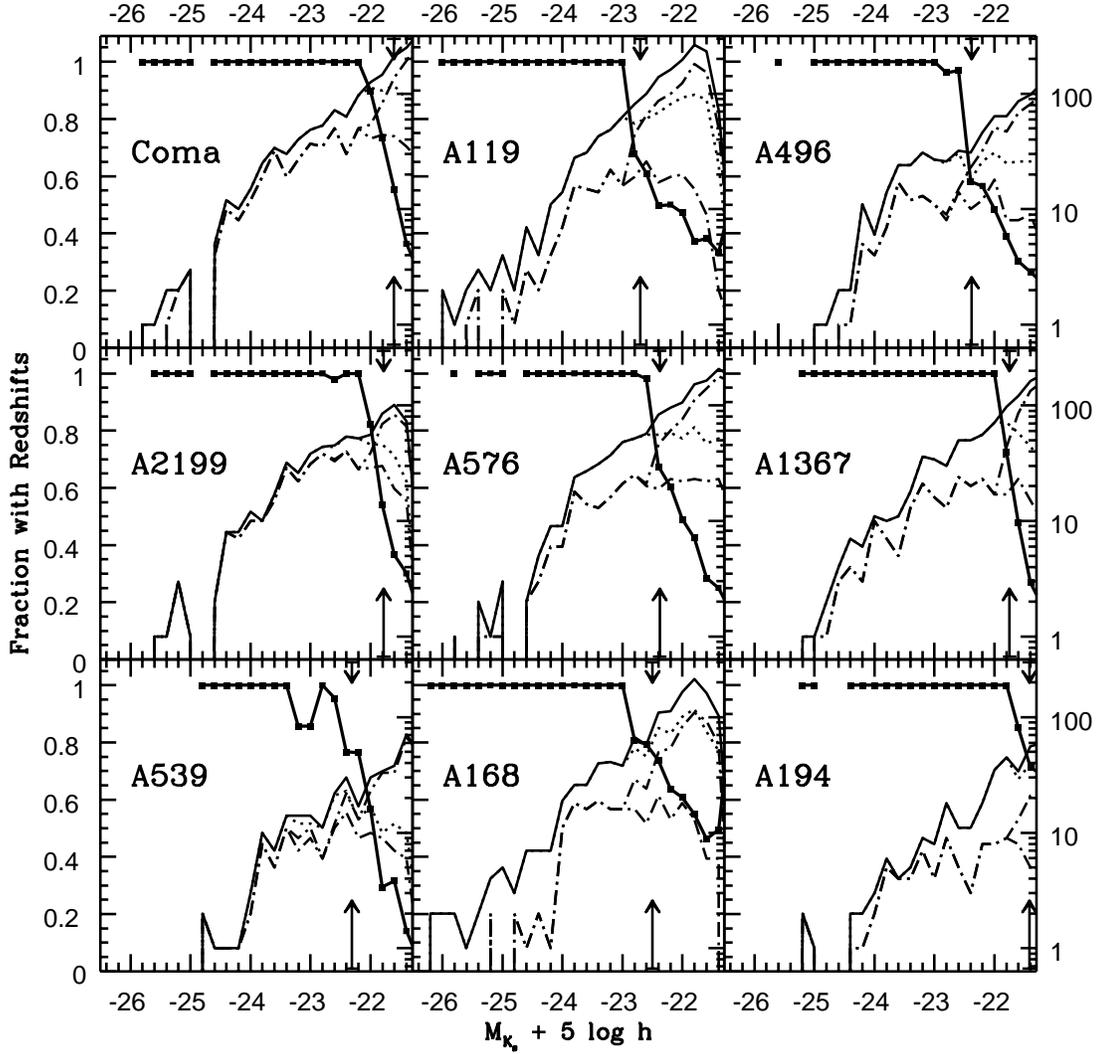}
\caption{Completeness of the CAIRNS spectroscopic catalogs versus
absolute magnitude $M_{K_s}$ (falling thick solid lines and scales on
left).  Vertical bars indicate the spectroscopic completeness limits.
The scales on the right show the number of galaxies per 0.2 magnitude
bin (rising thin solid lines). The dashed lines show the number of
galaxies with redshifts and the dash-dot lines show the number of
cluster galaxies with the upper and lower lines indicating the maximum
and minimum number of members.  The scales are identical in all
panels.  Clusters are ordered left to right, top to bottom, in
decreasing X-ray temperature.}
\end{figure*}

Figure \ref{kcomplete} shows the redshift completeness as a function
of apparent and absolute magnitude ($K_s$ extrapolated magnitude)
along with the total number of galaxies, the number with redshifts,
and the number of members versus magnitude.  Note that, as in Paper I,
we order the clusters by decreasing X-ray temperature from left to
right and from top to bottom in this and all similar later figures.
The catalogs are complete for cluster galaxy candidates brighter than
$M_{K_s}=-23$ except for five candidates in the outskirts of A539
which lie at high Galactic extinction.  It is not clear whether these
objects are galaxies or extended Galactic infrared sources.  The
brightest of these sources, IRAS 05155+0707, is an embedded Class 1
protostar and likely the source of Herbig-Haro objects HH114 and HH115
\citep{1997AJ....114.2708R}.  We exclude IRAS 05155+0707 from the
photometric catalog and the calculation of $f_{noz}$ in Table
\ref{ksamples}.

Figure \ref{kcomplete} also shows constraints on the luminosity
functions in the clusters.  The sets of dash-dotted lines show the
limits from assuming that (1) all galaxies without redshifts are
members or (2) none are.  We discuss the luminosity functions in more
detail in $\S 4$, but we note here that the faint-end slope of
the luminosity function in infall regions is poorly constrained
without deep, complete spectroscopy.

\section{Properties of Galaxies Inside and Outside the Virial Region}

Galaxy properties such as morphology and star formation rate are
strongly correlated with their local and global environments
\citep[e.g.,][and references
therein]{ellingson01,lewis02,gomez03,treu03,balogh03}.  Differences in
galaxy properties with environment may lead to apparent changes in the
observed mass-to-light ratio even if the ratio of dark matter to
stellar mass remains constant \citep[e.g.][]{bahcall2000}.  The CAIRNS
2MASS selected galaxies provide a well-defined population with which
to investigate these environmental effects.  The environments
considered range from cluster centers with densities
$\sim$1000$\rho_c$ to the edges of infall regions with densities
$\sim$3$\rho_c$ at the turnaround radius $r_t$.  These environments
are all denser than the universal average density $\Omega_m \rho_c$,
but they cover the density range where galaxy morphologies, optical
colors, and star formation rates change dramatically
\citep{ellingson01,lewis02,gomez03,treu03,balogh03}.  We
investigate the near-infrared photometric properties of galaxies
inside and outside the virial regions of the CAIRNS clusters and
compare them to field galaxies.

\subsection{Luminosity Functions}

Many investigators have sought to determine the environmental
dependence of the luminosity function \citep[e.g.,][and references
therein]{bczz01,2002MNRAS.329..385B,depropris03}.  Using the 2dF
Galaxy Redshift Survey,
\citet{depropris03} fit their cluster data to the \citet{schechter76} 
luminosity function (LF), 
\beqn
N(M) \propto 10^{0.4(\alpha+1)(M^*-M)}e^{-10^{0.4(M^*-M)}}
\eeqn
and find that the cluster LF in the $b_J$ band has a brighter
characteristic magnitude $M^*$ and steeper faint-end slope $\alpha$
than the field LF.  Although differences between cluster and field
luminosity functions exist at other wavelengths
\citep[e.g.,][]{1998MNRAS.293...71T,1998MNRAS.294..193T,2003ApJ...587..605M,2003MNRAS.341..981S},
the cluster LF in the $K_s$ band is quite similar to the field LF
\citep[e.g.,][]{mt98,dpes98,ap00,tustin,bczz01}, perhaps indicating a
universal stellar mass function \citep{andreon04}.  \citet{bczz01}
combine data from 2MASS and the Las Campanas Redshift Survey and find
that the cluster LF in the $J$ band has a brighter characteristic
magnitude and a steeper faint-end slope than the field LF; similar
differences are seen at $K_s$ band but the parameters differ by
$<$3-$\sigma$.  \citet{andreon04} finds that the cluster and field LFs
are indistinguishable at red wavelengths in the optical \citep[see
also][]{2003ApJ...591..764C}, suggesting that much of the difference
at bluer wavelengths is due to star formation.

\begin{figure*}
\figurenum{2}
\label{klfn}
\plotone{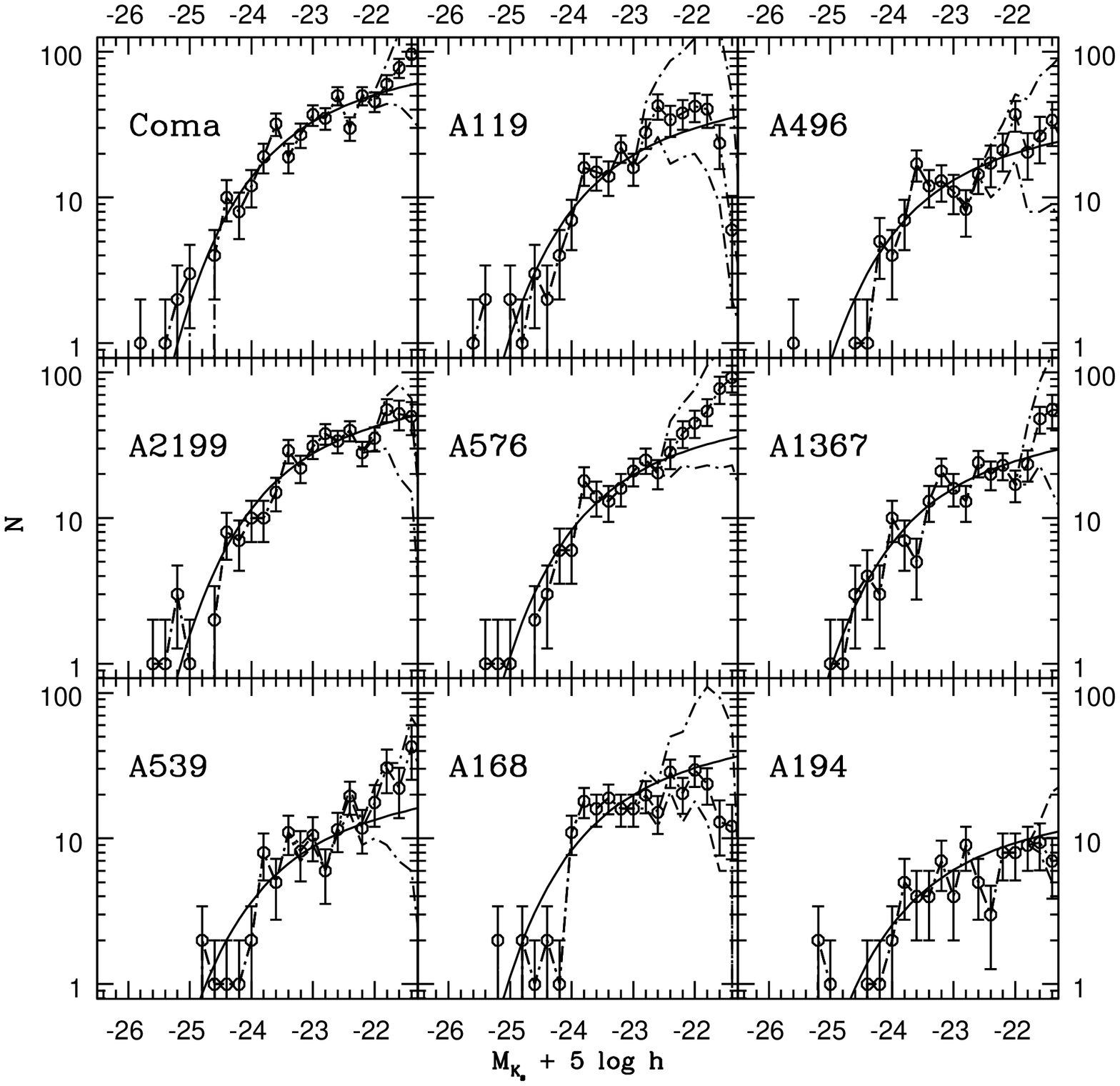}
\caption{Galaxy luminosity functions for all galaxies within the
infall regions of the CAIRNS clusters in $K_s$ band.  The solid line
is the field LF for comparison with arbitrary absolute normalization
but relative normalization scaled by $N_{tot}$, the number of galaxies
brighter than $M_{K_s}=-22.77$ within the limiting radius of the
caustics.}
\end{figure*}

Figure \ref{klfn} shows the near-infrared luminosity functions of each
of the CAIRNS clusters including all galaxies within the infall
regions.  We use the caustics from Paper I to define membership.  In
magnitude bins without complete redshifts, we compute a completeness
correction by assuming that the membership fraction of galaxies
without redshifts is the same as the membership fraction of galaxies
with redshifts.  Galaxies without redshifts tend to be at larger
projected clustrocentric distances than those with redshifts.  One
might thus expect that these galaxies are more likely to be
non-members because the ratio of cluster members to background
galaxies decreases with radius.  Counteracting this effect, galaxies
without redshifts tend to have lower surface brightnesses than those
with redshifts (because of observational bias towards higher surface
brightness galaxies); because of the correlation between absolute
magnitude and surface brightness, galaxies of a given apparent
magnitude with lower surface brightnesses should be intrinsically
fainter and are thus more likely to be cluster members
\citep{2002AJ....123.2246C,kg2000}.  

We count the number of bright
galaxies (those with $M_{K_s}\leq M_{K_s}^*+1$) in each cluster and
use this number to calculate relative normalizations for each cluster.
Figure \ref{klfn} shows the Schechter LF for field galaxies from
C01 scaled by this number of bright galaxies with an
arbitrary overall normalization.

\begin{figure*}
\figurenum{3}
\label{klfnin}
\plotone{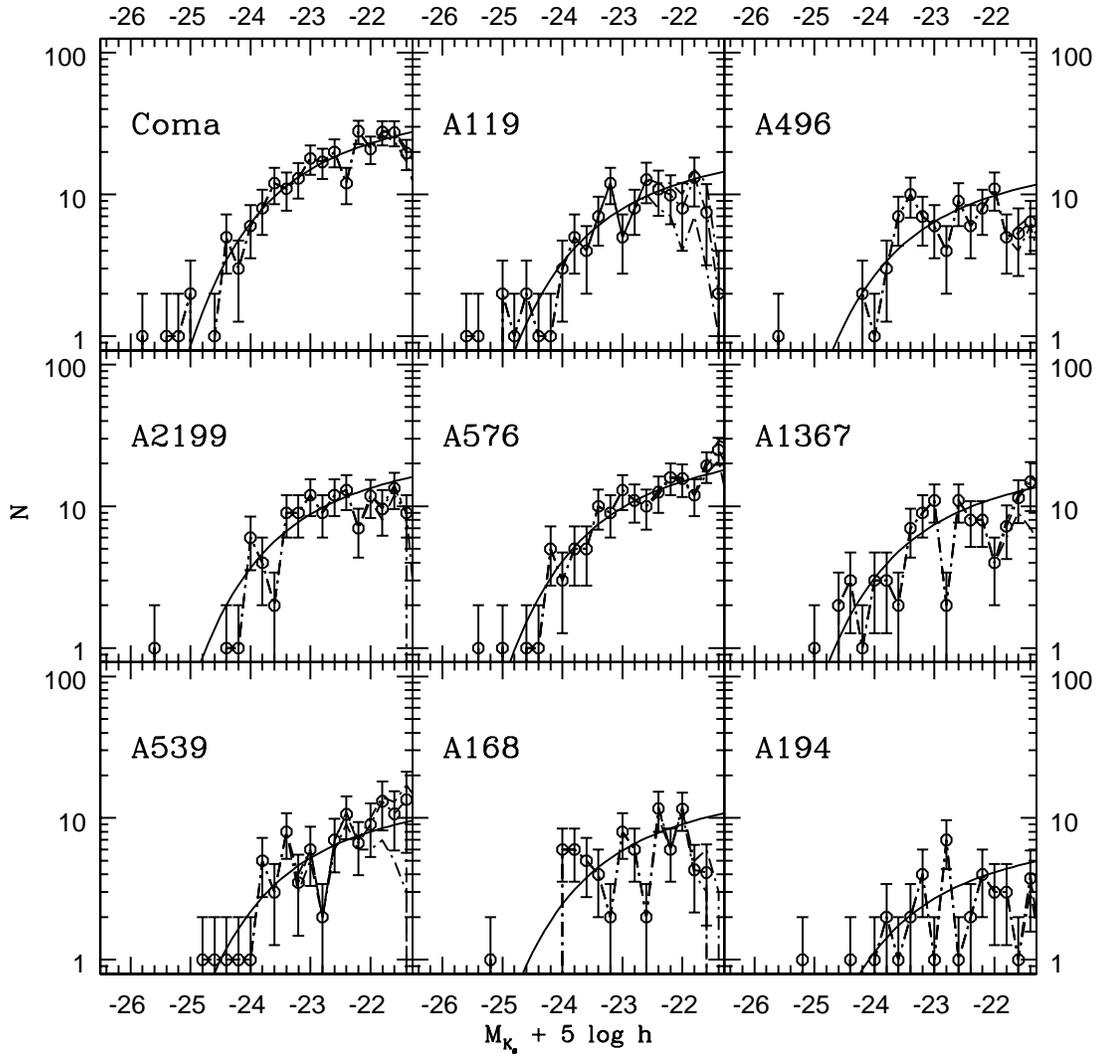}
\caption{Same as Figure \ref{klfn} for galaxies within $R_{200}$.  
Relative normalizations are made using $N_{200}$, the number of bright
galaxies projected within $R_{200}$.}
\end{figure*}

\begin{figure*}
\figurenum{4}
\label{klfnout}
\plotone{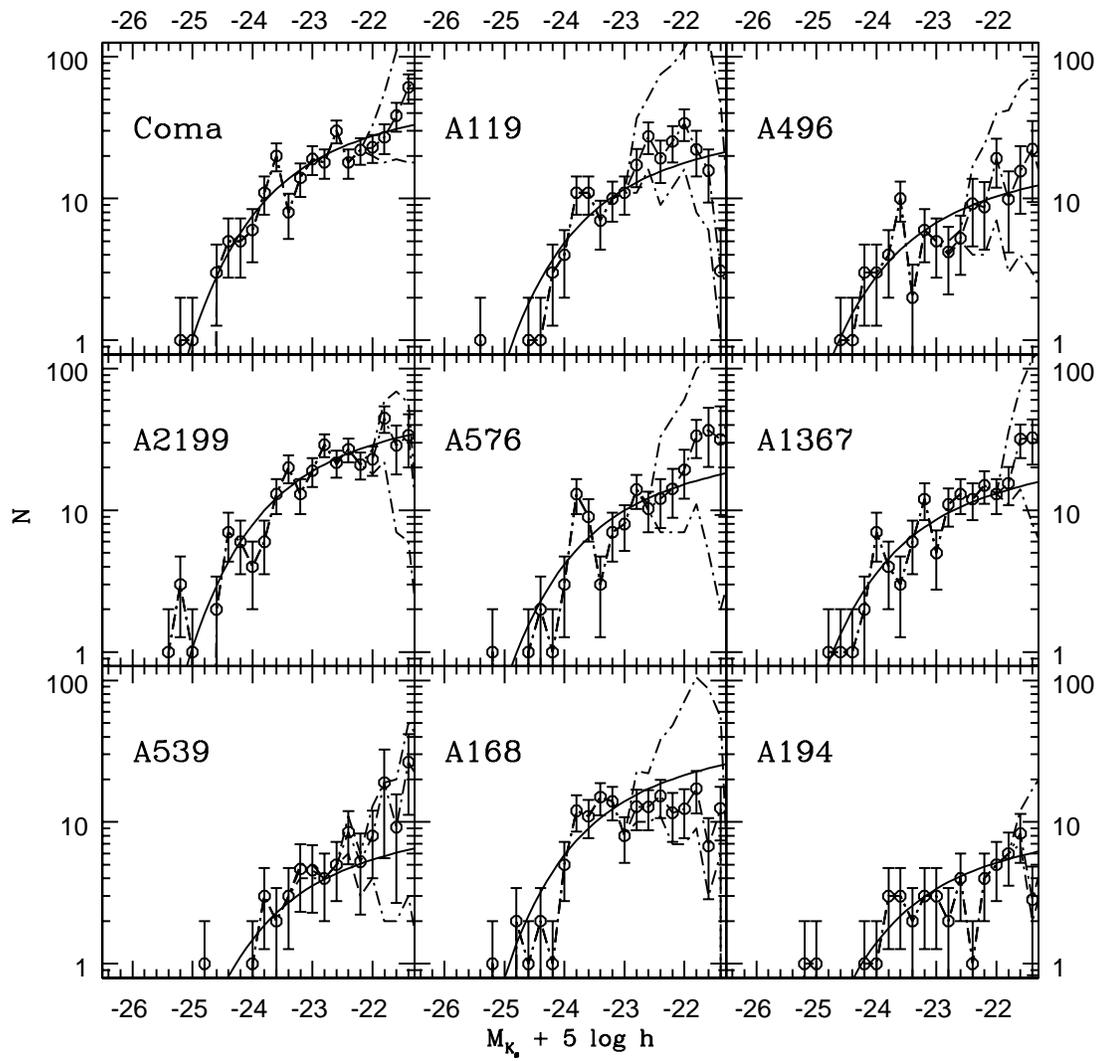}
\caption{Same as Figure \ref{klfn} for galaxies outside $R_{200}$.  
Relative normalizations are made using $N_{inf}$, the number of bright
galaxies projected outside $R_{200}$.}
\end{figure*}

\begin{figure*}
\figurenum{5}
\label{klfncomp}
\plotone{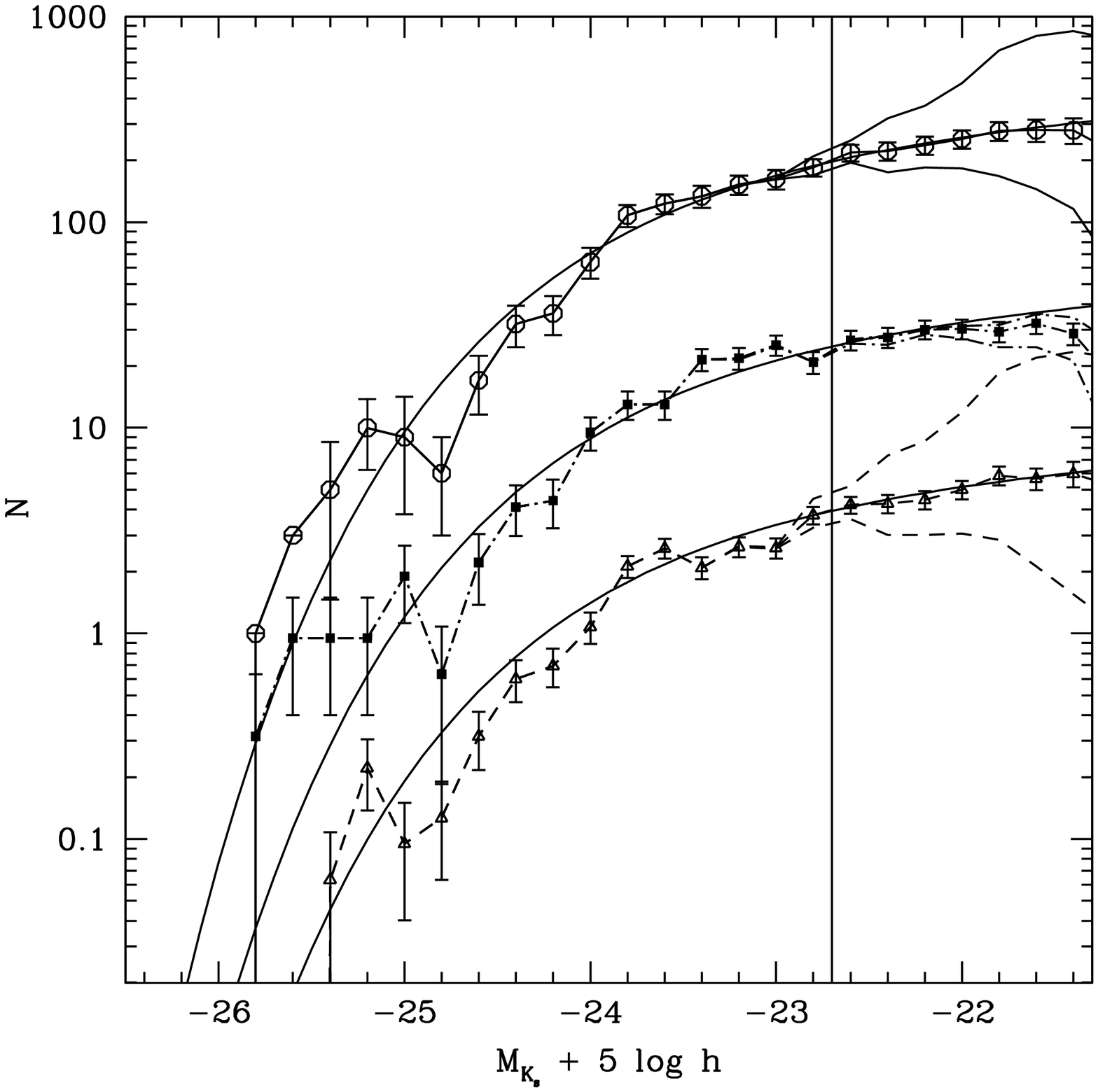}
\caption{Combined $K_s$-band luminosity functions for the CAIRNS clusters.  
Open circles show the total LF, squares show the LF inside $R_{200}$,
and triangles show the LF outside $R_{200}$.  Solid lines show
the shape of the 2dF/2MASS $K_s$-band LF with arbitrary normalization.
The LFs inside and outside $R_{200}$ are offset for clarity.  The
vertical line indicates $M_{K_s}^*+1$, the completeness limit of the
survey.}
\end{figure*}

We compute the luminosity functions separately for the virial regions
and the infall regions taking $R_{200}$ ($r_\delta$ is the radius
within which the enclosed mass density is $\delta$ times the critical
density, $R_\delta=r_\delta$ is the projected radius) as the dividing
radius.  Some galaxies projected inside $R_{200}$ lie outside
$r_{200}$, but no galaxies projected outside $R_{200}$ lie inside
$r_{200}$; thus the luminosity functions inside $R_{200}$ will be
contaminated by galaxies outside the virial region.  Figure
\ref{klfnin} shows the luminosity functions within $R_{200}$; Figure
\ref{klfnout} shows the luminosity functions outside this radius.  In
each panel, we plot the best-fit \citet{schechter76} luminosity
function for field galaxies from C01 scaled by the number of bright
galaxies with an arbitrary overall normalization.  Figure
\ref{klfncomp} shows the combined CAIRNS LFs inside and outside
$R_{200}$ as well as the total LF.  The LFs in the virial regions and
infall regions are very similar.

\begin{figure*}
\figurenum{6}
\label{klfnratio}
\plotone{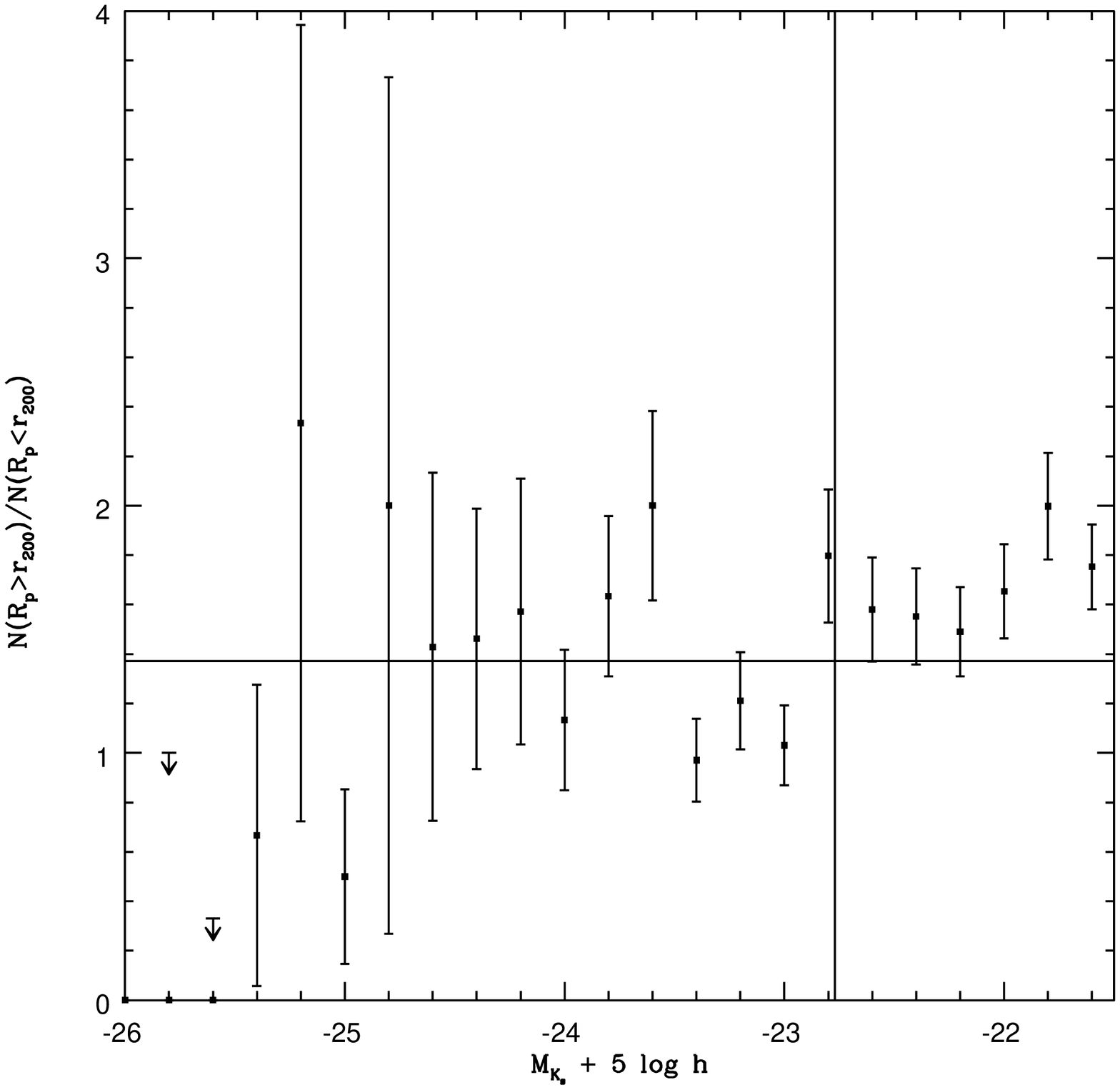}
\caption{Ratio of the combined CAIRNS $K_s$-band luminosity function 
of galaxies projected outside $R_{200}$ to that of galaxies projected
inside $R_{200}$.  Upper limits show the ratio if one galaxy were
present outside $R_{200}$.  The vertical solid line shows the 
approximate limit of the survey $M_{K_s}=-22.77$, one magnitude fainter 
than $M_{K_s}^*$ for field galaxies.  The horizontal line shows the 
ratio of all galaxies ($M_{K_s}\leq-22.77$) projected outside $R_{200}$ 
to all galaxies ($M_{K_s}\leq-22.77$) inside $R_{200}$.  Errorbars 
indicate 1-$\sigma$ Poissonian uncertainties.}
\end{figure*}

At the bright end, the LFs in both the virial regions and the infall
regions are poorly fit by a Schechter function (Figure
\ref{klfncomp}); the observed LFs contain more galaxies brighter than
$M_{K_s}=-25$ and fewer galaxies at $-25<M_{K_s}\leq -24$ than a
Schechter function which fits the faint-end slope.  This difference
may result from the existence and evolution of cD galaxies
\citep[e.g.,][]{schombert88,tonry87} present only in cluster
environments.  Figure \ref{klfnratio} shows the ratio of the LF
outside $R_{200}$ to that inside $R_{200}$.  The infall region LF
contains fewer extremely bright galaxies ($M_{K_s}\lesssim -25$) than
the virial region LF, but there is very little difference within the
Poissonian uncertainties.  Also, it is worth noting that extremely
bright galaxies are present in six of the nine infall regions (Coma,
A119, A2199, A576, A168, and A194), demonstrating that these bright
galaxies do not reside exclusively in cluster centers.  Many of these
galaxies likely occupy the centers of galaxy groups in the infall
regions \citep{rines01b}.  A $\chi ^2$ test shows that the LF ratios
for all bright galaxies ($M_{K_s}\leq -23.7$) are consistent with a
constant value at the 95\% confidence level.

At magnitudes fainter than the completeness limit, the LF in the
infall region (excluding the virial region) consistently exceeds that
inside the virial region, suggesting that the faint-end slope might be
steeper in the infall region.  The uncertainties in Figure
\ref{klfnratio} are Poissonian.  Because the correction for galaxies
without redshifts may be biased, these uncertainties may be
significantly underestimated.  A deeper complete spectroscopic survey
of the infall regions is necessary to determine the reality of effects
at these faint luminosities.

We calculate the best-fit luminosity function of the
\citet{schechter76} form for $M_{K_s}\leq -22.1 + 5 \mbox{log} h$ 
for all the clusters combined.  This limit corresponds to the 2MASS
completeness limit of $K_s=13.5$ for the most distant CAIRNS clusters.
We fit the LF for galaxies within $R_{200}$, outside $R_{200}$, and
all galaxies combined.  We do not account for measurement
uncertainties in the fits.  Table \ref{lfnfits} lists the best-fit
parameters (from minimizing $\chi^2$) as well as determinations of the
field luminosity function (K01,C01).  The uncertainties are 68\%
confidence limits for two interesting parameters.  We list two
different estimates from C01, one using extrapolated magnitudes (as
used here) and one using 2IDR Kron magnitudes converted to 'total'
magnitudes by subtracting -0.20 mag (see C01).  The LF parameters
differ by 2-3$\sigma$ from the field values, and agree well with
previous determinations \citep[][L03]{bczz01}.  However, the fits to
the CAIRNS LFs are not very good; the probability of obtaining a
larger value of $\chi^2$ from a sample drawn from the Schechter LF is
$<$0.7\% for the total LF.  The best-fit characteristic magnitude of
the virial region LF is brighter than the field LF, similar both in
sign and magnitude to the difference found by \citet{bczz01}; the
faint-end slope of the CAIRNS virial regions is slightly steeper than
the field values.  The LFs in the infall regions are intermediate
between the field LFs and the virial region LFs.

We repeat the fits using the completeness limit
of the redshift catalogs $M_{K_s}<-22.7$ and obtain consistent
parameters with larger uncertainties due to the weaker constraints on
the faint ends of the LFs.  We experimented with different cuts in
absolute magnitude both at the bright end (excluding cD-like galaxies
that could skew the LF parameters) and the faint end.  The best-fit LF
parameters are fairly sensitive to the limiting magnitude adopted,
perhaps because the cluster LF is not well described by a Schechter
function.  However, these parameters are generally within the
2-$\sigma$ range of the values listed in Table \ref{lfnfits}.  

It is interesting that the characteristic magnitude of the CAIRNS
virial region LF agrees well with that of the cluster LF constructed
by L03 without spectroscopy.  This agreement suggests that statistical
background subtraction produces little bias in the resulting LF
parameters.  Both L03 and CAIRNS use 2MASS photometry which provides
only a limited probe of the faint-end slope.  It would be instructive
to compare the LFs of individual clusters constructed with
spectroscopic membership to those constructed with statistical
background subtraction.  A detailed comparison is outside the scope of
the present work, but in $\S 5$ we will show that LFs constructed with
statistical background subtraction (L03) yield mass-to-light ratios
consistent with our results for clusters with complete spectroscopy.

The best-fit LF parameters significantly affect the estimates of the
fraction of light $f_L$ contained in faint galaxies (see Table
\ref{ksamples}).  However, for fixed LF parameters, the ratio of the
maximum to the minimum values of $f_L$ for the clusters varies by less
than 10\%; thus, the relative values of $f_L$ are robust.  Because the
CAIRNS LF parameters are consistent with the field LF but have larger
uncertainties, we continue to use the field LF to estimate the
fraction of light contributed by faint galaxies.  Note that the field
LF we adopt (C01 extrapolated magnitudes) has both a brighter
characteristic magnitude and a steeper faint-end slope than the LF of
C01 from Kron magnitudes.

\begin{figure*}
\figurenum{7}
\label{jlfn}
\plotone{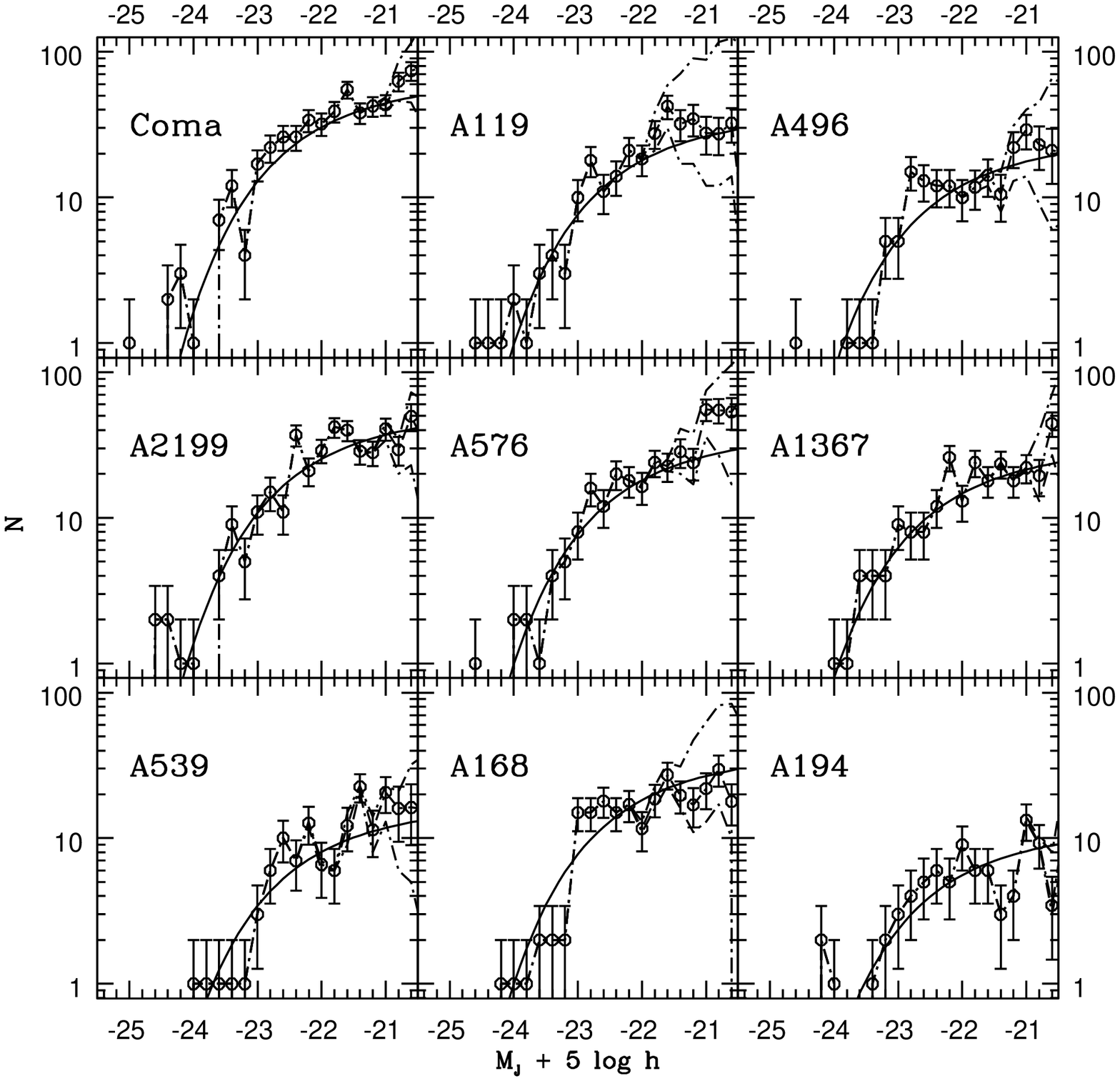}
\caption{Galaxy luminosity functions for all galaxies within the
infall regions of the CAIRNS clusters in $J$ band.  The solid line
is the field LF for comparison with arbitrary absolute normalization
but relative normalization scaled by $N_{tot}$, the number of galaxies
brighter than $M_{J}=-21.70$ within the limiting radius of the
caustics.}
\end{figure*}

\begin{figure*}
\figurenum{8}
\label{jlfnin}
\plotone{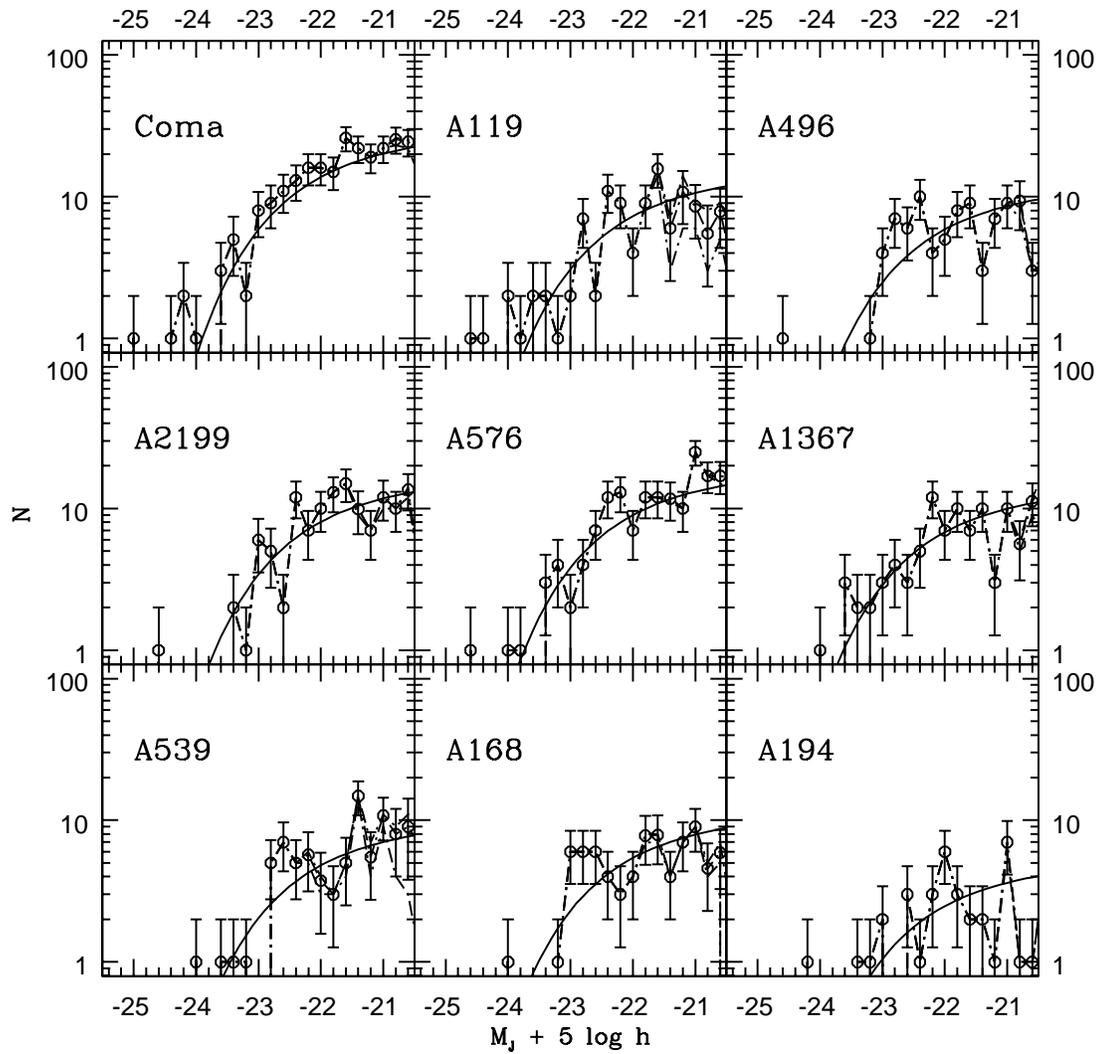}
\caption{Same as Figure \ref{jlfn} for galaxies within $R_{200}$.}
\end{figure*}

\begin{figure*}
\figurenum{9}
\label{jlfnout}
\plotone{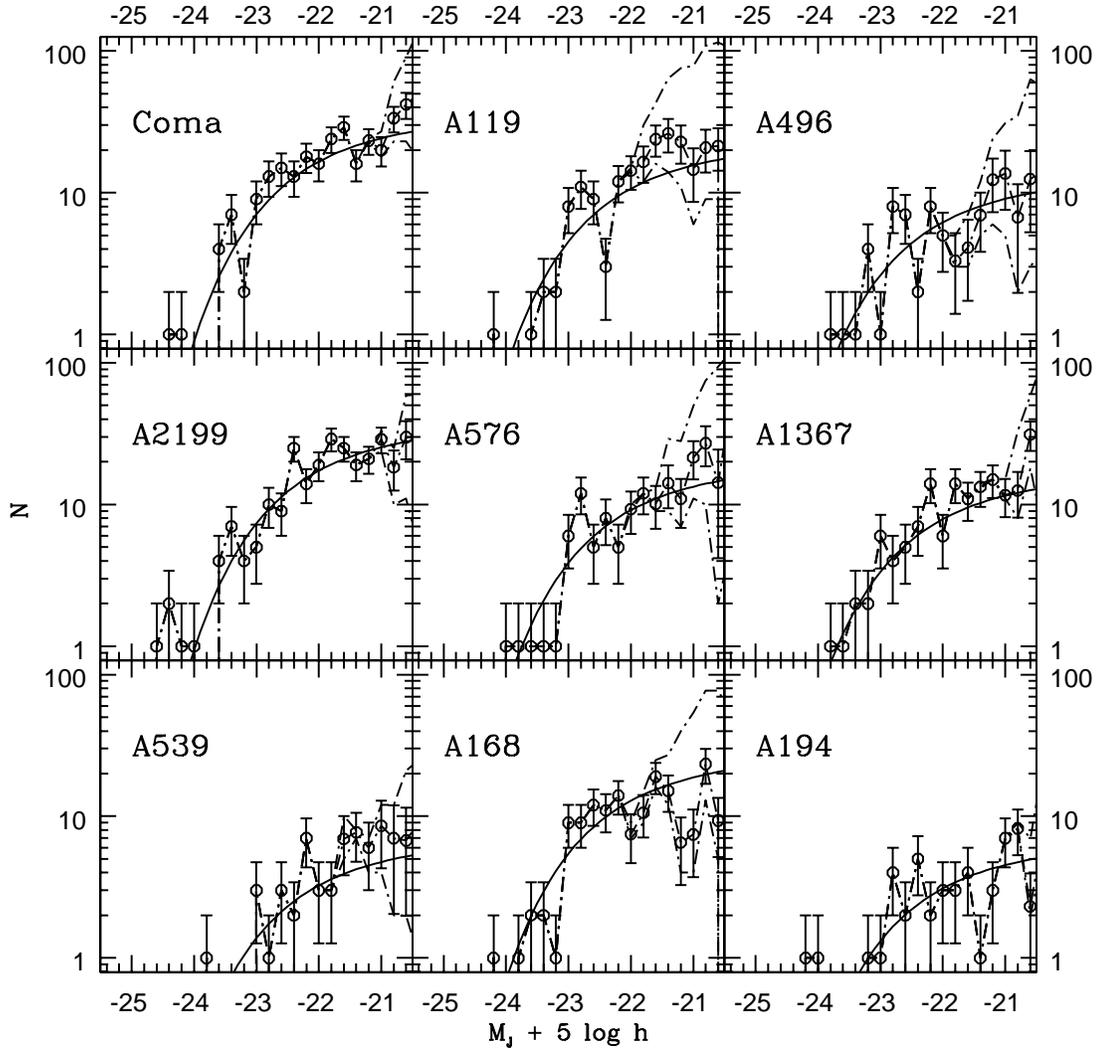}
\caption{Same as Figure \ref{jlfn} for galaxies outside $R_{200}$.}
\end{figure*}

\begin{figure*}
\figurenum{10}
\label{jlfncomp}
\plotone{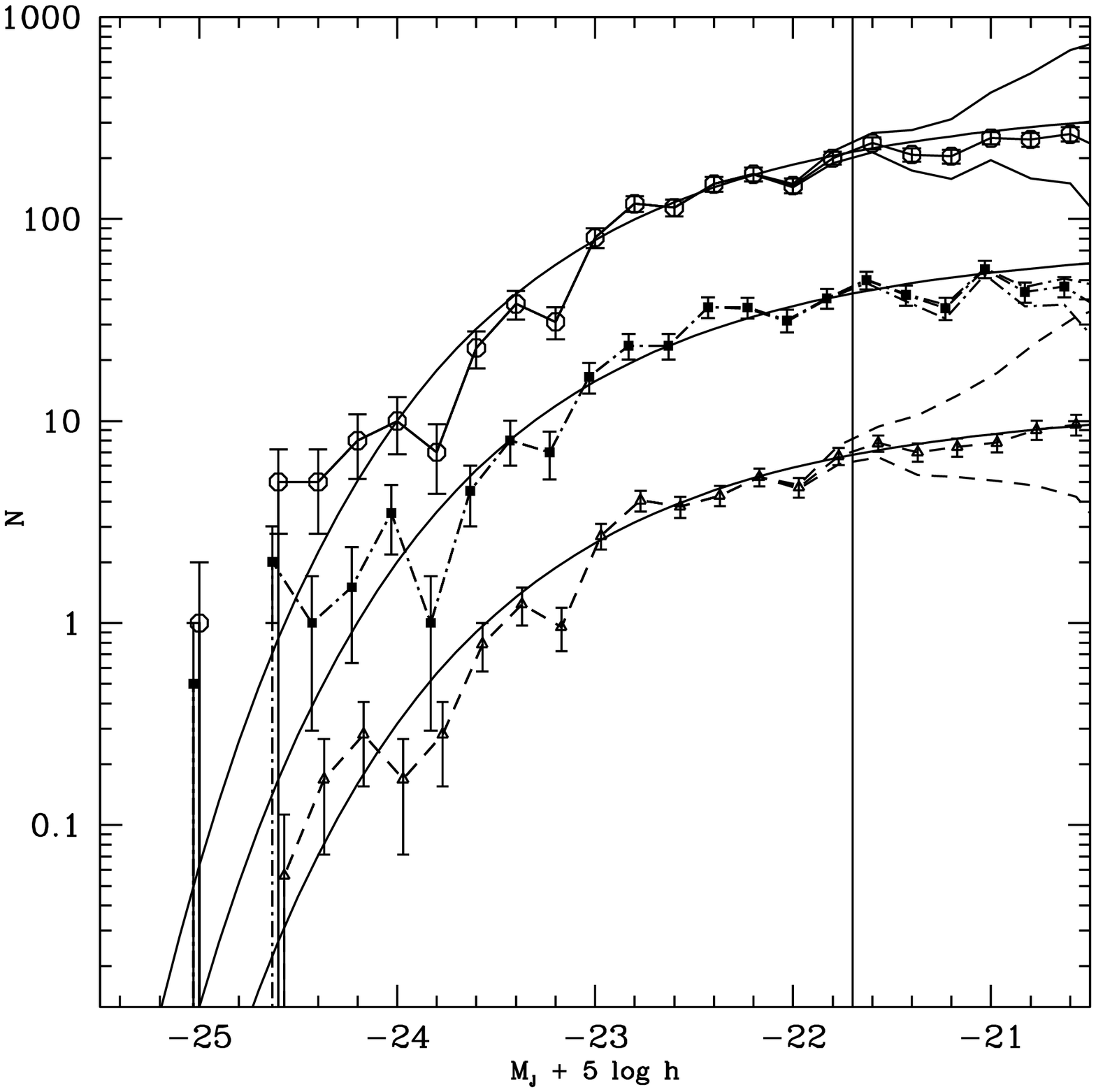}
\caption{Combined $J$-band luminosity functions for the CAIRNS clusters.  Open
circles show the total LF, squares show the LF inside $R_{200}$, and
triangles show the LF outside $R_{200}$.  Solid lines show the
shape of the 2dF/2MASS J-band LF with arbitrary normalization.  The
LFs inside and outside $R_{200}$ are offset for clarity.  The vertical
line indicates $M_{J}^*+1$, the approximate completeness limit of the
survey.}
\end{figure*}

\begin{figure*}
\figurenum{11}
\label{jlfnratio}
\plotone{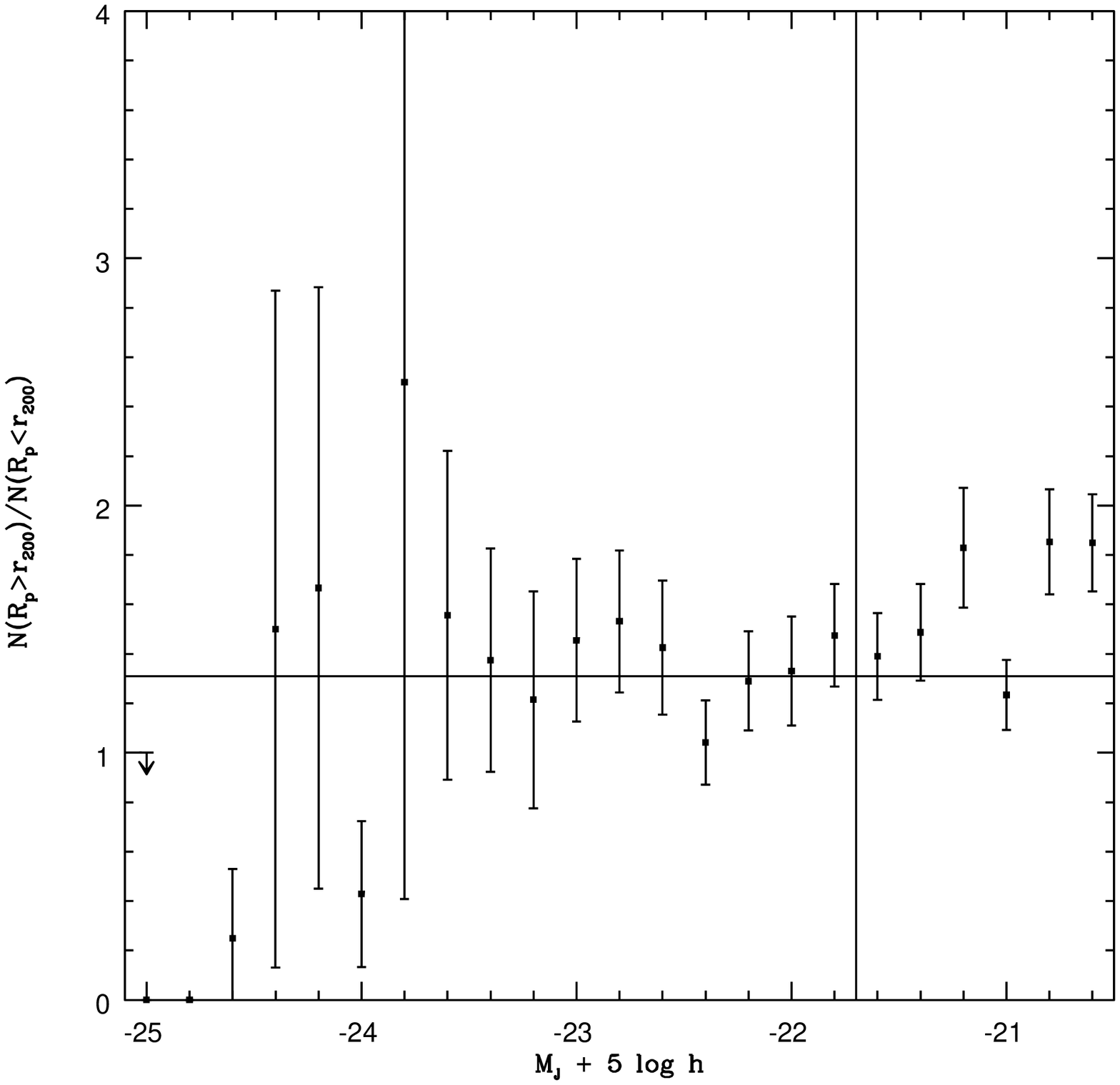}
\caption{Ratio of the combined CAIRNS $J$-band luminosity function 
of galaxies projected outside $R_{200}$ to that of galaxies projected
inside $R_{200}$.  Upper limits show the ratio if one galaxy were
present outside $R_{200}$.  The vertical solid line shows the 
approximate limit of the survey $M_{J}=-21.7$, one magnitude fainter 
than $M_{J}^*$ for field galaxies.  The horizontal line shows the 
ratio of all galaxies ($M_{J}\leq-21.7$) projected outside $R_{200}$ 
to all galaxies ($M_{J}\leq-21.7$) inside $R_{200}$.  Errorbars 
indicate 1-$\sigma$ Poissonian uncertainties.}
\end{figure*}

We repeat this analysis in the $J$ band, which extends deeper in 2MASS
and thus has smaller statistical uncertainties.  Figure \ref{jlfn}
shows the J band LF for all galaxies within $r_t$, Figure \ref{jlfnin}
shows the luminosity functions within $R_{200}$, and Figure
\ref{jlfnout} shows the luminosity functions outside this radius.  We
combine the LFs to produce an average cluster LF in Figure
\ref{jlfncomp}.  We scale the LF inside and outside $R_{200}$ to have
the same normalization at $M_J^*$ for field galaxies. As in $K_s$
band, the cluster LF has a very similar shape to the field LF except
for an excess of bright galaxies.  We repeat the non-parametric test
of computing the LF ratios (Figure \ref{jlfnratio}).  Table
\ref{lfnfits} lists the best-fit Schechter function parameters.  These
parameters differ by no more than 3-$\sigma$ from the field values
determined by C01.  The characteristic magnitude $M_J^*$ for cluster
virial regions is brighter than the field value by about 0.5
magnitudes, consistent with the results of \citet{bczz01}.  There is
remarkably little difference between the two LFs across the entire
range of magnitudes, although at faint magnitudes there is room for
significant differences which could be explored with deep, complete
spectroscopy.

To summarize, we see marginal evidence for differences between the
cluster LF and the field LF.  The cluster LF is slightly brighter and
has a steeper faint-end slope than the field LF.  We obtain similar
results in both $J$ and $K_s$ bands.  Our data sample only giant
galaxies, so significant differences may exist in the cluster and
field LFs in the dwarf galaxy regime.  For the purposes of computing
mass-to-light ratios, the systematic uncertainty introduced by
possible differences in the cluster and field LFs is $\lesssim 15\%$.
Note that, as expected, the LF in the infall region is intermediate
between the field LF and the cluster LF.

\subsection{Luminosity Segregation}

\begin{figure*}
\figurenum{12}
\label{lumseg}
\plotone{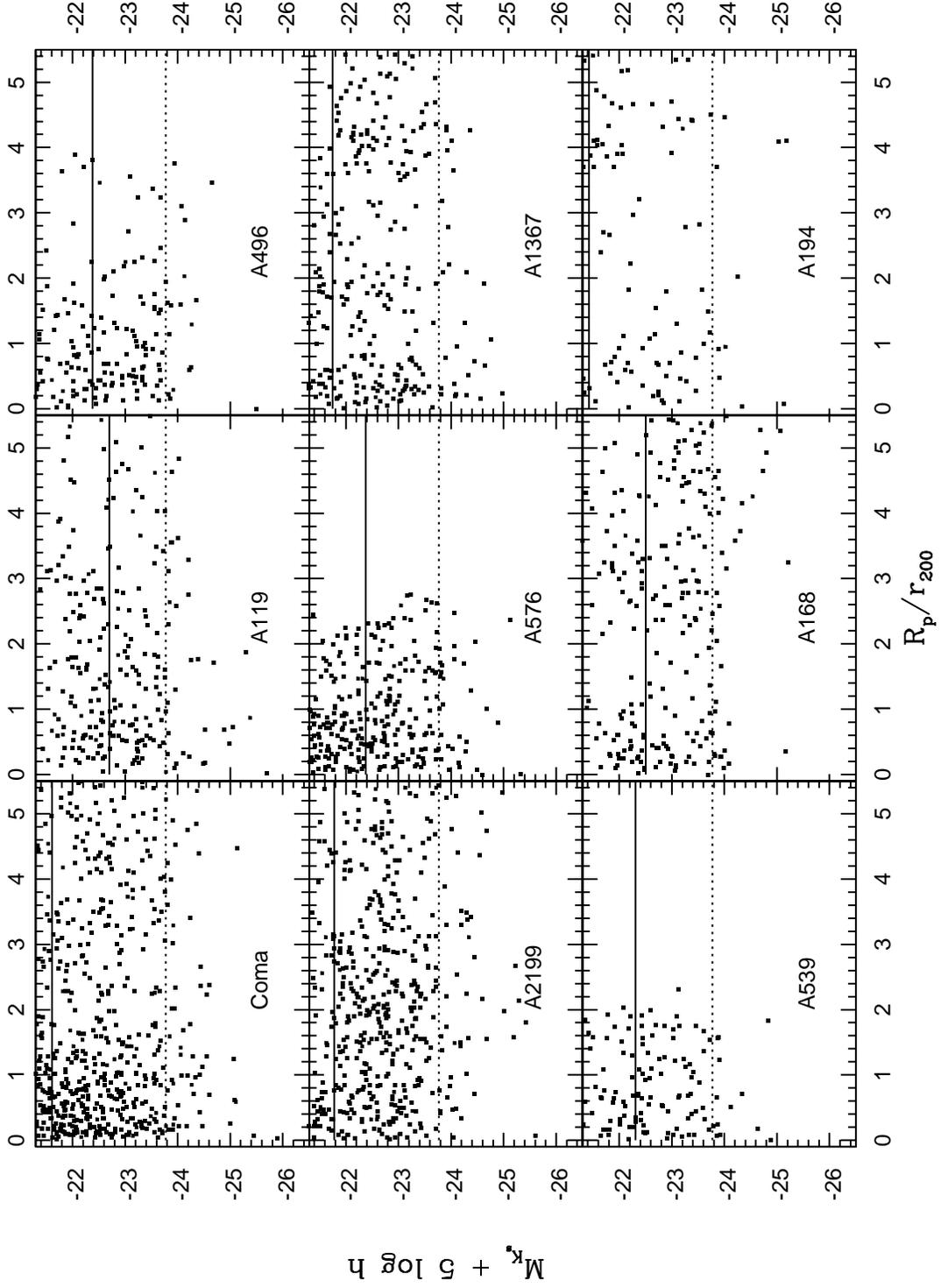}
\caption{Absolute magnitude versus (projected) clustercentric distance 
in units of $r_{200}$.  Dotted lines indicate $M_{K_s}^*$ for field
galaxies and solid lines indicate the spectroscopic completeness
limits.}
\end{figure*}

Dynamical friction could lead to luminosity segregation in galaxy
clusters.  Some investigators have claimed evidence for luminosity
segregation in compilations of cluster data \citep[e.g.][and
references therein]{1998A&A...331..439A,2002A&A...382..821A}.  Figure
\ref{lumseg} shows the distribution of absolute magnitude versus
(projected) distance $R_p$ from the cluster center.  If luminosity
segregation were significant, we would see more bright galaxies near
cluster centers.  The brightest cluster galaxy is typically very close
to the cluster center, consistent with a bright central cD galaxy
increasing in mass through accretion of smaller galaxies.  However,
there are also many comparably bright galaxies in the outskirts of the
clusters.  In A2199, many of the extremely bright galaxies outside the
virial region are at the centers of infalling groups
\citep{rines01b,rines02}.  There is little evidence for
luminosity segregation in the CAIRNS clusters, consistent with earlier
results for A576 \citep{rines2000}.  This result is not surprising
given the similarity of the LFs inside and outside $R_{200}$ (Figure
\ref{klfncomp}).  Again, note that the CAIRNS samples do not extend
into the dwarf galaxy regime, where luminosity segregation might be
present \citep{2002A&A...382..821A}.

\subsection{Broadband Galaxy Colors}

Star formation rates depend on environment \citep[e.g.,][and
references therein]{ellingson01,lewis02,gomez03,treu03,balogh03}.
Because stellar populations in field galaxies are on average younger
than those in cluster galaxies, more blue light is emitted per unit
mass in field-like environments than in cluster environments.  As a
consequence, mass-to-blue-light profiles might decrease with radius
\citep{bahcall2000} even if the ratio of gravitational mass to stellar
mass were constant.  

Because young stars are both hotter and bluer than older stars, the
difference in stellar mass-to-light ratios decreases toward longer
wavelengths \citep[see the synthesized stellar population models
of][]{bc03}.  For example, studies of near-infrared mass-to-light
ratios in galaxies suggest that the mass-to-light ratio at these
wavelengths is insensitive to the current star formation rate in
either disk galaxies \citep{gpb96} or early-type galaxies
\citep{zibetti02}.  Unfortunately, the color differences between J and
K bands are not very large because these wavelengths primarily trace
Population II stars \citep{lga}, making this effect difficult to
detect with 2MASS data alone.

\begin{figure*}
\figurenum{13}
\label{rkprofile}
\plotone{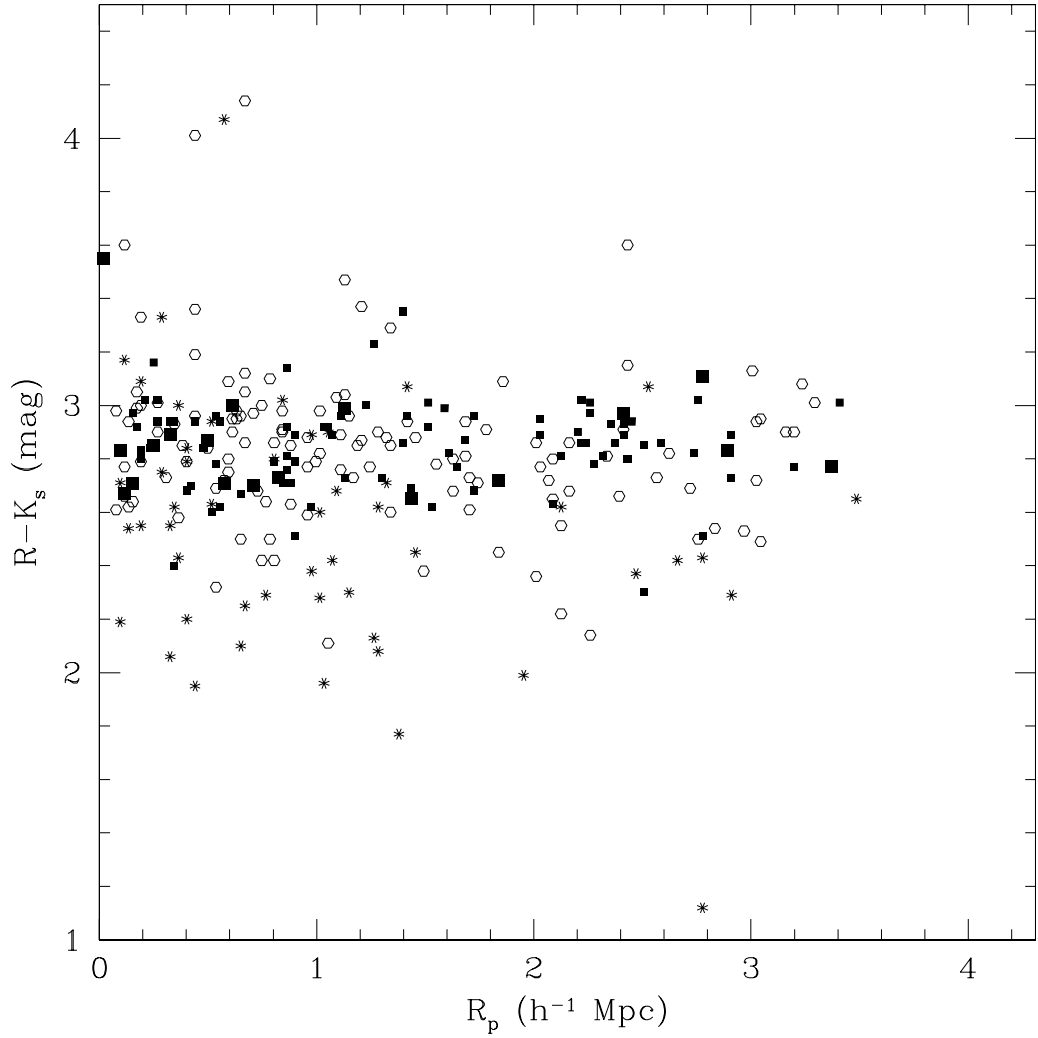}
\caption{A576 $R-K_s$ color versus projected distance from the cluster. 
The symbols show galaxies of different luminosities.  Large filled
squares, small filled squares, hexagons, and stars represent galaxies
in the magnitude bins $M_{K_s}\leq -23.77$, $-23.77<M_{K_s}\leq
-22.77$, $-22.77<M_{K_s}\leq -21.77$, and $-21.77<M_{K_s}\leq -20.77$
respectively.  Note that the latter two bins are not complete.}
\end{figure*}

For A576, our 9 square degrees of photometric CCD observations in R
band \citep{rines2000} allow a measurement of $R-K_s$ colors.
Optical-infrared colors enable us to investigate stellar population
effects.  Although both the 2MASS magnitudes used here and the R band
magnitudes in \citet{rines2000} are supposed to be close to total, a
systematic difference in the magnitude definitions could introduce an
artificial color gradient.  We reprocess the R band images using
SExtractor
\citep{sex} to obtain aperture magnitudes within a circular aperture
of radius 14\arcs~ in R band and radius 15\arcs ~in 2MASS.  This
slight mismatch in apertures produces a small bias towards redder
colors, but clustrocentric gradients, if any, should still be evident.
We calculate $R-J$ and $R-{K_s}$ colors for all of the galaxies in
both catalogs.  Figure
\ref{rkprofile} displays the $R-{K_s}$ colors versus projected radius.  
There is no obvious radial gradient in either $R-J$ or $R-{K_s}$
colors for bright galaxies.  There may be a radial gradient in
$R-{K_s}$ colors for galaxies fainter than $M_{K_s}=-22.77$, but we
lack complete spectroscopy at these magnitudes.
In a photometric study of clusters using SDSS, \citet{goto04} find
small but significant radial gradients in the fraction of blue
galaxies with radius (the fraction increases with radius).  The trends
are weakest in the most nearby clusters which are the most similar to
the CAIRNS clusters.  Note that the CAIRNS catalogs are selected at
$K_s$ rather than $r$, which may account for the lack of a gradient in
A576.  Also, the trends may be weaker in $R-{K_s}$ colors than in,
e.g., $u-r$ colors, which are much more sensitive to the presence of
young stars.  A multiwavelength study of several clusters with
spectroscopically determined membership would clarify the importance
of color gradients in clusters.

\begin{figure*}
\figurenum{14}
\label{krk}
\plotone{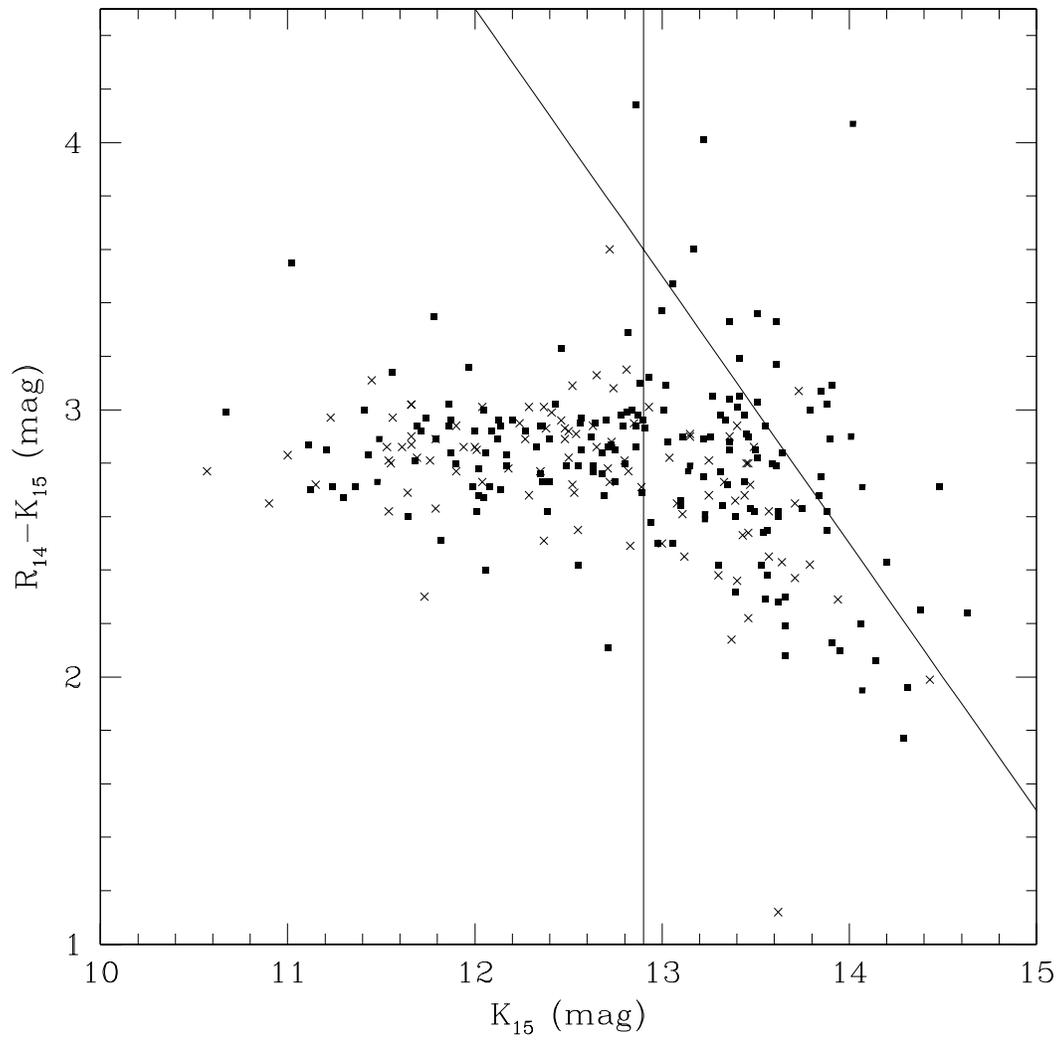}
\caption{A576 $R-K_s$ color (in a circular aperture of radius  $15''$) 
versus absolute $K_s$ band magnitude within this aperture.  Squares
are galaxies inside $R_{200}$ and crosses are galaxies outside
$R_{200}$. The vertical line shows the magnitude limit of the 2MASS
spectroscopic catalog while the slanted line shows the limit of the
$R$ band spectroscopic catalog.}
\end{figure*}

We plot the $R-{K_s}$ color versus $K_s$ magnitude in Figure
\ref{krk}.  There is little evidence for a color-magnitude relation.
Galaxies inside and outside $R_{200}$ occupy the same parts of the
diagram, indicating that there is no large difference in the two
populations.  The galaxies appear to have very similar stellar
populations.  Comparing the colors to the models of \citet{bc03}
indicates metallicities greater than solar.  The degeneracy between
age and metallicity effects prevents further conclusions.

\subsection{Near-Infrared Colors and the Color-Magnitude Relation}

\begin{figure*}
\figurenum{15}
\label{djk}
\plotone{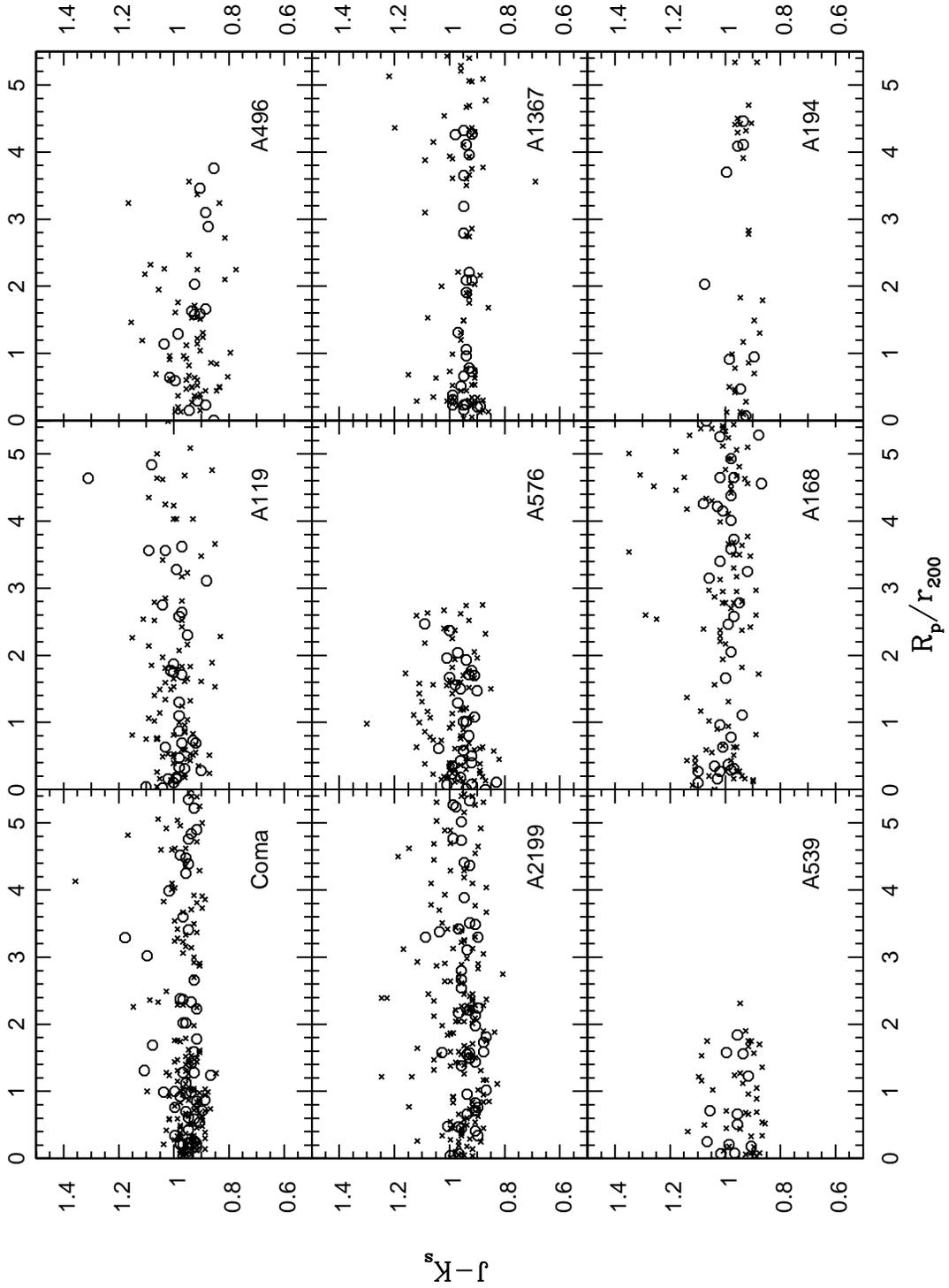}
\caption{Near-infrared $J-K_s$ color versus projected distance from 
the cluster. Circles and crosses represent galaxies
in the magnitude bins $M_{K_s}\leq -23.77$ and $-23.77<M_{K_s}\leq
-22.77$ respectively.}
\end{figure*}

\begin{figure*}
\figurenum{16}
\label{jkhist}
\plotone{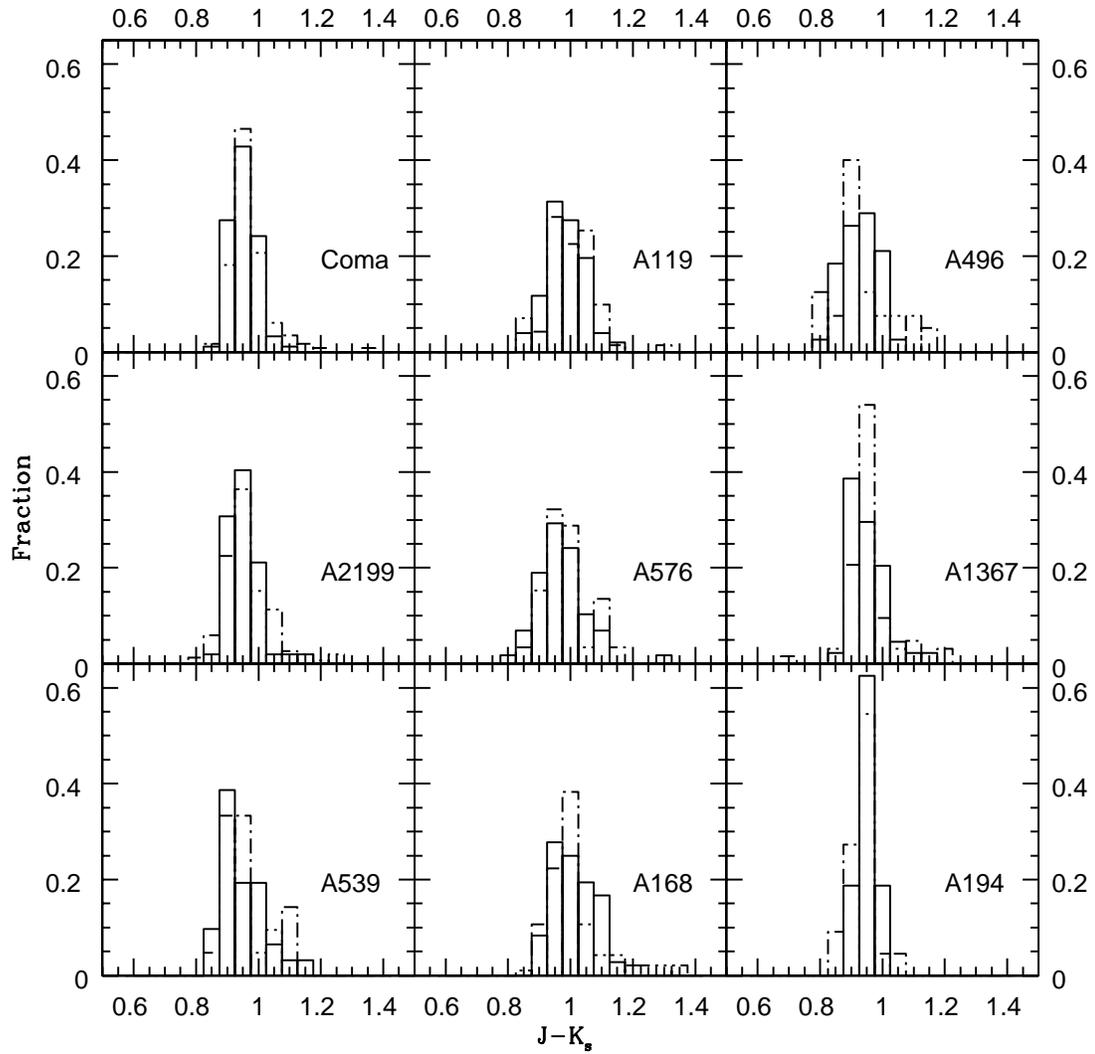}
\caption{Normalized histograms of near-infrared $J-K_s$ colors inside (solid lines) and outside (dash-dotted lines) $R_{200}$.}
\end{figure*}

Significant variations in the stellar mass-to-light ratio might be
indicated by radial gradients in $J-K_s$ colors.  Unfortunately,
identifying such gradients is difficult because variations in galaxy
$J-K_s$ colors are relatively small and because unlike optical colors,
galaxies with the reddest $J-K_s$ colors may contain younger stellar
populations and are red as a result of emission from hot dust
\citep{2002A&A...394..873H,barton03}.  We observe no radial color
gradients in $J-K_s$ (see Figure \ref{djk}, which highlights the lack
of trends in the outlying points).  Figure
\ref{jkhist} shows that the distributions of $J-K_s$ colors of bright
galaxies inside and outside $R_{200}$ are extremely similar.  There is
a possible excess of galaxies in the red tail of the distribution in
the sample outside $R_{200}$ (Figure \ref{djk}); some of these are
edge-on disk galaxies while others are probably AGN \citep{j00,lga}.
Because of the morphology-density relation, we expect more disk
galaxies in cluster outskirts.

\begin{figure*}
\figurenum{17}
\label{kjk}
\plotone{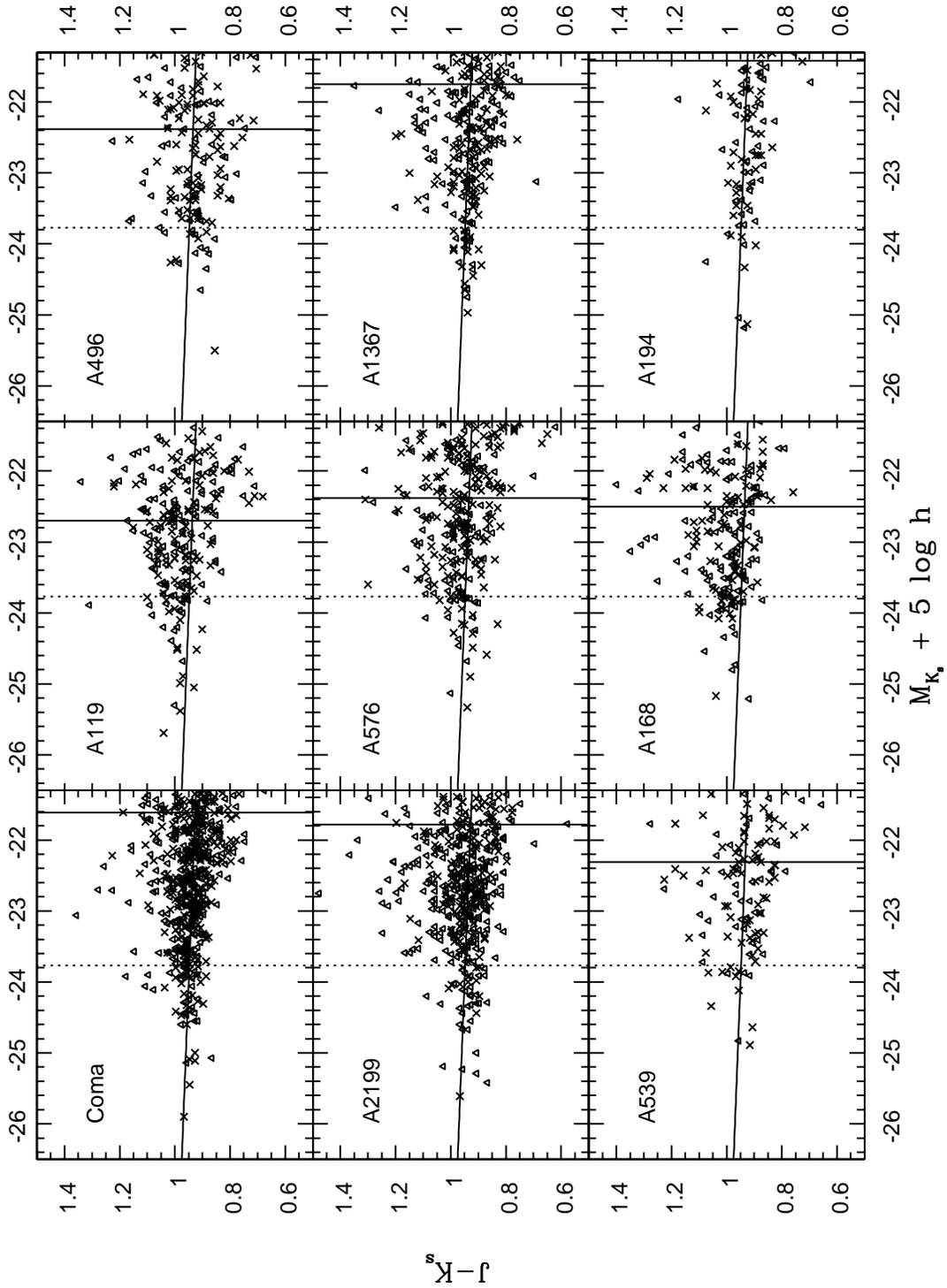}
\caption{The near-infrared color-magnitude relation for the CAIRNS 
clusters. Crosses are galaxies projected inside $R_{200}$ and
triangles are those projected outside $R_{200}$.  The solid and dotted
vertical lines indicate the spectroscopic completeness limits and
$M_{K_s}^*$.  We overplot a fiducial color-magnitude relation with
slope $-0.01
\mbox{mag}~\mbox{mag}^{-1}$. }
\end{figure*}

We plot $J-K_s$ color (computed within the elliptical isophote
$K_s$=20 mag arcsec$^{-2}$) versus absolute magnitude $M_{K_s}$ (from
the extrapolated $K_s$ magnitude) of CAIRNS members in Figure
\ref{kjk}.  The most striking result is that the outlying data points
are galaxies both inside and outside $R_{200}$, which suggests that
the stellar populations of galaxies in these regions are similar.
There is tentative evidence for a color-magnitude relation
\citep[i.e., fainter galaxies are bluer, see, e.g.,][and references
therein]{2001MNRAS.326.1547T} in the near-infrared, but the slope
($\approx -0.01 \mbox{mag}~\mbox{mag}^{-1}$) is much shallower than at
optical wavelengths, e.g.,the slope is $-0.14\pm0.01$ in $U-V$ versus
$V$ in Coma \citep{2001MNRAS.326.1547T}.  We obtain a similar
color-magnitude relation when the colors and magnitudes are determined
from aperture photometry, suggesting that the relation does not result
from systematic effects in 2MASS.  The variations in the colors can be
explained by variations in the metallicities of the stellar
populations.  The most recent stellar population models of
\citet{bc03} indicate that 10 Gyr old stellar populations (formed
instantaneously according to a \citet{chabrier03} initial mass
function) with metallicities $[Fe/H]=-0.64\rightarrow +0.56$ have
rest-frame $J-K_s=0.75\rightarrow 1.1$.  As at optical wavelengths,
there is degeneracy between age and metallicity effects; bluer colors
result from either lower metallicities or younger ages
\citep{1994ApJS...95..107W}.  Accurate spectral information is required
to break this degeneracy \citep[e.g.,][]{2000ApJ...536L..19C}.

The scatter in the observed near-infrared color-magnitude relation is
larger for fainter galaxies; the fainter galaxies have more varied
stellar populations and/or larger uncertainties.  Note that galaxies
in clusters at low galactic latitude (A539 and A496) have larger
scatter than those in clusters near the galactic poles (Coma and
A1367).  This observation suggests that a significant part of the
scatter may result from uncertainties in Galactic extinction.
A full accounting of the
color-magnitude relation is beyond the scope of this paper.  Instead,
we note that the near-infrared properties of galaxies do not change
dramatically with radius.  This result implies that the stellar
mass-to-light ratios do not change dramatically with radius; thus,
measuring near-infrared mass-to-light ratios is a good approximation
to a measurement of the ratio of total mass to stellar mass.

\section{Near-Infrared Luminosity and Number Density Profiles}

\subsection{Number Density Profiles}

Because our catalogs are essentially complete within their respective
magnitude limits, we can count the number of bright galaxies to
compare cluster richness.  We adopt $M_{K_s}=-22.77 + 5 \mbox{log} h$
as our limiting magnitude because all clusters are complete to this
depth (Table \ref{ksamples}).  This limit is equivalent to $M_{K_s}^*
+ 1$ for field galaxies.  Table \ref{rich} lists the number of
galaxies inside and outside $R_{200}$ (``outside $R_{200}$'' means the
projected radius $R_p$ satisfies $r_{200}<R_p\leq r_{max}$).  In all
clusters but A539, there are more cluster members outside $R_{200}$
than inside $R_{200}$.  We suggested this result in Paper I but lacked
the uniform photometry necessary to establish it.

\begin{figure*}
\figurenum{18}
\label{sdens}
\plotone{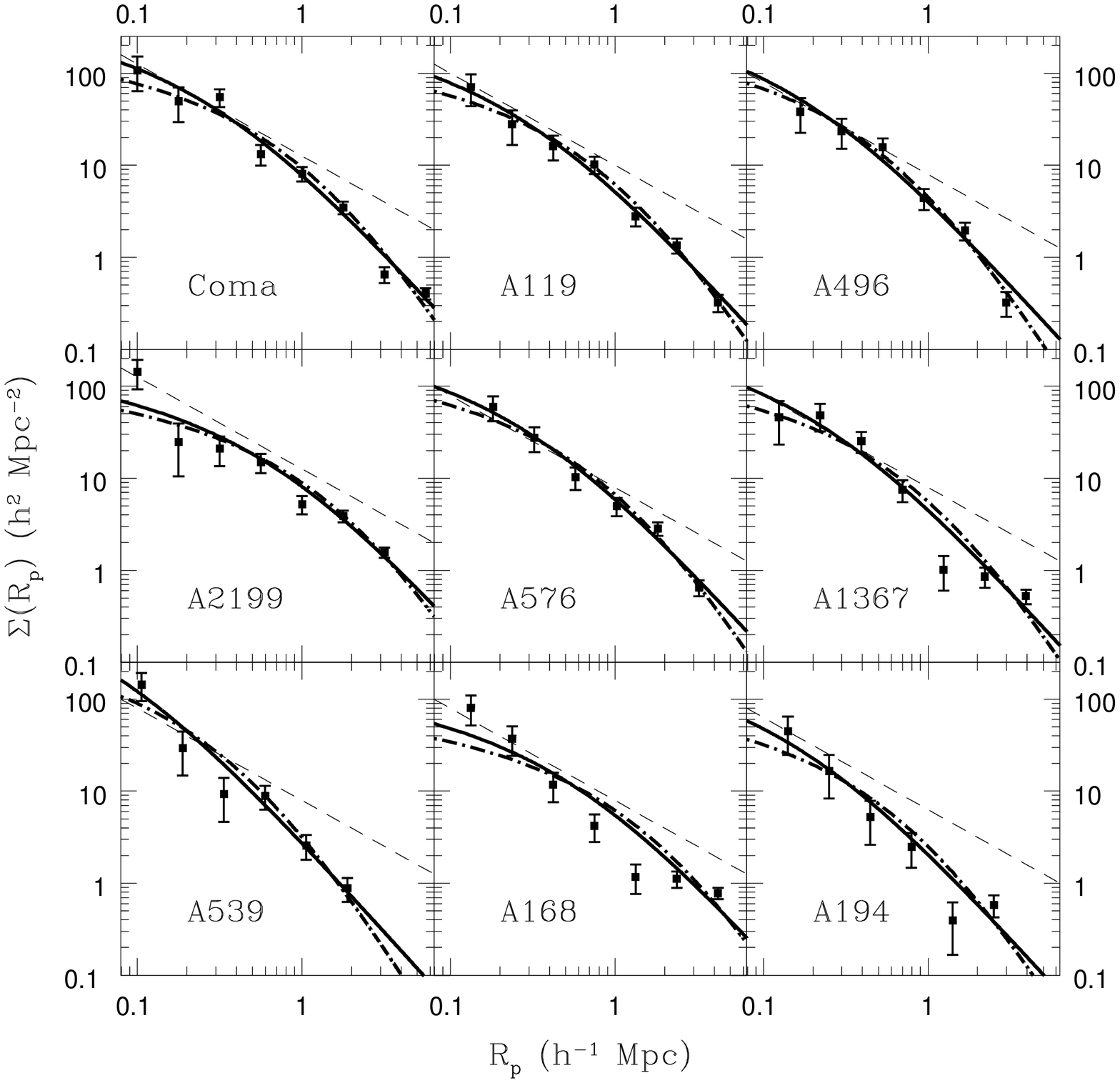}
\caption{Surface number density profile of cluster members 
brighter than $M_{K_s}=-22.77 + 5 \mbox{log}h$, equivalent to
$M_{K_s}^*+1$ for field galaxies.  The dashed line shows a singular
isothermal sphere.  The solid and dash-dotted lines show the best-fit
NFW and Hernquist profiles.}
\end{figure*}

Figure \ref{sdens}  shows the surface number density
profiles of the CAIRNS clusters.  We choose radial bins spaced
logarithmically by 0.25; the outermost bin contains the
maximum radius $r_{max}$ of the caustics.  We fit the number density
profiles of the CAIRNS clusters to three simple analytic models.  The
simplest model of a self-gravitating system is a singular isothermal
sphere (SIS). The volume density of the SIS decreases with radius
according to $\rho \propto r^{-2}$; the projected number density of
objects $\Sigma$ decreases as $\Sigma(R_p) \propto R_p^{-1}$.
\citet{nfw97} and \citet{hernquist1990} propose two-parameter models
based on CDM simulations of haloes.  These density profiles are 
\beqn
\rho (r) \propto [\frac{r}{a}(1+\frac{r}{a})^{-\alpha}]
\eeqn
where $a$ is a scale radius and $\alpha$=2 for the NFW profile and
$\alpha$=3 for the Hernquist profile.  At large radii, the NFW density
profile decreases as $r^{-3}$ and the density of the Hernquist model
decreases as $r^{-4}$ (implying a finite total mass).  The NFW surface
number density profile is
\beqn
\Sigma(<s) = \frac{N(a)}{\pi \mbox{ln}(4/e) a^2 (s^2-1)}[1-X(s)]
\eeqn
where $s=R_p/a$ is the projected radius in units of the scale radius,
$N(a)$ is the number of galaxies within the sphere of radius $a$, and
\beqn
X(s) = \frac{\mbox{sec}^{-1} s}{\sqrt{s^2-1}}.
\eeqn
We fit the parameter $N(a)$ rather than the core
density $n_a=3 N(a)/(4\pi a^3)$  because $N(a)$
and $a$ are much less correlated than $n_a$ and $a$
\citep{mahdavi99}.  The Hernquist surface density profile is
\beqn
\Sigma(<s) = \frac{2 N(a)}{\pi a^2 (s^2-1)^2}[(2+s^2)X(s)-3]
\eeqn
where $a$ is the scale radius and $M$ is the total mass. Note that
$M(a) = M/4$.   We minimize $\chi ^2$ and list the best-fit
parameters $a_N$ for the NFW and Hernquist models
in Table \ref{radii}.  We perform the fits on all data points within
the maximum radii listed in Table \ref{ksamples}.  We plot the surface
number density profiles and the best-fit NFW (solid lines) and
Hernquist (dash-dotted lines) models in Figure \ref{sdens}.  The SIS
(dashed lines) is not normalized and is shown only for comparison.

The best-fit scale radii $a_N$ for both the NFW and Hernquist models
are larger than the best-fit scale radii $a_M$ of the mass profiles in
Paper I for all clusters except A539, where the NFW scale radius is
the same.  In individual clusters, $a_N$ and $a_M$ differ only at
1-3$\sigma$ significance.  However, a K-S test indicates that the
distributions of $a_N$ and $a_M$ are not drawn from the same
population at the 99.6\% confidence level for the NFW model and at
99.95\% confidence for the Hernquist model.  These differences suggest
that mass is more concentrated than light.

Two clusters, A168 and A1367, are poorly fit by Hernquist and NFW
profiles.  They both could be fit by these profiles within $R_{200}$,
but the surface number density outside $R_{200}$ exceeds the predicted
profile.  This result suggests that these clusters are not isolated
from surrounding large-scale structure and that we may be observing
them at an early stage of their evolution.  In fact, both these
clusters contain major mergers.  Furthermore, they are the {\it only}
CAIRNS clusters currently undergoing major mergers.  Excluding these
clusters from the comparison of scale radii only reduces the
significance of the K-S test to 99.5\% for both models.  Thus, the
difference in distribution of scale radii is not solely a result of
these merging systems.  Similarly, A2199 has a large core component.
Excluding the innermost bin slightly increases the best-fit value of
$a_N$ and slightly decreases the best-fit value of $N(a)$.

\citet{bg03} find similar results from a Jeans analysis of an
ensemble cluster constructed from the 2dFGRS: the ratio of mass
density to galaxy (deprojected) number density decreases with radius.
Similarly, \citet{lin04} find that the concentration of galaxies is
smaller than the expected concentration of mass, i.e., the galaxies
are more extended than expected.  From our results and the independent
analyses of these other authors, we thus conclude that the difference
is a real physical effect.

\subsection{Luminosity Profiles}

Because the cluster LF is not significantly different from the field
LF, the estimates of the fraction $f_L$ of the total cluster light
contained in galaxies brighter than the magnitude limits in Table
\ref{ksamples} (which assume the LF parameters of the field LF) are
justified.  We estimate the total light by adding the luminosity in
galaxies brighter than the magnitude limits, then dividing by $f_L$.
We make no corrections for the small incompleteness in our
spectroscopic catalogs ($f_{noz}$ in Table \ref{ksamples}).  This
omission could lead to slight underestimates of the luminosity in the
outskirts of the clusters.  The photometric uncertainties in the
luminosity profiles are $\lesssim$10\%.  Because we compute the
luminosity profiles only from relatively bright galaxies, the
uncertainties are dominated by counting statistics
\citep{kochanek03}.

\begin{figure*}
\figurenum{19}
\label{ldens}
\plotone{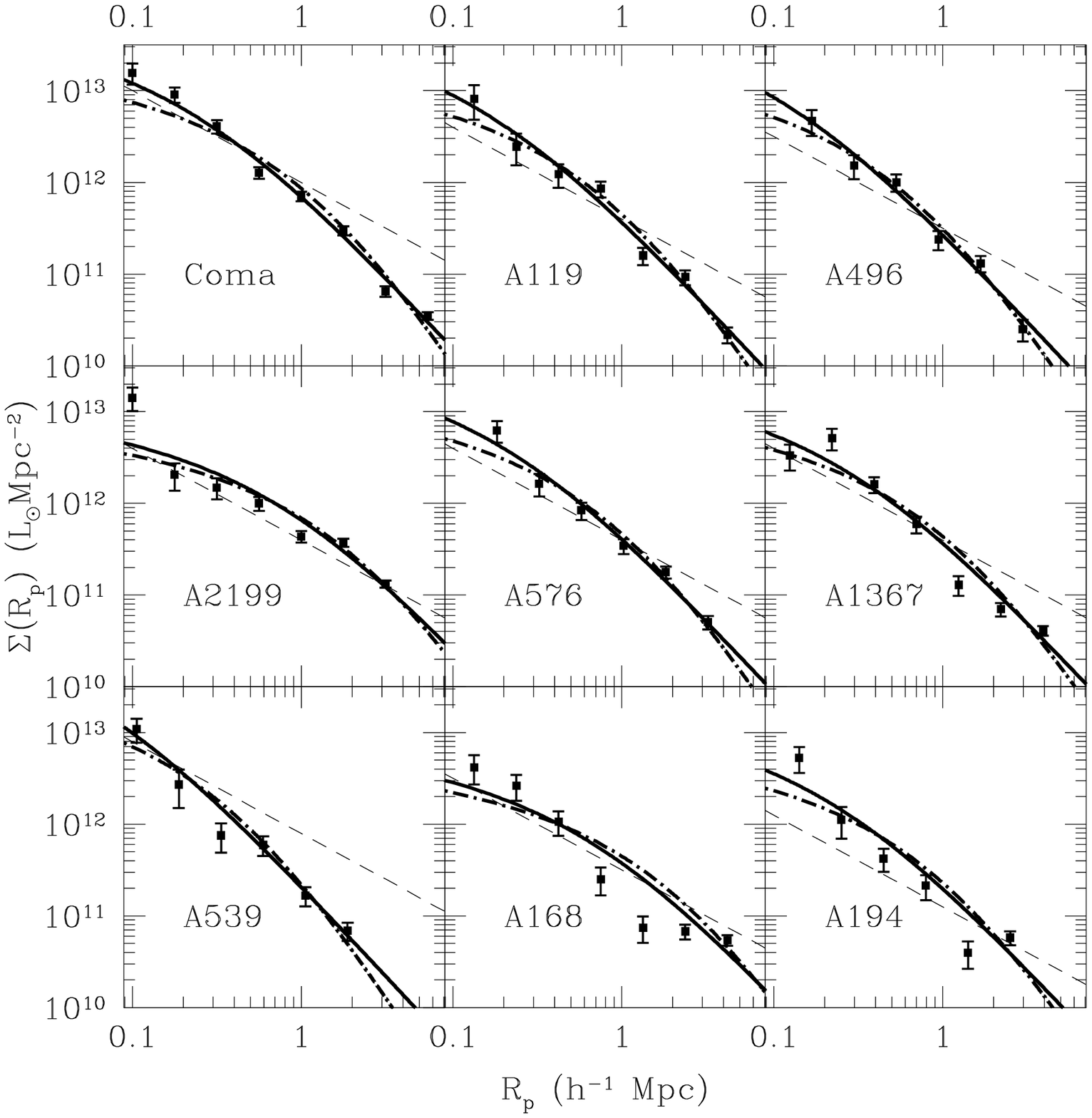}
\caption{Surface luminosity density profiles for cluster members 
brighter than $M_{K_s}=-22.77 + 5 \mbox{log}h$, equivalent to
$M_{K_s}^*+1$ for field galaxies.  The dashed line shows a singular
isothermal sphere.  The solid and dash-dotted lines show the best-fit
NFW and Hernquist profiles.}
\end{figure*}

We fit the luminosity density profiles of the CAIRNS clusters to the
simple analytic models described in the previous section, replacing
$N(a)$ with $L(a)$, the luminosity contained within the sphere of
radius $a_L$.  Figure \ref{ldens} shows the surface luminosity density
profiles and the best-fit NFW (solid lines) and Hernquist (dash-dotted
lines) models.  The scale radii $a_L$ of the light distributions are
close to $a_N$ and larger than $a_M$, again implying that the light in
galaxies is more extended than the mass.  A K-S test indicates that
the distributions of $a_L$ and $a_M$ are not drawn from the same
population at the 98.1\% confidence level for the NFW model and at
99.95\% confidence for the Hernquist model.  A K-S test detects no
differences in the distributions of $a_L$ and $a_N$ for either model.
Again we conclude that the mass is more concentrated than the light.

\section{Near-Infrared Mass-to-Light Profiles}

\begin{figure*}
\figurenum{20}
\label{mlkprof}
\plotone{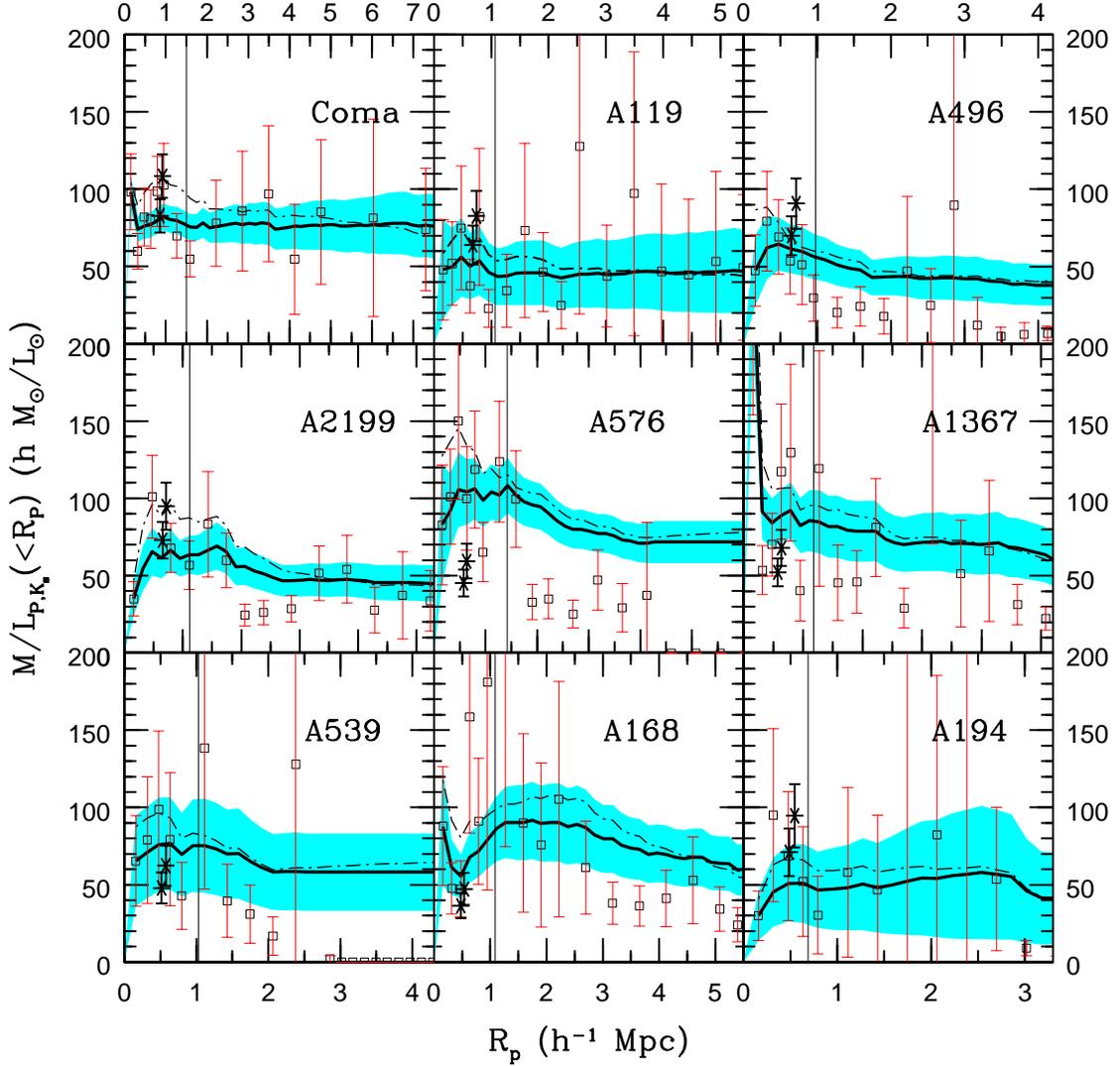}
\caption{Mass-to-light ratio in $K_s$ band as a function of radius for
the CAIRNS clusters. The solid lines show the caustic mass profile
$M(<r)$ divided by the projected luminosity profile $L_{K_s}(<R_p)$
and the shaded regions indicate the associated 1-$\sigma$
uncertainties.  The open squares show the mass-to-light ratio
$M(r,r+dr)/L_{K_s}(R_p,R_p+dR_p)$ in radial shells.  The dash-dotted
line shows the projected best-fit Hernquist mass profile $M_H(<R_p)$
divided by $L_{K_s}(<R_p)$. The stars show the mass-to-light
ratio calculated using the X-ray temperature and the mass-temperature
relation to estimate the mass.  The lower point shows
$M_X(<r_{500})/L_{K_s}(<R_{500})$ and the upper point shows
$M_X(<R_{500})/L_{K_s}(<R_{500})$ assuming $M_X(<R_{500}) = 1.3
M_X(<r_{500})$ as is true for an NFW mass profile with $c$=5.}
\end{figure*}

We next compute $M(<r)/L_{K_s}(<R)$ as a function of radius (in units
of $M_\odot/L_\odot$) using the caustic mass profiles and the
luminosity profiles from the previous section (solid lines in Figure
\ref{mlkprof} with uncertainties shown in shading). The luminosity
profiles $L_{K_s}(<R)$ are projected in two dimensions; the mass
profiles $M(<r)$ are radial profiles.  Thus, these mass-to-light
profiles are the mass in spheres divided by the light in cylinders.
Table \ref{mlktable} summarizes the mass-to-light ratios inside and
outside $r_{200}$ calculated by dividing the caustic masses (in
spheres) by the light profiles (in cylinders).  Without correcting for
this geometric effect, the mean mass-to-light ratio inside $r_{200}$
($70\pm7 h$) is a factor of $1.8\pm0.3$ larger than the mean
mass-to-light ratio outside $r_{200}$ ($38\pm6 h$).  The mean
mass-to-light ratio inside the maximum radius probed, $r_{max}$, is
$53\pm5 h$.

One notable feature of Figure \ref{mlkprof} is the variety in the
shapes of the mass-to-light profiles in individual clusters.  Some
clusters have flat profiles while others have strongly peaked
profiles.  There is no obvious cause of these differences (e.g.,
presence of a bright cD galaxy, presence of a major merger).  Indeed,
significant variation in the shapes of mass-to-light profiles obtained
with the caustic technique is expected from projection effects along
different lines of sight (D99).

Obviously, it is preferable to compute both the mass and the light in
either spheres or cylinders but not one of each.  Because mass and
light are never negative, $L_{K_s}(<r) \leq L_{K_s}(<R_p)$ and $M(<r)
\leq M(<R_p)$.  Thus, this geometric effect should artificially
decrease the ``observed'' mass-to-light ratios in the centers of the
clusters.  To correct for this geometric effect, one must make
assumptions about the shapes of the profiles and their symmetries.  We
prefer to present the data with few manipulations.  We thus project
the mass profiles into cylinders rather than attempting to deproject
the noisy luminosity profiles.  In particular, we assume that the mass
distribution is well-described by one of the simple mass models.
Because the Hernquist profile has a finite total mass, the best-fit
Hernquist profiles have smaller densities than the best-fit NFW
profiles at large radii.  Thus, the projected Hernquist profile is
more centrally concentrated than the NFW profile.  These projection
effects will thus be larger if the true profile is a Hernquist
profile.  We show these projected $M_H(<R_p)/L_{K_s}(<R)$
mass-to-light profiles as dash-dotted lines in Figure \ref{mlkprof}.
As expected, these profiles have larger mass-to-light ratios at small
radii than the spheres-by-cylinders profiles.  If the decreasing
shapes of these profiles are correct, the deprojected mass-to-light
profiles $M(<r)/L(<r)$ should decrease slightly faster than in
projection.

We quantify the size of this effect for NFW mass profiles.  For NFW
profiles with concentrations $c=5-20$ (in Paper I we measured $c=5-17$
for the CAIRNS clusters), the projected mass within a cylinder of
radius $R_{200}$ is a factor of 1.15-1.25 larger than the mass in a
sphere of radius $r_{200}$ (the factor increases with decreasing
concentration).  Projection effects are less dramatic for cylindrical
shells compared to spherical shells at large radii because some
light/mass outside the spherical shell is projected into the
cylindrical shell; some light/mass within the spherical shell is
projected into cylindrical shells at smaller radii.  For NFW halos
with concentrations $c=5-20$, the projected mass in the cylindrical
shell bounded by $R_{200}$ and $R_{max}$ is 1--5\% greater than the
mass in the spherical shell bounded by $r_{200}$ and $r_{max}$.  Thus,
if the CAIRNS clusters are well-described by NFW profiles with
$c\approx$5, the mass-to-light ratio inside the cylinder $R_{200}$ is
larger by a factor of 1.15-1.25 than the measured quantity
$M(<r_{200})/L(<R_{200})$.  Similarly, the mass-to-light ratio in the
cylindrical shell bounded by $R_{200}$ and $R_{max}$ is larger by a
factor of 1.01-1.05 than the measured quantity
$[M(<r_{max})-M(<r_{200})]/[L(<R_{max})-L(<R_{200})]$.  The difference
in mass-to-light ratios inside and outside $R_{200}$ is therefore
larger than calculated above;
\begin{equation}
\frac{(M/L)(<R_{200})}{(M/L)(>R_{200})} \approx 1.2 \frac{M(<r_{200})/L(<R_{200})}{M(>r_{200})/L(>R_{200})} \approx 2.2\pm0.4.
\end{equation}
The projection effects therefore aggravate the difference in
mass-to-light ratios between cluster virial regions and their
outskirts.

The preceding calculation used the nonparametric mass profiles from
Paper I.  We repeat the calculations of Table \ref{mlktable} using the
best-fit (parametric) NFW mass profiles from Paper I projected into
cylinders.  The mean mass-to-light ratio inside $r_{200}$ is
$M_{NFW}/L|_{r_{200}}=77\pm7 h$ and the mean mass-to-light ratio
outside $r_{200}$ is $48\pm 5 h$.  The ratio of these is $1.6\pm0.2$,
very similar to the ratio $1.8 \pm 0.3$ calculated above with no
corrections for geometric projection effects.  We note here that the
best-fit NFW parameters do not vary significantly if the fits are
restricted to $r<1.5~\Mpc$; at these radii, the mass profiles agree
with X-ray and virial mass estimates (see Paper I).

We now calculate the mass-to-light profile in individual shells.  The
enclosed mass-to-light profiles calculated above decrease with radius.
Because these profiles are dominated by the mass-to-light ratio in the
core, the mass-to-light ratio in the outer shells must generally be
smaller than the enclosed mass-to-light ratio at that radius.  The
uncertainties in individual shells are sufficiently large that we must
bin several shells to obtain a significant signal.  These
mass-to-light ratios are the mass in spherical shells divided by the
light in cylindrical shells.  As noted above, for spherically
symmetric NFW models, the projection effects decrease with radius and
lead to underestimates of the mass-to-light ratio inside $r_{200}$.
Open squares (Figure \ref{mlkprof}) show the mass-to-light ratios of
these shells.  Indeed, there is a general trend for lower
mass-to-light ratios in shells at larger radii, but the uncertainties
are quite large.  Note that the enclosed mass-to-light profiles
$M(<r)/L(<R_p)$ are weighted by mass and light and therefore differ
from the (unweighted) profiles of mass-to-light ratios in individual
shells.  For some clusters (e.g., Coma and A119), the mass-to-light
ratios of shells at large radii agree well with the enclosed
mass-to-light profile, while for others (e.g., A496 and A576) the
mass-to-light ratios in shells at large radii are significantly lower
than the enclosed mass-to-light profile.  This variety is likely due
to projection effects, namely the large variety in the appearance of
the caustic pattern for an individual cluster viewed from different
lines of sight (D99).  This variety suggests that the shapes of
individual mass-to-light profiles should not be taken too seriously;
however, the average mass-to-light profile should be unbiased (D99).

X-ray mass estimates provide an independent check of our mass-to-light
ratios within $r_{500}$.  As shown in Paper I, the caustic mass
profiles evaluated at $r_{500}$ agree quite well with X-ray mass
estimates from the mass-temperature relation.  Thus, it is no surprise
that the mass-to-light ratios calculated from the X-ray mass $M_{500}$
and the luminosity evaluated at $L_{500}$ yield similar values.  Note
that, again, the luminosity estimates include all galaxies projected
within $R_{500}$.  An NFW profile with concentration $c$=5 appropriate
for clusters \citep{nfw97} would have a deprojected mass-to-light
ratio $(M/L)(<r_{500}) \approx 1.3 M_{500}/L(<R_{500})$ (a higher
concentration of $c$=20) reduces this factor from 1.3 to 1.2). We show
both the projected and deprojected estimates as stars in Figure
\ref{mlkprof}.  Note that $(M/L)(<r_{500})\approx
(M/L)(<R_{500})$; these points may be compared with the profiles of
mass in cylinders divided by light in cylinders.  The mean deprojected
mass-to-light ratio is $(M/L)(<r_{500})=78\pm 7 \mbox{(statistical)} h
M_\odot/L_\odot$.  This result agrees with the deprojected
mass-to-light ratio $(M/L)(<r_{200})=88\pm 9 h$ (statistical) at
$r_{200}$ taken from the caustic mass profiles assuming a correction
of 1.25 appropriate for a $c$=5 NFW halo.  The agreement between the
mass-to-light ratios within $r_{500}$ and within $r_{200}$ suggests
that mass-to-light ratios are reasonably constant throughout the
virial region of a cluster (see also $\S 6.2$).  That is, the observed
decrease in mass-to-light ratios with radius is not monotonic but may
begin only at roughly $r_{200}$.

Our results are in excellent agreement with \citet[][hereafter
L03]{lin03}, who find a mass-to-light ratio $M/L_{K_s}=76\pm 4 h$
(statistical) at $r_{500}$ for hot ($T_X\geq 3.7$keV) clusters using
X-ray masses and 2MASS photometry from a larger cluster sample.  We
include only galaxies within the caustics in our luminosity estimates;
L03 use statistical background subtraction to correct their luminosity
estimates.  The close agreement shows that the methods L03 use to
subtract background galaxies do not introduce a bias in the luminosity
estimates.

Previously, we have used the caustic technique to calculate
mass-to-light profiles in R band for A576 \citep{rines2000} and in
$K_s$ band for Coma \citep{rines01a}.  In A576, we found a steeply
decreasing mass-to-light profile in $R$ band.  In Coma, we found a
flat profile but noted that the systematic effects allowed for a
decreasing profile.  The results we derive here for these clusters are
consistent with these earlier determinations.

Other investigators have applied Jeans analysis to ensemble clusters
to test for variations in the mass-to-light ratio.  This effort is
complicated by the fact that one needs to assume an orbital
distribution to measure variations in $M/L$.  \citet{cye97} and
\citet{2000AJ....119.2038V} find that light traces mass in the CNOC1
ensemble cluster (composed of massive clusters at $z=0.2-0.5$) to a
radius of $2r_{200}$.  \citet{bg03} construct an ensemble cluster from
poor clusters in the 2dFGRS.  They find that the ratio of the mass
density to the galaxy number density decreases with radius to
$2r_{200}$, similar to our result for the CAIRNS clusters.  When they
exclude late-type galaxies from the galaxy number density, the ratio
is roughly constant. \citet{katgert03} construct an ensemble cluster
from the ESO (European Southern Observatory) Nearby Abell Cluster
Survey and find that the mass-to-light ratio decreases with radius in
the range $0.2-1.5 r_{200}$, although the mass-to-light ratio is
roughly constant when late-type galaxies are excluded.

Weak lensing provides an independent estimate of mass-to-light ratios
on large scales that does not depend on the dynamical state of the
system. \citet{kaiserxx} and \citet{gray02} estimate the
mass-to-light ratios of superclusters with weak lensing.
\citet{kaiserxx} find $M/L_B= 280 \pm 40 h$ for light in early-type
galaxies. Assuming a typical early-type color of $B-K_s=3.7$
\citep{j00}, this value corresponds to $M/L_{K_s}\approx 64\pm9 h$.
Including late-type galaxies would decrease this
ratio. \citet{wilson01} find similar results from weak lensing in
blank fields. \citet{gray02} find $M/L_B \sim 200 h$ (early-type light
only) for individual clusters; when they cross-correlate mass and
light they find $M/L_B=130 h$ (early-type light only), but they caution
that there are many systematic uncertainties in this estimate.
Recently, \citet{kneib03} used weak lensing to estimate the mass
profile of CL0024+1654 to a radius of 3.25$\Mpc$.
\citet{kneib03} conclude that the $K$ band mass-to-light ratio is roughly
constant on these scales.  Assuming passive evolution, their
mass-to-light ratio corresponds to $65\pm 9 h$ ($74\pm10 h$ for red
sequence galaxies only) at $z=0$, intermediate between our estimates
of the mass-to-light ratio inside and outside $r_{200}$.

\citet{bahcall2000} use simulations to show that cluster
mass-to-light ratios in B band exceed the global value due to the
older, less luminous stellar populations found in clusters.  Cluster
mass-to-light ratios measured in $K_s$ band should then be much closer
to the global value because $K_s$ band light has a much weaker
dependence on stellar population ages \citep[e.g.,][]{bell01}.  If
stellar populations are the primary cause of the decreasing
mass-to-light profiles in the simulations of \citet{bahcall2000}, the
CAIRNS clusters should have roughly flat $K_s$ band mass-to-light
profiles.  Thus, the similarity of the {\it decreasing} $K_s$ band
mass-to-light profiles of the CAIRNS clusters to the simulations of
\citet{bahcall2000} shows that the decreasing profiles in their
simulations may not result primarily from differences in the stellar
populations but from differences in the relative distribution of dark
matter and galaxies.

The CAIRNS sample is unique in both the completeness of the individual
cluster catalogs and in the near-infrared digital photometry used to
avoid stellar population effects.  The mass-to-light ratios of the
CAIRNS clusters decrease with radius and the mass-to-light ratios
inside the virial regions agree with other estimates at optical and
near-infrared wavelengths \citep[see also][]{rines2000,rines01a}.  The
decreasing mass-to-light profiles are consistent with results from
other cluster studies.  We discuss these results in more detail in $\S
7$.

\section{Properties of the Virial Regions}

\subsection{The Halo Occupation Distribution}

The halo occupation distribution \citep[see the review
by][]{2002PhR...372....1C} is an important input for converting the
results of numerical simulations into observables
\citep[e.g.,][and references therein]{peacock00,berlind02,berlind03}.
The simplest prediction is that the number of galaxies formed is
directly proportional to the available baryonic mass.  Thus, the
number of galaxies $N$ (brighter than some minimum mass or luminosity)
contained in a halo of mass $M$ is given by $N\propto M$ (i.e., the
efficiency of galaxy formation is a universal constant for
sufficiently massive haloes).  If galaxy formation is more efficient
in the most massive haloes, then the relation might be $N\propto
M^\mu$ with $\mu>1$.  Conversely, if galaxy formation is less
efficient in massive haloes (e.g., if the gas is heated by the halo
potential and is unable to collapse into galaxies) or galaxy
disruption is more efficient (e.g., dynamical friction and tidal
stripping), then the relation might be $N\propto M^\mu$ with $\mu<1$.
Models of the halo occupation distribution suggest that, for cluster
mass halos, the relation is close to a power law with slope $\mu<1$.
Semi-analytic models predict $\mu \sim 0.8-0.9$
\citep{sheth01,berlind03}.  A smoothed particle hydrodynamics
simulation of a $\Lambda$CDM cosmological model predicts halo
occupation distributions with $\mu \sim 0.56-0.74$ for cluster mass
halos, similar to the values for a different set of semi-analytic
models \citep{berlind03}.  \citet{2003MNRAS.339..312S} show that
numerical simulations predict suppression of galaxy formation in the
most massive halos because gas cannot cool and collapse into galaxies.

One of the few previous determinations of the relation between virial
masses and galaxy numbers is that of \citet{marinoni02}, who compute
masses and (blue) luminosities of virialized objects in the Nearby
Optical Catalog.  \citet{marinoni02} find $N \propto M^{0.55\pm0.03}$,
similar to the semi-analytic models.  \citet{kochanek03} use a
constrained numerical simulation of 2MASS to develop a matched filter
algorithm to study cluster properties in the 2MASS catalog and
heterogeneous auxiliary observations from the literature (e.g.,
redshifts and X-ray properties).  Their best-fit relation between
cluster mass and number of members is $N \propto M^{1.10\pm0.09}$.
\citet{pisani03} find $N \propto M^{0.70\pm0.04}$ in a sample of
groups, although mass estimates of groups are very uncertain.
Recently, \citet{lin04} analyzed the halo occupation distribution for
clusters with 2MASS photometry and X-ray mass estimates.  They find $N
\propto M^{0.84\pm0.04}$, steeper than
\citet{marinoni02} but still in reasonable agreement with models.

\begin{figure*}
\figurenum{21}
\label{hof}
\plotone{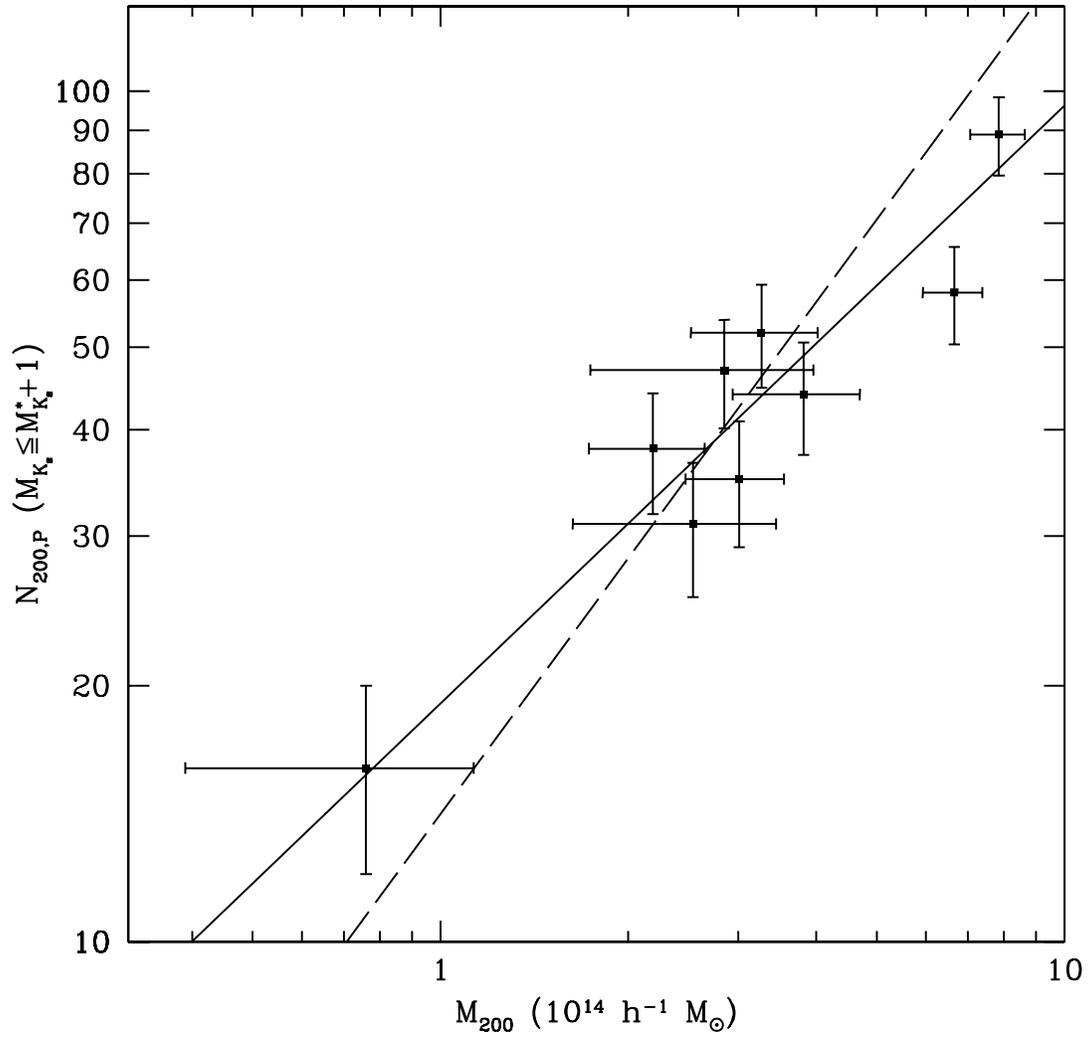}
\caption{Number of bright galaxies $N_{200,P}$ projected within 
$R_{200}$ versus $M_{200}$ for the CAIRNS clusters.  The solid line
shows the bisector of the two ordinary least-squares fits to the data.
The dashed line shows $N_{200} \propto M_{200}$.}
\end{figure*}

We can constrain the halo occupation distribution with the CAIRNS
clusters, which cover roughly an order of magnitude in mass and have
both accurate photometry and complete spectroscopy.  Our results have
the advantages of uniform sky coverage, greater redshift completeness,
and galaxy selection at near-infrared wavelengths, which is a better
tracer of stellar mass and suffers less dust extinction than blue
light.  Conveniently, the magnitude limit we adopt ($M_{K_s}\leq
M_{K_s}^*+1$) is very similar to the luminosity threshold used in
\citet{berlind02} and one of the thresholds used in \citet{berlind03},
$L\gtrsim 0.5 L_*$.  Figure \ref{hof} shows the number of galaxies
$N_{200}$ projected within $R_{200}$ versus $M_{200}$, the mass of the
halo.  We do not attempt to deproject the number density profiles to
obtain a deprojected estimate of $N_{200}$ because the number density
profiles are too noisy.  If all haloes have similar concentrations,
then the fraction of interlopers should be constant with mass.  If the
halo concentration decreases with mass (as expected for NFW models),
then the fraction of interlopers should increase with mass. In this
case, the fit to $\mu$ would be an overestimate.  The bisector of the
two ordinary least-squares fits \citep{1992ApJ...397...55F} yields
$N_{200} \propto M_{200}^{0.70\pm0.09}$, 3.3$\sigma$ shallower than a
linear relation (shown by a dashed line in Figure \ref{hof}).  This
result is not driven by A194, the least massive cluster; excluding
this cluster yields a least squares fit $N_{200} \propto
M_{200}^{0.74\pm0.15}$.  This result agrees well with previous
determinations as well as with expectations from semi-analytic models
for galaxy formation
\citep[e.g.,][]{kauffmann1999a,sheth01,marinoni02,berlind03,pisani03,lin04}.
We speculate that the significant difference from \citet{kochanek03}
is due to the systematic uncertainties from the process of matching
their simulation to the observations. \citet{kochanek03} use a matched
filter algorithm which is finely tuned to reproduce the expected
properties of clusters based on simulations (where galaxies trace the
dark matter distribution).  Systematic effects can arise both from
mismatches in the assumed and true cosmology and recipes for galaxy
formation as well as unknown systematics in the heterogeneous
auxiliary observations.

The comparison with \citet{lin04} is especially interesting because
both datasets use 2MASS photometry.  \citet{lin04} use a much larger
sample of clusters but they use only statistical background
subtraction whereas we study fewer clusters but use complete
spectroscopic information to assign cluster membership.  A detailed
comparison of these two methods would be instructive but it lies
beyond the scope of this paper.  In particular, there are few clusters
in both samples, so cluster-to-cluster variations could significantly
affect the comparisons.  We refer the reader to \citet{lin04} for an
excellent discussion of the physical significance of a non-linear HOF
as well as the observational implications for clusters.

\subsection{Mass Dependence of Mass-to-Light Ratios}

\begin{figure*}
\figurenum{22}
\label{mlk}
\plotone{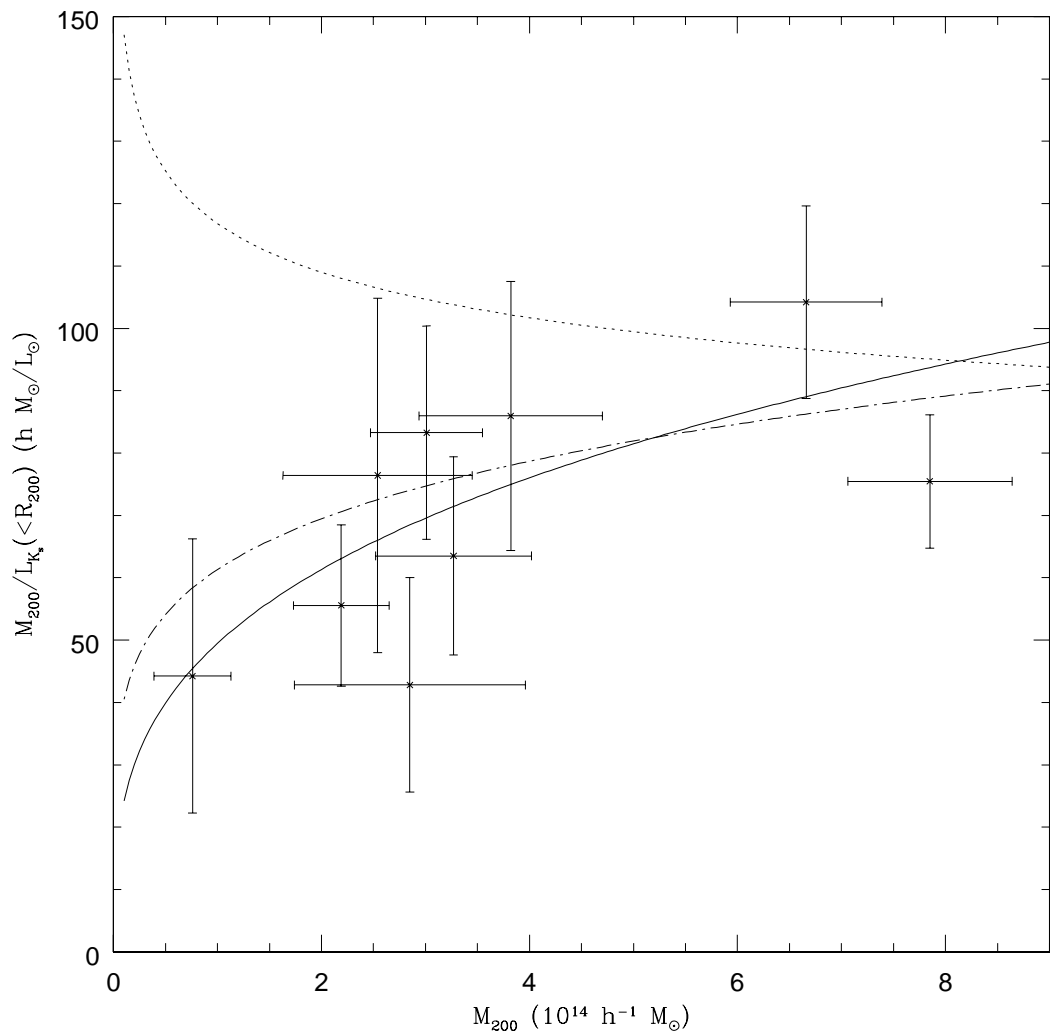}
\caption{Mass to light ratio versus $M_{200}$ for the CAIRNS clusters. 
 The solid curve shows the relation for X-ray clusters in L03
 (converted to $M_{200}$), the dotted line the relation for 2MASS
 clusters with the model of \citet{kochanek03}, and the dash-dotted
 line shows the $M/L-T_X$ relation of \citet{bahcall02} converted to
 $M/L-M$ using the mass-temperature relation of \citet{frb2001}.}
\end{figure*}

Figure \ref{mlk} shows the mass-to-light ratio within $r_{200}$ versus
$M_{200}$ for the CAIRNS clusters.  The scatter is large, but the
CAIRNS clusters show an increase in M/L (evaluated at $r_{200}$) with
increasing mass.  \citet[][hereafter L03]{lin03} found a similar
correlation between X-ray mass and near-infrared mass-to-light ratios;
more massive clusters have larger mass-to-light ratios with a best-fit
relation
\begin{equation}
(M/L_{K_s})(<r_{500}) = (67 \pm 4) h 
(\frac{M_{500}}{2.1 \times 10^{14} \msun})^{0.31\pm0.09}.  
\end{equation}
Figure \ref{mlk} shows this relation assuming that the mass-to-light
ratios within $r_{500}$ and $r_{200}$ are identical, $M_{200}\approx 3
M_{500}$, and multiplying by 0.8 to convert to the
spheres-by-cylinders measured here.  The CAIRNS clusters follow this
relation quite closely, showing that the mass estimator used (X-ray
versus virial/caustic mass) does not affect the correlation of
mass-to-light ratio with cluster mass.  The close agreement also
demonstrates that cluster mass-to-light ratios do not change
dramatically between $r_{500}$ and $r_{200}$ (see also $\S$ 5).

A compilation of virial masses and luminosities by
\citet{g2000} yields $(M/L_{B_j})\propto M^{0.17-0.23}$ Similarly,
\citet{bahcall02} use a heterogeneous catalog to derive a dependence
of (optical) M/L on X-ray temperature which they attribute to
differences in the ages of stellar populations.  One can convert their
relation into a $M/L-M$ relation with the X-ray mass-temperature
relation \citep{frb2001}. Specifically, $(M/L) \propto
T_X^{0.30\pm0.08}$ and $M \propto T_X^{1.64\pm0.04}$ yield
$(M/L)|r_{200} \propto M_{200}^{0.18\pm0.05}$, slightly shallower than
but in agreement with the L03 relation and the CAIRNS clusters (Figure
\ref{mlk}).  Note, however, that the CAIRNS relation has little
dependence on the ages of the stellar populations, counter to the
conclusion of \citet{bahcall02}.  If differences in stellar
populations produce the $M/L-M$ relation, the slope of the relation
should be steeper for $M/L_B$ than for $M/L_{K_s}$.

In contrast, \citet{kochanek03} find that mass-to-light ratios are
smaller in more massive clusters; they find a best-fit relation of
$(M/L_{K_s})(<r_{200}) = 116\pm46 (M_{200}/10^{15}
\msun)^{-0.10\pm0.09}$.  We multiply this relation by 0.8 to convert
to the spheres-by-cylinders measured here (Figure \ref{mlk}).  The
CAIRNS clusters follow the relation found by L03, \citet{g2000}, and
\citet{bahcall2000}, and exclude the relation of \citet{kochanek03}.
The disagreement with \citet{kochanek03} is perhaps not surprising
given the disagreement between their halo occupation function and that
of the CAIRNS clusters found in the previous section.

\begin{figure*}
\figurenum{23}
\label{nl}
\plotone{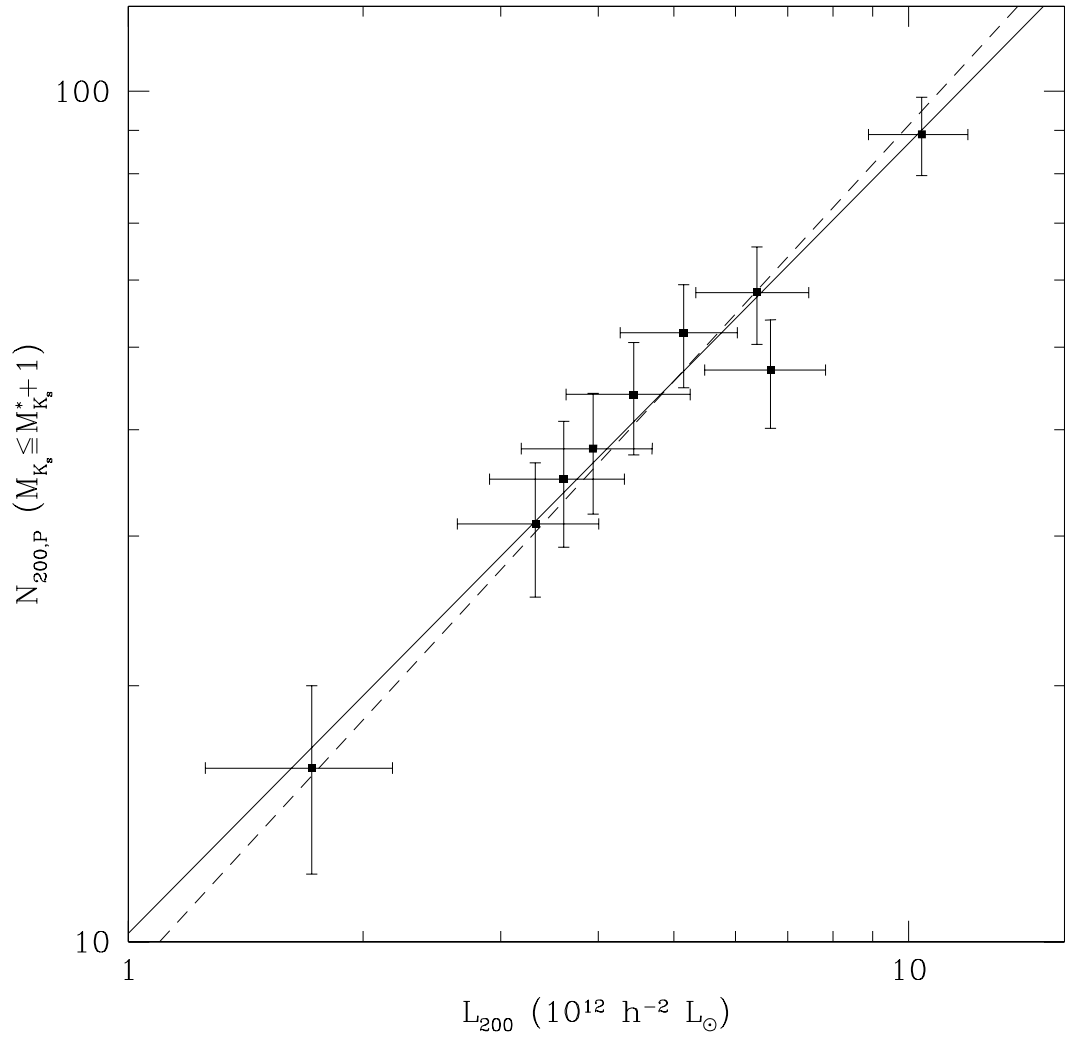}
\caption{Luminosity within $R_{200}$ versus number of bright galaxies 
within $R_{200}$.  The solid line shows the bisector of the two
ordinary least-squares fits and the dashed line shows $N_{200}\propto
L_{200}$. }
\end{figure*}

Figure \ref{nl} shows the relation between $L_{200}$, and $N_{200}$,
the number of bright galaxies ($M_{K_s}<-22.77 + 5 \mbox{log} h$)
projected within $R_{200}$.  The bisector of the ordinary
least-squares fits is $N_{200}\propto L_{200}^{0.93\pm0.07}$,
consistent with a slope of unity.  This result underscores the result
of $\S 3.1$; the cluster-to-cluster variations in the LF are small.
Galaxy formation is suppressed (and/or that the efficiency of galaxy
disruption is enhanced) in massive clusters, with greater suppression
for more massive clusters.  The correlation of mass-to-light ratio
with mass is then a natural byproduct of the correlation of
$M_{200}/N_{200}$ with mass and a universal LF.

These results are consistent with the decreasing mass-to-light
profiles found in $\S 5$.  These decreasing profiles imply that
cluster infall regions, which contain galaxies formed in environments
with smaller virial temperatures than galaxies in the virial regions,
have smaller mass-to-light ratios.  The presence of X-ray groups in
cluster infall regions \citep{rines01b,rines02} demonstrates the
overlap between cluster infall regions and low-mass clusters.  In $\S
6.1$, we show that the number of galaxies within $R_{200}$ increases
more slowly than the cluster mass.  These results all suggest that the
efficiency of galaxy formation is suppressed \citep[see the numerical
simulations by][]{2003MNRAS.339..312S} and/or that the efficiency of
galaxy disruption is enhanced \citep[see the numerical simulations
by][]{kk1999,1999ApJ...523...32C} in environments with larger virial
temperatures.  In the latter case, the contribution of intracluster
stars to the cluster light budget can be substantial (5-50\%, see $\S
7.4$).  Neglecting this contribution (the normal procedure and the one
adopted here) may result in a severe underestimate of the total light
in the cluster.

\section{Discussion}

\subsection{Predicting Mass Profiles From the Galaxy Distributions}

\begin{figure*}
\figurenum{24}
\label{allvdp2}
\plotone{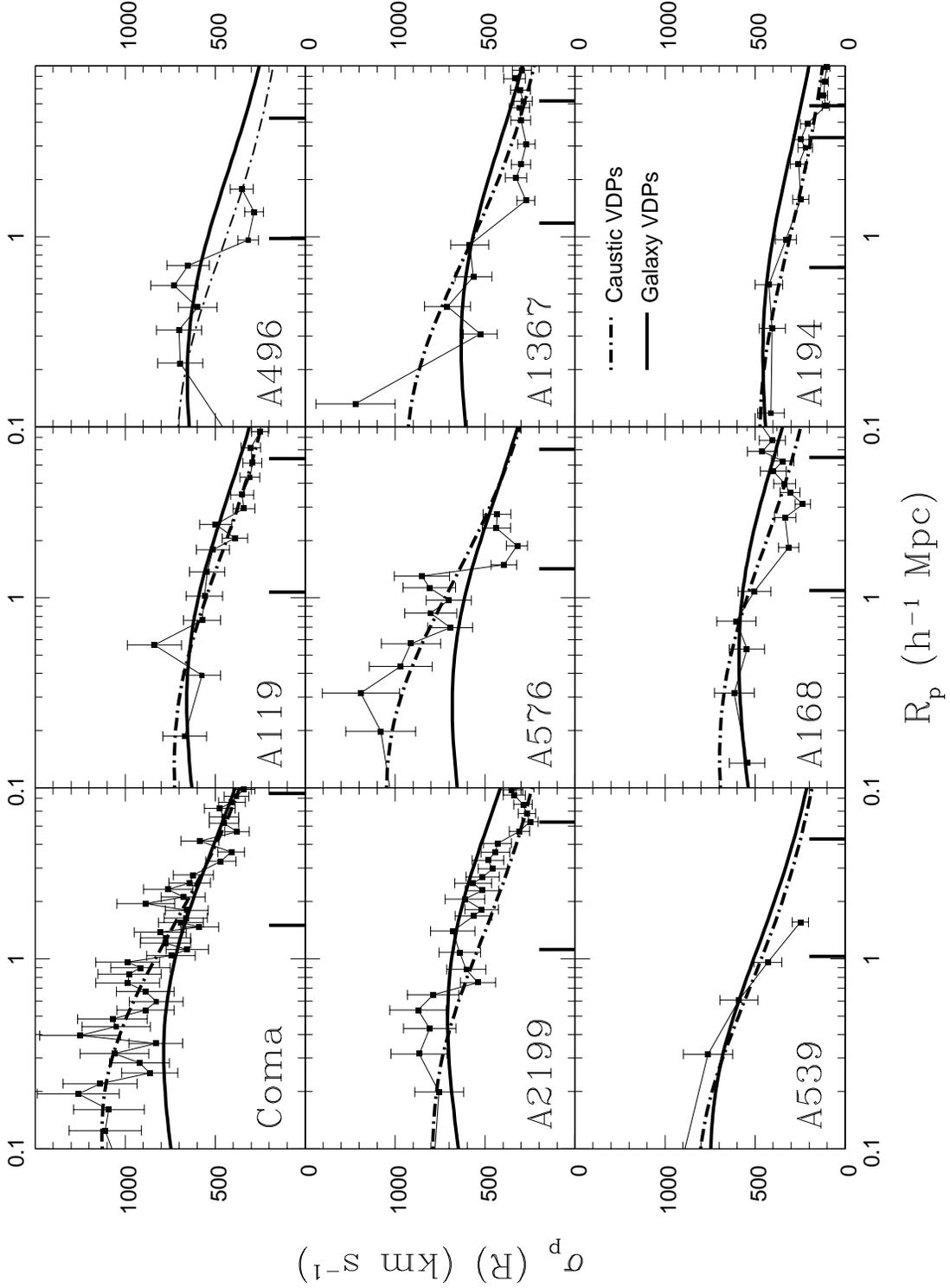}
\caption{Velocity dispersion profiles for the CAIRNS clusters taken
from Paper I.  The bars on the abscissa indicate $r_{200}$ and $r_t$.
The dash-dotted lines are the VDPs of the Hernquist mass profiles that
best fit the caustic mass profiles assuming isotropic orbits.  The
solid lines show the Hernquist VDP predicted from the surface number
density profiles assuming isotropic orbits and a constant ratio of
mass to number density.}  
\end{figure*}

Because the caustic technique is relatively new, we test the
consistency of our results with velocity dispersion profiles (VDPs), a
more traditional tool of galactic dynamics.  Here we predict
the mass profiles based on the observed distribution of galaxies
assuming that they trace the mass.  If the radial variations in the
mass-to-light ratio ($\S 5$) are real, then the mass profiles
calculated from the galaxy distributions (either number density or
luminosity density) assuming a constant mass-to-light ratio should
differ demonstrably from those in Paper I.  We reproduce the VDPs from
Paper I in Figure \ref{allvdp2} (dash-dotted lines).

In $\S 4.1$ above, we fit the surface number density profiles of the
bright galaxy distribution.  The scale and normalization of the
profiles predict the shape of the velocity dispersion profiles (VDPs)
of the clusters under the assumption of isotropic orbits and a
(globally) constant ratio $M/N$ of mass to number of bright galaxies
($M_{K_s}<-22.77 + 5 \mbox{log} h$).  We calculate the value of this ratio
within $r_{200}$ and find $(M/N)|_{r200} = (7.8\pm0.8)
\times 10^{12} \msun \mbox{galaxy}^{-1}$.  Figure \ref{allvdp2}
displays these predicted VDPs (solid lines) along with the observed
VDPs and the VDPs of the best-fit Hernquist mass profiles from Paper I
(dash-dotted lines).  As noted in Paper I, the VDPs predicted by the
caustic mass profiles agree well with the observed VDPs.  The VDPs
predicted from the surface density profiles, however, do not agree
with the observed VDPs, especially in Coma and A576, where the
predicted VDPs lie substantially below the observations.

This disagreement may result from anisotropic orbits or from
cluster-to-cluster variations in $M/N$ (we find such variations in $\S
6.1$).  Anisotropic orbits would affect both the shapes and
normalizations of the predicted VDPs, but variations in $M/N$ only
affect the normalizations (assuming that $M/N$ is constant with radius
in each cluster).  Figure \ref{nl} shows that $(M/N)|_{r200}$
increases with $M_{200}$.  Correcting for this trend reconciles some
of the differences in Figure \ref{allvdp2}, i.e., the most discrepant
clusters are those with the highest masses.  Substituting a higher
$(M/N)|_{r200}$ increases the amplitude of the predicted VDPs and
brings the predicted and observed VDPs into better agreement.
Outside $r_{200}$, galaxies are not relaxed. At these radii,
VDPs do not necessarily contain information about the orbital
distribution.  Thus, the VDPs outside $r_{200}$ should not be
considered strong constraints on the mass and orbital distributions.


The most straightforward predictions of VDPs based on the assumption
that light traces mass disagree with the observed VDPs.  In contrast,
the VDPs predicted by the caustic mass profiles agree well with the
observed VDPs.  Thus, the decrease in the efficiency of galaxy
formation (and/or the increase in the efficiency of galaxy disruption)
for haloes with larger virial temperatures is not an artifact of the
caustic technique.  Velocity dispersion profiles, a more traditional
tool of galactic dynamics, also indicate a discrepancy between the
distribution of galaxies and mass (although subject to possible biases
from the orbital distribution and/or the dynamical state of galaxies
outside $r_{200}$).  It is interesting that \citet{katgert03} also
find a decreasing mass-to-light profile for an ensemble cluster using
Jeans analysis to compute the mass profile.  Their results strengthen
our conclusion that the decreasing mass-to-light profiles are physical
effects.

\subsection{Morphological Gradients}

Because of the well-known correlation between morphology and density,
we expect larger fractions of late-type galaxies with increasing
clustrocentric radius (decreasing density).  If the LFs of early-type
and late-type galaxies differ significantly, the total LF
should vary with clustrocentric radius.  That is, the LF in cluster
centers should closely resemble the early-type LF, whereas at larger
radii it should resemble the late-type LF.  K01 separate the LF into
early-type and late-type LFs and find that fits to Schechter functions
yield a brighter $M_{K_s}^*$ for the early-type LF; the
faint-end slope is slightly shallower for the late-type LF \citep[but
see][who find that 2MASS misses many blue low surface brightness
galaxies present in SDSS]{bell03}.  

Because our spectroscopic surveys extend to fixed absolute magnitudes,
the correction for light in galaxies fainter than our limiting
magnitudes changes with radius.  Adopting the type-dependent LFs of
K01, the correction becomes larger with radius provided the limiting
magnitude is $M_{K_s}\lesssim -10$ (note that the type-dependent LFs
are only constrained for $M_{K_s,iso}\leq -20.5$).  That is, a
magnitude-limited survey misses more light at large clustrocentric
radii.  If the fraction of early-type galaxies changes from 1 to 0 (a
huge overestimate), the correction changes by $\sim$20\% ($\sim$10\%)
for a magnitude limit of $M_{K_s,iso}$=-22.5 (-21.5), approximately
the magnitude limits for the CAIRNS clusters.  Correcting for this
effect in $\S 5$ would add even more light to the cluster outskirts
and lead to more steeply decreasing profiles.  Thus, the decreasing
mass-to-light profiles in $\S 5$ cannot be explained by the
morphology-density relation.

\subsection{Stellar Populations and Correcting for Faint Galaxies}

Profiles of the ratio of dark matter to stellar mass can be used both
to estimate $\Omega_m$ and to constrain prescriptions for galaxy
formation.  If $K_s$-band light traces stellar mass exactly, the
results of $\S 5$ indicate that the efficiency of star formation is
reduced in dense cluster environments.  However, the properties of
galaxies change rapidly with increasing distance from cluster centers
\citep[][and references therein]{balogh03}.  In particular, the
stellar mass-to-light ratio is smaller in late-type galaxies than in
early-type galaxies by up to a factor of two.  Late-type galaxies are
much more common in the field than in clusters.  Thus, the mean
stellar mass-to-light ratio should decrease with radius.

Because stellar populations are younger at larger clustrocentric
radii, mass-to-optical-light profiles might decrease with radius
\citep{bahcall2000} even if the ratio of gravitational mass to stellar
mass is constant.  Thus, the total-to-stellar mass profiles of the
CAIRNS clusters may decrease less steeply than the mass-to-light
profiles.  Here, we test for radial gradients in the stellar
mass-to-light ratio in A576 and estimate the magnitude of this effect
in $K_s$ band.

\subsubsection{A Test in A576 and the Importance of Faint Galaxies}

We showed above ($\S 3.3$) that there are no obvious gradients in the
$R-K_s$ colors of galaxies in A576.  We test for gradients in the
stellar mass-to-light ratio directly by comparing the mass-to-light
profiles in an optical band (R) and near-infrared bands.  If there were
a significant gradient in stellar mass-to-light ratios, the
near-infrared profile would be flatter than at optical wavelengths.

\begin{figure*} 
\figurenum{25}
\label{mlcomp} 
\plotone{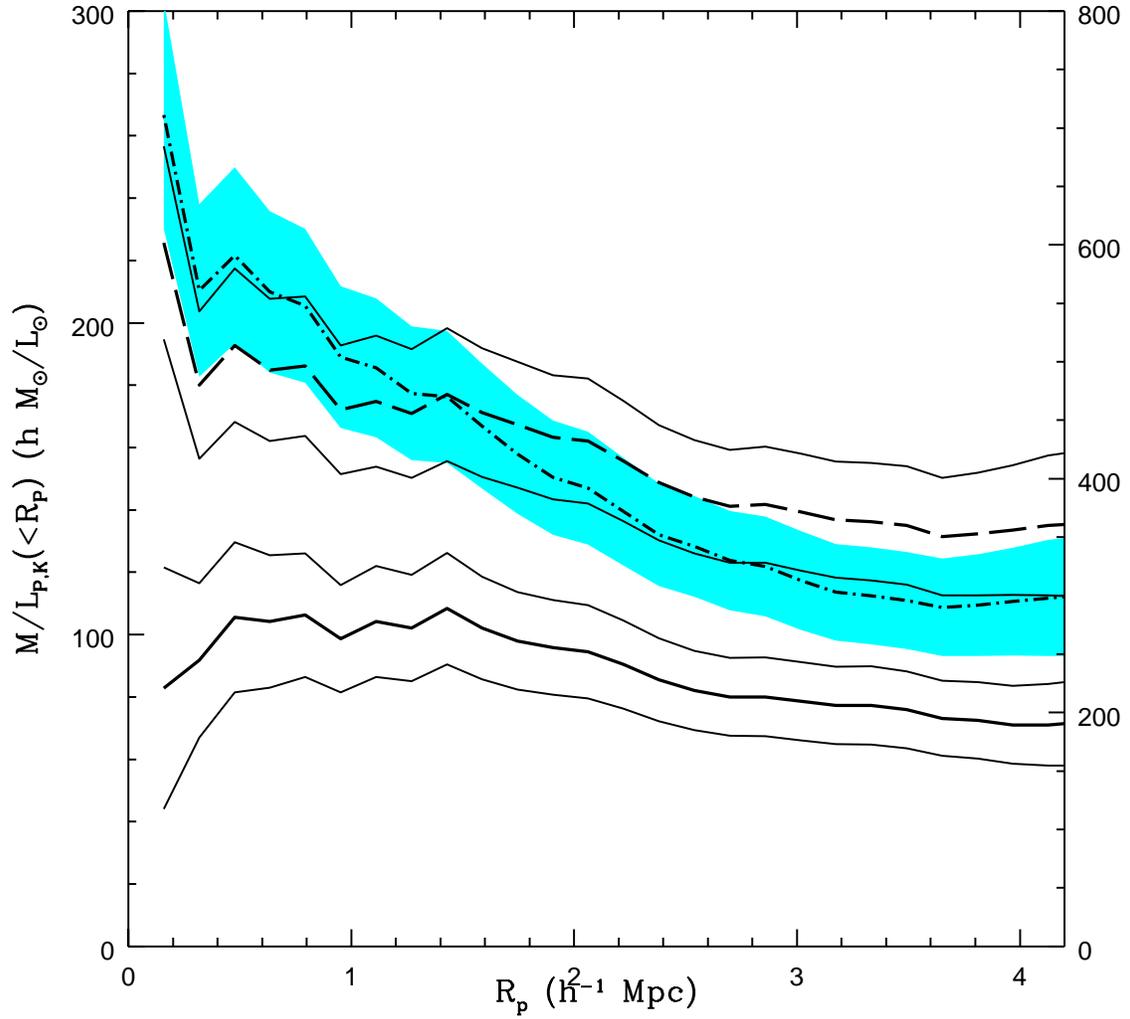}
\caption{Comparison of $K_s$ band and $R$ band mass-to-light profiles
for A576.  The vertical scale on the right is for $R$ band.  The $K_s$
band mass-to-light profile is the lower set of solid lines with 1-$\sigma$
uncertainties.  The dash-dotted line and shaded region show the $R$ band
mass-to-light profile and 1-$\sigma$ uncertainties from
\citet{rines2000}. The dashed line shows the $R$ band
mass-to-light profile and surrounding solid lines show the 1-$\sigma$
uncertainties calculated from bright galaxies assuming a constant
fraction of light in fainter galaxies.}
\end{figure*}

Indeed, the ${K_s}$ band mass-to-light profile (thick solid line in
Figure \ref{mlcomp}) decreases more slowly than the R band profile
(dash-dotted line).  The cumulative mass-to-light ratio decreases by
a factor of $\sim$2 in R band and by a factor of $\sim$1.4 in $K_s$
band.  This result suggests that the effect of star formation
gradients on mass-to-light profiles is significant in the R band as
well as in the B band \citep[e.g.,][]{bahcall2000}.  However, there is
no obvious change in the average $R-{K_s}$ color with radius in A576
($\S 3.3$).  Thus, the steeper decrease in the cumulative
mass-to-light profile in the $R$ band is not readily explained by a
simple color gradient.

Another explanation for the difference in $R$ and $K_s$ band $M/L$
profiles is the corrections for faint galaxies without redshifts.  The
R band catalog has complete spectroscopy to R=16.5 and complete
photometry to R=18.0.  \citet{rines2000} used several techniques to
estimate the (assumed constant) flux surface density contributed by
background galaxies.  Here, by contrast, we correct for faint galaxies
by assuming a universal luminosity function in all environments.
Under this assumption, the total luminosity in all galaxies is simply
a constant factor multiplied by the luminosity contained in bright
galaxies.

Because 2MASS is a shallow survey, it is difficult to estimate
magnitudes (and hence number counts) of galaxies fainter than our
spectroscopic completeness limit \citep[these number counts are
necessary to make a background correction similar to][]{rines2000}.
However, it is interesting that the 2MASS galaxy counts indicate a
steeply rising LF in the outskirts of A576 fainter than the
spectroscopic completeness limit (Figure \ref{klfnout}).  The number
counts therefore suggest that applying a constant background
subtraction to the 2MASS data would lead to better agreement between
the two mass-to-light profiles.  The presence of a background group or
cluster behind A576 \citep{rines2000} shows that the true background
is non-uniform.

Fortunately, it is straightforward to apply the assumption of a
universal luminosity function to the mass-to-light profile in the R
band.  Under this assumption, the two mass-to-light profiles are in
quantitative agreement; the decrease in the mass-to-light ratio
between the inner $1~\Mpc$ and the outskirts is a factor of $\sim$1.4
at both wavelengths.  Figure \ref{mlcomp} shows that the shapes of the
mass-to-light profiles (thick solid and dashed lines) agree well under
this assumption (except at $R_p\lesssim 0.5~\Mpc$ where the photometry
of two bright galaxies is uncertain; see $\S 2.2$).  Thus, the
apparent disagreement between the R-band and $K_s$-band shapes of the
cumulative mass-to-light profiles in A576 is {\it not} due to radial
gradients in the stellar mass-to-light ratio but is simply a result of
using different methods to account for the luminosity contributed by
faint galaxies.  Using consistent methods produces both qualitative
and quantitative agreement in the mass-to-light profiles calculated at
optical and near-infrared wavelengths.  Because the same caustic mass
profile is used for both wavelengths, this comparison tests the
relative shapes of the light profiles calculated at different
wavelengths.  The agreement between the two profiles should therefore
generalize to all clusters with similar galaxy populations; the fact
that A576 has one of the most strongly peaked mass-to-light profiles
(Figure \ref{mlk}) should not affect this generalization.  This result
demonstrates the importance of using consistent corrections for faint
galaxies when comparing mass-to-light ratios or profiles at different
wavelengths and/or for different clusters.  This result also suggests
that a complete census of cluster light requires deep, complete
spectroscopy.

\subsubsection{Estimating the Total-to-Stellar Mass Profiles}

Although we showed above ($\S 3$) that there is no obvious evidence
from near-infrared photometry or $R-K_s$ colors for dramatic changes
in the stellar populations with clustrocentric radius, the degeneracy
between age and metallicity and the weak dependence of near-infrared
colors on these properties might obscure a real gradient.  Here we
estimate the potential size of this effect.



Because the galaxies we sample are relatively bright, the
color-magnitude relation implies that these galaxies should have red
colors.  We estimate the fraction of light in early-type galaxies at
large radii using the type-dependent LFs of K01.  Using their
estimates of the type-dependent LFs, early-type galaxies contribute
roughly half of the total light in bright galaxies ($M_{K_s}\leq
-22.77 + 5\mbox{log}h$) averaged over all environments.  The $K_s$
band mass-to-light ratio is a factor of $1.8\pm0.3$ times larger in
virial regions than in infall regions ($\S 5$).  Thus, the ratio of
total matter to stellar matter (in galaxies) could be roughly constant
on scales up to $\sim 10~\Mpc$ if the $K_s$ band stellar mass-to-light
ratio in early-type galaxies is $\sim$2.6 times larger than in
late-type galaxies.  \citet{bell01} and \citet{bell03} find that the stellar
mass-to-light ratio measured in $K_s$ band varies by no more than a
factor of 2 over a wide range of star formation histories.
Such a large difference would probably produce
significant $R-K_s$ color gradients, contradicting Figure
\ref{rkprofile}.  The age-metallicity degeneracy and/or complicated
star formation histories could conceivably mask these gradients, but
these effects are generally small at $K_s$ band.

A second method of quantifying the changes in stellar mass-to-light
ratios is to use the relation between stellar mass-to-light ratio and
galaxy color \citep{bell01,bell03}.  The range of galaxy colors in the
SDSS is $0.4\lesssim g-r \lesssim 1.0$, although little stellar mass
is contained in the bluest galaxies \citep[e.g.,][]{kauffmann03}.  If
the average galaxy $g-r$ color changed from 0.9 in cluster centers to
0.4 in the outskirts (an extreme assumption), the corresponding change
in stellar mass-to-light ratio is only a factor of 1.3.  A more
realistic estimate is that early-type red galaxies comprise roughly
half the light in bright galaxies.  Then, the average $K_s$ band
stellar mass-to-light ratio in cluster outskirts is at most 20\%
smaller than in cluster centers.  Thus, gradients in stellar
populations do not account for the radially decreasing mass-to-light
ratios of the CAIRNS clusters.

\subsection{Intracluster Light}

Another explanation for the decreasing mass-to-light profiles is that
we do not account for light outside of galaxies.  The existence of
intracluster red giant branch stars \citep{2002ApJ...570..119D},
planetary nebulae
\citep{1998ApJ...492...62C,1998ApJ...503..109F,2002ApJ...570..119D,2003AJ....125..514A,2003ApJS..145...65F},
globular clusters \citep{1995ApJ...453L..77W,2003AJ....125.1642J},
diffuse light
\citep{zwicky1951,1977MNRAS.180..207M,ub91,1995AJ....110.1507B,1998Natur.396..549G,1998MNRAS.293...53T,2000ApJ...536..561G,2002ApJ...575..779F},
and supernovae \citep{1981AJ.....86..998S,2003AJ....125.1087G} not
associated with individual galaxies all suggest that stars are
stripped from cluster galaxies and form diffuse intracluster light
\citep{1999MNRAS.304..465M,2003ApJ...589..752G}.  Numerical
simulations of clusters in $\Lambda$CDM cosmologies show that
processes such as tidal stripping and dynamical friction disrupt
cluster galaxies \citep{kk1999,1999ApJ...523...32C}.

Surveys of the above tracers of intracluster stars indicate that
intracluster light constitutes $\sim$5-50\% of the total light in the
virial regions.  The decreasing mass-to-light profiles found here may
be flat if intracluster light is taken into account.  Thus, the most
secure conclusion we can draw is that the number of galaxies per unit
mass is smaller in cluster virial regions than in infall regions.  The
star formation efficiency could be constant in all environments with
the observed dependence resulting from more efficient galaxy
disruption in environments with larger virial temperatures.  A
prediction of this scenario is that the fraction of intracluster light
should increase with cluster mass.  A similar trend has recently been
noted in simulated clusters \citep{murante04}.

\subsection{Uncertainties in the Caustic Mass Profiles}

D99 used numerical simulations to investigate the systematic
uncertainties in the caustic technique.  D99 found that the
uncertainties in individual cluster profiles are large ($\sim 50\%$)
but unbiased.  We extend this work to observations in Paper I.
Surprisingly, the contrast between the caustic envelope and the
background is larger in the CAIRNS clusters than in the simulations of
D99.  This difference may indicate a mismatch between the cosmological
model used in D99 (standard $\Lambda$CDM) and the true model and/or
deficiencies in the recipe for star formation and galaxy formation
used in the simulations.

D99 anaylze only Coma-size clusters.  It is
possible that the caustic technique is less accurate and/or biased for
less massive halos.  Further high-resolution simulations with
different cosmological models and/or different recipes for star and
galaxy formation may clarify this issue.

The masses obtained with the caustic technique agree very well with
virial masses and X-ray estimates at small radii (Paper I).  Thus, the
mass-to-light ratios at these radii are reasonably secure (if
corrected for projection effects).  The caustic mass profiles in Paper
I agree very well with both NFW and Hernquist models, with each model
providing a slightly better fit on roughly half of the clusters.  The
NFW profile predicts more mass at large radii than the Hernquist
profile, and it produces better fits to halos in CDM simulations
\citep{nfw97}.  The extrapolation of the NFW profile beyond the virial
radius provides a reasonable description of clusters in simulations
\citep{tasitsiomi03}.  We therefore calculate the mass-to-light ratios assuming
that the NFW profiles inside $r_{200}$ extend to $r_t$ and that the
caustic diagrams indicate cluster/infall region membership.  This
calculation yields results similar to those found using the caustic
mass profiles.

Caustics are a good but not perfect indicator of cluster/infall region
membership.  Galaxies outside the caustics are outside the infall
region, but there may be interlopers in the caustic diagram.  In
cluster cores, only $\sim$1\% of galaxies are interlopers
\citep{2000AJ....119.2038V}.  The number of interlopers should increase
at roughly the same rate as the area sampled.  Indeed, detailed
numerical simulations of clusters indicate that the fraction of
interlopers increases with radius \citep{1997ApJ...485...39C}.  To
estimate this effect in Coma, we estimate the contribution of
interlopers to be the background luminosity density times the volume
within the caustics, which scales roughly with the area on the
sky. The luminosity contained in possible interlopers in each radial
bin is less than 10\% of the luminosity in that bin.  Thus, although
the luminosity of interlopers could lead to an overestimate of the
luminosity of infall region members at large radii, the overestimate
is likely $\lesssim$10\%.

Finally, $\S 7.1$ shows that the observed VDPs (a better established
tool of galactic dynamics) differ from those predicted by the galaxy
distributions under the assumption of isotropic orbits.  The above
discussion indicates that the decreasing mass-to-light profiles are
probably not caused by (currently unknown) systematic effects in the
caustic technique.

\subsection{Cosmological Implications}

Two groups have used the Second Incremental Data Release (2IDR) of
2MASS to measure the near-infrared galaxy luminosity function
(K01,C01).  Both groups find acceptable fits with the functional form
proposed by \citet{schechter76}.  Further, their estimates of the
near-infrared luminosity density and the best-fit parameters of the LF
agree within the uncertainties.  Assuming that the cluster luminosity
function is identical to the field luminosity function of C01
(calculated using 2MASS $K_s$ extrapolated magnitudes, the same
magnitude definition used here), the average mass-to-light ratio
within the turnaround radius $(M/L_{K_s})_{tot}$ implies $\Omega_m =
0.18\pm0.04$ (statistical).  If the global value of $M/L_{K_s}$ is
closer to the value in cluster infall regions than the value in
cluster virial regions, the best estimate of $\Omega_m$ is from
$(M/L_{K_s})_{inf}$ (the average mass-to-light ratio between $r_{200}$
and $r_t$), which yields $\Omega_m = 0.13\pm0.03$ (statistical).  Note
that these estimates become $\sim$20\% smaller if we adopt the C01
2IDR Kron magnitude LF with a -0.20 magnitude adjustment for
converting 2IDR Kron magnitudes to 2MASS extrapolated magnitudes (the
best-fit LF parameters for the Kron LF have a fainter $M_{K_s}^*$ and
a shallower faint-end slope).

We explicitly use the luminosity function of C01 (for extrapolated
magnitudes) to estimate the completeness corrections for faint
galaxies.  This constraint means that our results are independent of
the faint-end slope of the luminosity function in clusters.  Provided
the luminosity functions are similar at the bright end (as shown in
$\S 3.1$), the estimate of $\Omega_m$ is independent of the properties
of dwarf galaxies.  Similarly, any unusual systematic effects in the
measured photometric properties of the galaxies are present in both
our sample and that of C01.  Thus, any such effects should cancel out
in the estimate of $\Omega_m$.  In particular, 2MASS misses faint, low
surface brightness (LSB) galaxies \citep{andreon02,bell03}.  Because
these galaxies are missing in both the CAIRNS catalogs and in the
estimates of the field LF, the omission of these galaxies leads to an
overestimate of $M/L_{K_s}$ but does not affect the estimate of
$\Omega_m$.  If these LSB galaxies were substantially more numerous in
low-density environments than in cluster environments, a bias could
result, but \citet{andreon02} shows that these LSB galaxies are
present in clusters.  Furthermore, most of these LSB galaxies are
fainter than the portion of the LF sampled in the CAIRNS 2MASS
catalogs ($M_{K_s}\lesssim -22$).

Some investigators suggest that the local universe is substantially
underdense with respect to the global average density
\citep[e.g.,][and references therein]{busswell03,frith03}.  In
particular, galaxy number counts indicate that the region surveyed by
the 2dFGRS is significantly underdense \citep{frith03}.  The presence
of such an underdensity obviously has important implications for
estimating $\Omega_m$ using the mass-to-light ratio.

\citet{wright01} suggests that the near-infrared luminosity density
estimated by C01 is a factor of 2.3 smaller than the value obtained by
extrapolating the $z$ band Sloan Digital Sky Survey LF using typical
spiral galaxy colors.  However, \citet{blanton03} recently released a
corrected version of the SDSS LF (including evolutionary
K-corrections) which yields different LF parameters and a $z$ band
luminosity density smaller by a factor of 1.29.  From a comparison of
galaxies in both SDSS and 2MASS, \citet{blanton03} find that the mean
difference between the $^{0.1}i$ SDSS band (the notation means that
the bandpass is the rest-frame bandpass of a galaxy at $z=0.1$ as
observed in the SDSS $i$ band, i.e., $^{0.1}i$ is slightly blueward of
$^{0.0}i$, the observed bandpass) and the 2MASS $K_s$ band is
$^{0.1}i-K_s\approx 2.52$.  The $^{0.1}i$ band luminosity density can
then be extrapolated to $K_s$ band (using $M_{\odot,^{0.1}i}=4.58$ and
$M_{\odot,K_s}=3.39$) to obtain $j (K_s) \approx 7.22 \times 10^8 h
L_\odot \mbox{Mpc}^{-3}$.  We thus obtain $\Omega_m \approx 0.14\pm0.05$
using the total (virial plus infall region) mass-to-light ratio
$(M/L_{K_s})_{tot}$, and $\Omega_m \approx 0.10\pm0.03$ using the
mass-to-light ratio $(M/L_{K_s})_{inf}$ in the infall region only.


\citet{huang03}, using a smaller but deeper
survey, suggests that the infrared luminosity density is significantly
larger \citep[but see][who note that this estimate ignores
evolutionary corrections]{bell03}.  Using his luminosity density
yields $\Omega_m = 0.23\pm0.04$ from $M/L_{K_s}(<r_t)$ and $\Omega_m =
0.16\pm0.03$ from $M/L_{K_s}(r_{200}\rightarrow r_t)$.  Including
evolution corrections reduces the estimate of $\Omega_m$ by $\sim$20\%
\citep{bell03}, yielding estimates similar to those for SDSS
and 2dFGRS.  Clearly, the normalization of the
infrared luminosity density is a significant source of uncertainty in
using the cluster mass-to-light ratio to determine the matter density.

Table \ref{omegam} list several recent estimates of $\Omega_m$ from a
variety of techniques.  The estimates in Table \ref{omegam} typically
assume a flat universe dominated by dark energy.  A detailed
discussion of the systematic uncertainties and potential biases in the
various techniques lies outside the scope of this paper.  In general,
estimates of $\Omega_m$ from cluster abundances and dynamics and weak
lensing yield low values; estimates from supernovae and from the
combination of microwave background with large-scale structure yield
higher values of $\Omega_m$.  Our estimates are smaller than the
currently popular value of $\Omega_m \approx 0.27$, but within the
range of estimates from other techniques.  It is curious that our
estimates agree with other estimates based on mass-to-light ratios
both inside and outside of clusters.

However, we find a significantly smaller value than estimates based on
the cluster baryon fraction.  In particular, L03 \citep[see
also][]{mme} calculate the baryon fraction within $r_{500}$ and
estimate $\Omega_m =0.28\pm0.03$ (statistical).  Comparing this
estimate with the mass-to-light ratios of hot clusters in their
sample, they conclude that the mass-to-light ratio in hot clusters
($kT_X\geq 3.7\mbox{keV}$) is a factor of $0.68\pm0.10$ {\it smaller}
than the global value.  This conclusion disagrees with our result that
the mass-to-light inside $r_{200}$ is a factor of $1.8\pm0.3$ {\it
larger} than the mass-to-light ratio outside $r_{200}$ (which should
better approximate the global value).  Note that departures from
hydrostatic equilibrium of intracluster gas due to nonthermal pressure
would aggravate this problem by decreasing the true cluster baryon
fraction \citep[e.g.,][]{sadat01}.  At least two explanations may
account for the discrepancy between L03 and the decreasing
mass-to-light profiles.  First, if baryons in hot gas avoid cluster
centers due to, e.g., shock heating, the baryon fraction within
$r_{500}$ may be smaller than the global value \citep[see the
comparison of many simulations by][]{1999ApJ...525..554F}.  Detailed
observations with {\em ROSAT} and {\em ASCA} showed that the gas mass
fraction increases with radius in some nearby clusters
\citep{djf95,1997ApJ...491..467M,ettori99}, but little data exist
beyond $r_{500}$.  Even with {\em Chandra} and {\em XMM-Newton},
warm/hot gas is presently not observable at large radii because its
temperature and density are too low.  If the baryon fraction continues
to increase outside $r_{500}$, the baryon fraction within $r_{500}$
leads to an overestimate of $\Omega_m$.  Thus, it may be possibile to
reconcile these results, but at present, the baryon fraction outside
$r_{500}$ remains unconstrained by observations.  Second, cluster
infall regions and low-mass clusters (which L03 and $\S 6.2$ show have
smaller mass-to-light ratios than more massive clusters) may provide a
uniquely favorable environment for star formation where the baryon
density is high enough to encourage gravitational collapse but not so
high that virial temperatures prevent collapse.  Under this scenario,
the mass-to-light profile of a cluster would peak in the center,
decrease in the infall region, then rise again to the global value.

The latter explanation is intriguing, and might lead to consistency
with \citet{turner02}, who notes that $\Omega_m = 0.33$ is signicantly
larger than previous determinations based on the mass-to-light ratios
in clusters and concludes that the difference results from variations
in the mass-to-light ratio with environment.  The mass-to-light
profiles presented here disagree with this conclusion both
qualitatively and quantitatively to densities as small as $\approx 3
\rho_c$.  The mass-to-light ratio decreases with radius, and there are
no obvious systematic effects in the caustic technique that can
reconcile these results.  Thus, the scale dependence of the
mass-to-light ratio on scales $\lesssim 10~\Mpc$ cannot be the
resolution of this profound problem.

Some recent simulations suggest that galaxies form preferentially in
overdense regions of the universe
\citep{1999ApJ...522..590B,2003ApJ...597....1O}.  
These simulations imply that the estimates of $\Omega_m$ from the
mass-to-light ratio in cluster virial regions may underestimate the
true value by a factor of $\sim$1.25.  We find instead that the
mass-to-light ratio decreases in cluster infall regions.  It is
possible, however, that the global mass-to-light ratio is
significantly higher than in cluster infall regions and low-mass
clusters.  Although somewhat arbitrary, such a scenario is consistent
with all the constraints found in this paper and other investigations.
Future studies of bulk flows are the most likely candidate to test
this scenario.

\section{Conclusions}

We discuss some of the first estimates of radial variations in
mass-to-light ratios on scales of 1-10$\Mpc$ using near-infrared
photometry from 2MASS and mass profiles from the kinematics of
infalling galaxies.  Because cluster infall regions contain the
transition from cluster galaxies to field galaxies
\citep[][and references
therein]{ellingson01,lewis02,gomez03,treu03,balogh03}, mass-to-light
ratios in infall regions should closely resemble the global value.

To summarize our results:
\begin{itemize}
\item{Infall regions contain more bright galaxies (to a fixed absolute 
magnitude limit) than cluster virial regions.}
\item{The near-infrared luminosity functions for bright galaxies 
($M_{K_s}\lesssim -22 + 5 \mbox{log} h$) in the CAIRNS cluster virial
regions and infall regions do not differ significantly from the field
galaxy luminosity function.  Clusters contain an excess of extremely
bright galaxies above the predictions of a Schechter function.}
\item{Optical-near-infrared colors in A576 show no radial dependence.  
This lack of a color gradient shows that the stellar populations do
not change dramatically with radius.  It is likely that the mild
gradients found in the photometric study of \citet{goto04} may be
enhanced in optically selected samples as compared to near-infrared
selected samples such as CAIRNS.}
\item{Galaxies in cluster virial regions and infall regions exhibit a 
near-infrared color-magnitude relation with a shallower slope than at
optical wavelengths.  These galaxies also exhibit little scatter in
$J-K_s$ colors, indicating that the stellar populations are fairly
homogeneous and that internal dust extinction and/or emission is
important for only a few galaxies. }
\item{Both the surface number density profiles and surface luminosity 
density profiles of CAIRNS members indicate that galaxies and stellar
light are more extended than mass. }
\item{Near-infrared mass-to-light
ratios generally decrease with radius by a factor of $1.8\pm0.3$ in
the infall regions of the CAIRNS clusters.  This result agrees with
previous results based on individual clusters and optical photometry.
The presence of decreasing mass-to-light profiles even at $K_s$ band
suggests that the decrease is not due to changes in stellar
populations.}
\item{Near-infrared mass-to-light ratios calculated at $r_{200}$ using 
caustic mass estimates agree quite well with mass-to-light ratios
calculated at $r_{500}$ from X-ray mass estimates.  This agreement
suggests that the decreasing mass-to-light profiles are not monotonic;
the mass-to-light ratio is roughly constant inside $r_{200}$.}
\item{We derive some of the first constraints on the halo occupation 
function using cluster masses and near-infrared selected galaxy
samples. The number of bright galaxies $N_{200}$ projected within
$R_{200}$ increases as $N_{200} \propto M_{200}^{0.70\pm0.09}$,
significantly shallower than $N_{200}\propto M_{200}$.  Earlier
studies of the halo occupation distribution suggest that a halo
occupation distribution shallower than $N_{200}\propto M_{200}$ is
necessary to reproduce the observed clustering properties of galaxies
\citep[e.g.,][]{berlind02,berlind03}.}
\item{No such non-linear relation is evident between $N_{200}$ and 
$L_{200}$, the $K_s$ band luminosity inside $R_{200}$.  This result
shows that the non-linearity of the halo occupation function is not
driven by variations in the luminosity function.}
\item{More massive virialized halos have larger mass-to-light ratios.  
This result follows logically from the two prior points.  Our $M/L-M$
relation agrees with previous determinations \citep[][L03]{bahcall02}.
These results signify that the efficiency of galaxy formation
decreases (and/or that the efficiency of galaxy disruption increases)
with increasing halo mass and/or virial temperature.}
\item{We investigate possible systematic effects and conclude that dark
matter is more concentrated than stellar mass {\it contained in
galaxies}.  This result could arise either from different clustering
properties of dark matter and baryonic matter or from variations in
the efficiency of converting baryonic matter into galaxies.
The cluster environment seems to be either less
efficient at converting baryons into galaxies or more efficient at
disrupting galaxies than less dense environments.  Such a difference
is predicted by simulations of $\Lambda$CDM cosmologies where
processes such as tidal stripping and dynamical friction disrupt
galaxies in clusters
\citep{kk1999,1999ApJ...523...32C} and supported by observations of
significant numbers of intergalactic stars in clusters.
Alternatively, the heating of the intracluster medium may cut off the
supply of cold material needed to form stars \citep[e.g.,][ and
references therein]{1999ApJ...522..590B,bnm2000}, thus lowering the
star formation efficiency in cluster galaxies.}

\item{Assuming the mass-to-light ratios at large radii are similar to the
global value, we estimate $\Omega_m = 0.10\pm0.03$ (1-$\sigma$
statistical uncertainty) using the SDSS luminosity density with
appropriate color corrections or $\Omega_m = 0.13\pm0.03$ (1-$\sigma$
statistical uncertainty) from the 2dFGRS.  We suggest that the 2dFGRS
and the CfA/SSRS2 surveys sample local underdensities.  Uncertainties
in the luminosity density, especially at infrared wavelengths,
contribute a significant amount of the systematic uncertainty in
estimating $\Omega_m$.  These estimates of $\Omega_m$ are small
compared with other recent estimates from the microwave background,
the galaxy power spectrum, and supernovae.  However, they agree well
with other estimates based on cluster mass-to-light ratios
\citep{cnoc96,cye97,bahcall2000,g2000,bahcall02}, cluster
abundances \citep[][but see Schuecker et
al.~2003]{hiflugcs,bahcall03a} and weak lensing
\citep{kaiserxx,wilson01,hoekstra01,gray02}.  We discuss possible
systematic effects that could cause our result to be anomalously low.
Reconciling these estimates of $\Omega_m$ by invoking bias requires
that the typical value of $M/L_K$ at the smallest densities we probe
$\approx 3 \rho_c$ is a factor of 2-3 smaller than the global value.
For instance, if galaxy formation occurs nearly exclusively above a
density threshold $\delta\sim 10$, the mass-to-light ratios in cluster
outskirts may underestimate the global value. }

\end{itemize}

One promising future direction is to study clusters at moderate
redshifts where weak lensing provides an independent mass estimate
\citep{kneib03}.  Comparing lensing mass profiles to caustic mass
profiles will constrain unknown systematics in both techniques.  For
instance, a sheet of mass of uniform density produces no lensing
signal (the mass-sheet degeneracy), but this mass should be evident
in the galaxy kinematics.  Conversely, foreground and background
structures may produce a weak lensing signal but would not affect the
kinematics of the infall region.

Infall regions are interesting environments for studying the
evolution of galaxy populations.  We show here that infall regions are
important in constraining models of galaxy bias or antibias and
$\Omega_m$.  If other methods yield a precise measurement of
$\Omega_m$, the changes in mass-to-light ratios with environment
provide important clues to the formation and evolution of galaxies.

\acknowledgements

We once again thank the hard work of Perry Berlind and Michael
Calkins, the remote observers at FLWO, and Susan Tokarz, who processed
the spectroscopic data. We thank Andi Mahdavi, Dan Fabricant, Jeff
Kenney, Scott Kenyon, Stefano Andreon, and Ken Nagamine for helpful
discussions.  We thank the referee for many suggestions which improved
the clarity of the paper.  MJG and MJK are supported in part by the
Smithsonian Institution.  We thank the Max-Planck-Institut f\"ur
Astrophysik in Garching for allowing us to use some of their computing
resources.  We thank the entire 2MASS team (in particular J.~Huchra,
M.~Skrutskie, T.~Chester, R.~Cutri, J.~Mader, and S.E.~Schneider).
This publication makes use of data products from 2MASS, a joint
project of the University of Massachusetts and the Infrared Processing
and Analysis Center, funded by NASA and NSF.

\bibliographystyle{apj}
\bibliography{rines}

\begin{table*}[th] \footnotesize
\begin{center}
\caption{\label{ksamples} \sc CAIRNS Near-Infrared Spectroscopic Completeness}
\begin{tabular}{lccccccccc}
\tableline
\tableline
\tablewidth{0pt}
Cluster & $r_{200}$ & $r_{t}$ & $r_{max}$ & $K_{lim}$ & $f_{noz}$ &
 $A_{K_s}$ & $KE_{K_s}(z)$ & $M_{K_s,lim}$ & $f_{L}\tablenotemark{1}$ \\ 
 &  $\Mpc$ &  $\Mpc$ &  $\Mpc$ & mag & & mag & mag & mag &   \\ 
\tableline
A119 & 1.07 & 5.4 & 5.4 & 12.9 & 0.096 & 0.014 & -0.125 & -22.70 & 0.616 \\ 
A168 & 1.09 & 5.5 & 5.5 & 13.1 & 0.092 & 0.013 & -0.127 & -22.50 & 0.664 \\ 
A496 & 0.98 & 4.2 & 4.0 & 12.6 & 0.042 & 0.050 & -0.094 & -22.38 & 0.690 \\ 
A539 & 1.03 & 4.3 & 2.5 & 12.4 & 0.086 & 0.062 & -0.083 & -22.31 & 0.704 \\ 
A576 & 1.42 & 6.0 & 4.3 & 12.9 & 0.036 & 0.028 & -0.110 & -22.38 & 0.690 \\ 
A1367 & 1.18 & 5.2 & 5.2 & 12.3 & 0.033 & 0.008 & -0.062 & -21.75 & 0.801 \\ 
Coma & 1.50 & 7.4 & 7.4 & 12.7 & 0.028 & 0.003 & -0.068 & -21.61 & 0.820 \\ 
A2199 & 1.12 & 5.3 & 4.1 & 13.0 & 0.027 & 0.004 & -0.087 & -21.78 & 0.800 \\ 
\tableline
A194 & 0.69 & 3.3 & 3.3 & 12.2 & 0.044 & 0.015 & -0.052 & -21.42 & 0.844 \\ 
\tableline
\tablenotetext{1}{Assuming a luminosity function with $M_{K_s}^*=-23.77$ and $\alpha=-1.14$ as in the 2dF/2MASS luminosity function for extrapolated $K_s$ magnitudes.}
\end{tabular}
\end{center}
\end{table*}

\begin{table*}[th] \footnotesize
\begin{center}
\caption{\label{lfnfits} \sc Luminosity Function Parameters}
\begin{tabular}{lccc}
\tableline
\tableline
\tablewidth{0pt}
LF & Filter & $M^*$ & $\alpha$ \\ 
\tableline
CAIRNS Virial & $K_s$ & $-24.24^{+0.21}_{-0.26}$ & $-1.35^{+0.15}_{-0.16}$ \\
CAIRNS Infall & $K_s$ & $-23.86^{+0.20}_{-0.21}$ & $-1.23^{+0.17}_{-0.15}$ \\
CAIRNS Total & $K_s$ & $-23.97^{+0.13}_{-0.16}$ & $-1.26^{+0.09}_{-0.12}$ \\
2dF/2MASS Extrap & $K_s$ & $-23.77\pm0.03$ & $-1.14\pm0.05$ \\
2dF/2MASS Kron\tablenotemark{1} & $K_s$ & $-23.64\pm0.03$ & $-0.96\pm0.05$ \\
CfA/2MASS\tablenotemark{2} & $K_s$ & $-23.63\pm0.05$ & $-1.09\pm0.06$ \\
\tableline
CAIRNS Virial & $J$ & $-23.25^{+0.19}_{-0.23}$ & $-1.26^{+0.15}_{-0.14}$ \\
CAIRNS Infall & $J$ & $-22.79^{+0.17}_{-0.18}$ & $-1.12^{+0.16}_{-0.14}$ \\
CAIRNS Total & $J$ & $-23.00^{+0.13}_{-0.13}$ & $-1.20^{+0.11}_{-0.09}$ \\
2dF/2MASS Extrap & $J$ & $-22.70\pm0.02$ & $-1.07\pm0.03$ \\
2dF/2MASS Kron\tablenotemark{3} & $J$ & $-22.56\pm0.02$ & $-0.93\pm0.04$ \\
\tableline
\tablenotetext{1}{We shift the $M_{K_s}^*$ of C01 by -0.20 to convert from Kron magnitudes (from J band Kron magnitudes and $J-K_s$ colors) to total (extrapolated) 2MASS $K_s$ magnitudes (see Figure 5 in C01).}
\tablenotetext{2}{We shift the $M_{K_s}^*$ of K01 by -0.05 to convert from isophotal to Kron magnitudes and then by -0.20 to convert from C01 Kron magnitudes to 2MASS $K_s$ total magnitudes.}
\tablenotetext{3}{We shift the $M_{J}^*$ of C01 by -0.20 to convert from Kron magnitudes  to total (extrapolated) 2MASS $J$ magnitudes.}
\end{tabular}
\end{center}
\end{table*}

\begin{table*}[th] \footnotesize
\begin{center}
\caption{\label{rich} \sc CAIRNS Richness}
\begin{tabular}{lcccccc}
\tableline
\tableline
\tablewidth{0pt}
Cluster & $r_{200}$ & $r_{max}$ & $L_{200}\tablenotemark{1}$ &
 $L_{tot}\tablenotemark{1}$ & $N_{200}\tablenotemark{2}$ & $N_{inf}\tablenotemark{2}$ \\ 
 &  $\Mpc$ &  $\Mpc$ & $10^{12} h^{-2} L_\odot$ & $10^{12} h^{-2} L_\odot$ &  &  \\ 
\tableline
A119 & 1.07 & 5.4 &  6.66 & 13.75 & 47 & 69 \\
A168 & 1.09 & 5.5 &  3.61 & 11.69 & 35 & 83 \\
A496 & 0.98 & 4.0 &  3.94 &  7.86 & 38 & 40 \\
A539 & 1.03 & 2.5 &  3.32 &  5.43 & 31 & 21 \\ 
A576 & 1.42 & 4.3 &  6.39 & 12.35 & 58 & 59 \\
A1367 & 1.18 & 5.2 & 4.44 &  9.75 & 44 & 51 \\
Coma & 1.50 & 7.4 & 10.40 &21.96 & 89 & 107 \\
A2199 & 1.12 & 4.1 & 5.15 & 17.43 & 52 & 112 \\
\tableline
A194 & 0.69 & 3.3 &  1.72 &  4.02 & 16 & 20 \\
\tableline
\tablenotetext{1}{Luminosities are corrected for faint galaxies by dividing by $f_L$ from Table \ref{ksamples}.}
\tablenotetext{2}{Richness is for galaxies brighter than
$M_{K_s}=-22.77 + 5 \mbox{log} h$, equivalent to one magnitude fainter than
$M_{K_s}^*$ for field galaxies.}
\end{tabular}
\end{center}
\end{table*}

\begin{table*}[th] \footnotesize
\begin{center}
\caption{\label{radii} \sc CAIRNS Scale Radii}
\begin{tabular}{lcccccccc}
\tableline
\tableline
\tablewidth{0pt}
Cluster & $r_{200}$ & $r_{max}$ & $a_M$ & $a_N$ & $a_L$ & $a_M$ & $a_N$ & $a_L$   \\ 
 &  $\Mpc$ &  $\Mpc$ & NFW & NFW & NFW & Hern & Hern & Hern  \\ 
\tableline
A119 & 1.07 & 5.4 & 0.17 &0.33 & 0.19 & 0.58& 1.24 & 1.00\\ 
A168 & 1.09 & 5.5 & 0.21 &0.63 & 0.75 & 0.65& 2.17 & 2.50\\ 
A496 & 0.98 & 4.0 & 0.07 &0.22 & 0.13 & 0.31& 0.81 & 0.75\\ 
A539 & 1.03 & 2.5 & 0.07 &0.08 & 0.07 & 0.25& 0.49 & 0.45\\ 
A576 & 1.42 & 4.3 & 0.13 &0.35 & 0.25 & 0.43& 1.22 & 1.12\\ 
A1367 & 1.18 & 5.2 & 0.07 &0.27 & 0.33& 0.29& 1.18 & 1.30\\ 
Coma & 1.50 & 7.4 & 0.15 &0.35 & 0.28& 0.50& 1.40 & 1.32\\ 
A2199 & 1.12 & 4.1 & 0.15 &0.75 & 0.90& 0.47& 2.13 & 2.58\\ 
\tableline
A194 & 0.69 & 3.3 & 0.11 &0.20 & 0.27 & 0.35& 0.93 & 1.14\\ 
\tableline
\end{tabular}
\end{center}
\end{table*}

\begin{table*}[th] \footnotesize
\begin{center}
\caption{\label{mlktable} \sc CAIRNS Mass-to-Light Ratios}
\begin{tabular}{lccccccc}
\tableline
\tableline
\tablewidth{0pt}
Cluster & $r_{200}$ & $r_{max}$ & $L_{200}$ & $L_{tot}$ & ($M/L_{K_s}$)$_{tot}$  & ($M/L_{K_s}$)$_{200}$  & ($M/L_{K_s}$)$_{inf}$  \\ 
 &  $\Mpc$ &  $\Mpc$ & $10^{12} h^{-2} L_\odot$ & $10^{12} h^{-2} L_\odot$ & $h M_\odot/L_\odot$ & $h M_\odot/L_\odot$ & $h M_\odot/L_\odot$ \\ 
\tableline
A119 & 1.07 & 5.4  & 6.66 & 13.75   & 46$\pm 27$ & 43$\pm18$ & 49$\pm 21$ \\
A168 & 1.09 & 5.5  & 3.61 & 11.69   & 56$\pm 16$ & 83$\pm 21$ & 44$\pm 10$ \\
A496 & 0.98 & 4.0  & 3.94 & 7.86   & 38$\pm 14$ & 56$\pm 15$ & 21$\pm 6$ \\
A539 & 1.03 & 2.5  & 3.32 5.43  & 59$\pm 25$ & 76$\pm 31$ & 31$\pm 9$ \\ 
A576 & 1.42 & 4.3  & 6.39 & 12.35  & 71$\pm 14$ & 104$\pm 19$ & 36$\pm 7$ \\
A1367 & 1.18 & 5.2 & 4.44 & 9.75  & 57$\pm 18$ & 86$\pm 24$ & 34$\pm 8$ \\
Coma & 1.50 & 7.4  & 10.40 & 21.96 & 75$\pm 21$ & 75$\pm 12$ & 75$\pm 20$ \\
A2199 & 1.12 & 4.1 & 5.15 & 17.43 & 33$\pm 11$ & 63$\pm 17$ & 21$\pm 5$ \\
\tableline
A194 & 0.69 & 3.3  & 1.72 & 4.02  & 37$\pm 28$ & 44$\pm 24$ & 32$\pm 19$ \\
\tableline
\end{tabular}
\end{center}
\end{table*}

\begin{table*}[th] \footnotesize
\begin{center}
\caption{\label{omegam} \sc Estimates of $\Omega_m$}
\begin{tabular}{lccc}
\tableline
\tableline
\tablewidth{0pt}
Technique & $\Omega_m$ &  Reference \\ 
\tableline
CAIRNS Total M/L + 2dF & $0.18\pm0.04$ & -- \\
CAIRNS Virial M/L + 2dF & $0.24\pm0.05$ & -- \\
CAIRNS Infall M/L + 2dF & $0.13\pm0.03$ & -- \\
\tableline
CAIRNS Total M/L + SDSS & $0.14\pm0.05$ & -- \\
CAIRNS Virial M/L + SDSS & $0.18\pm0.06$ & -- \\
CAIRNS Infall M/L + SDSS & $0.10\pm0.03$ & -- \\
\tableline
WMAP + 2dF & $0.27\pm0.04$ & 1 \\
WMAP + SDSS & $0.30\pm0.04$ & 2 \\
WMAP + Other CMB & $0.1-0.5$ (95\%) & 3 \\
Type Ia SNe & $0.28\pm0.05$ & 4 \\
Type Ia SNe & $0.25^{+0.07}_{-0.06}\mbox{(stat)}\pm0.04\mbox{(sys)}$ & 5 \\
Cluster Abundance & $0.12^{+0.07}_{-0.06}$ & 6 \\
Cluster Abundance & $0.17\pm0.05$ & 7,8 \\
Clus.~Abun.~+Clustering & $0.34\pm0.03\mbox{(stat)}\pm0.09\mbox{(sys)}$ & 9 \\
Weak Lensing (Groups) & $0.13\pm0.07$ & 10 \\
Weak Lensing (Superclusters) & $\lesssim 0.1$ & 11,12 \\
Weak Lensing (Blank) & $\approx 0.1$ & 13 \\
Gas Mass Fraction & $0.28\pm0.03$ & 14,15 \\
CMB + Power Spectrum + BBN + $f_g$ & $0.33\pm0.04$ & 16 \\
\tableline
\tablerefs{
(1) \citet{spergel03}; (2) \citet{tegmark03}; (3) \citet{bridle03}; (4) \citet{tonry03}; (5) \citet{knop03}; (6) \citet{hiflugcs}; (7) \citet{bahcall03a}; (8) \citet{bahcall03b}; (9) \citet{schuecker03}; (10) \citet{hoekstra01}; (11) \citet{kaiserxx}; (12) \citet{gray02}; (13) \citet{wilson01}; (14) \citet{ettori99}; (15) L03; (16)\citet{turner02}.}
\end{tabular}
\end{center}
\end{table*}

\clearpage
 
\clearpage
 
\clearpage

\clearpage

\end{document}